\documentclass[12pt]{article}
\usepackage{verbatim}
\usepackage{listings}
\usepackage{graphicx} 
\usepackage{array}
\usepackage{hyperref} 
\usepackage{graphicx}
\usepackage{amsmath,amssymb,amsthm}
\usepackage{rotating}
\usepackage{amssymb}
\usepackage{rotating}

\def\ges{$^{76}$Ge }
\def\xe{$^{136}$Xe }
\def\xen{$^{136}$Xe}
\def\tect{$^{130}$Te }

\def\bidq{$^{214}$Bi }
\def\udt{$^{238}$U }
\def\thdt{$^{232}$Th }
\def\tld{$^{208}$Tl }

\def\cs{$^{60}$Co }
\def\cpkky{counts/(keV$\cdot$kg$\cdot$yr)}
\def\cpkty{counts/(keV$\cdot$tonne$\cdot$yr)}
\def\tectn{$^{130}$Te}

\def\mel{$m_{\mathrm{e}}$~}

\def\mnu{$\langle m_{\mathrm{\nu}} \rangle$}
\def\amnu{$\vert \langle m_{\mathrm{\nu}} \rangle \vert$~}
\def\amnun{$\vert \langle m_{\mathrm{\nu}} \rangle \vert$}

\def\BBz{$\beta\beta(0\nu)$~}
\def\BBzn{$\beta\beta(0\nu)$}

\def\BBmn{$\beta\beta(0\nu,\chi)$}
\def\BBd{$\beta\beta(2\nu)$~}
\def\BBdn{$\beta\beta(2\nu)$}
\def\BB{$\beta\beta$~}
\def\BBn{$\beta\beta$}

\def\QBB{$Q_{\beta\beta}$~}
\def\Sz{$S^{0\nu}$~}
\def\SFz{$F^{0\nu}_{68 \% C.L.}$~}
\def\SFzz{$F^{0\nu~ZB}_{68 \% C.L.}$~}
\def\TH{$\tau_{1/2}$~}
\def\Tz{$T^{0\nu}_{1/2}$~}
\def\Td{$T^{2\nu}_{1/2}$~}

\def\ca{$\sim$}

\def\dot{$\cdot$}
\def\pom{$\pm$ }
\def\gm{$\gamma$}

\def\teod{TeO$_2$~}

\def\be{\begin{equation}}
\def\ee{\end{equation}}

\def\per{$\times$}
\def\al{$\alpha$}

\def\z+{0$^{+}$}
\def\d+{2$^{+}$}
\def\gt{$>$}
\def\Mbb{M$_{\beta\beta}$} 
\def\Abb{A$_{\beta\beta}$}

\def\FZ{\ref{eq:sensitivity}}
\def\FZ0{\ref{eq:0sensitivity}}
\def\dms{$\delta m^2$}
\def\dma{$\Delta m^2$}

\DeclareGraphicsExtensions{.pdf,.eps,.png,.jpg,.mps,.pdf}
\input macro.def    %<-useful dbd macros
\input epsf.tex    %<-If you need EPS figures to be
\input epsf.def   %<-If you need EPS figures to be
\begin{document}

\title{Challenges in Double Beta Decay}

\author{
Oliviero Cremonesi \\
{\it INFN - Sezione di Milano Bicocca, Milano I-20126 - Italy} \\
Maura Pavan \\
{\it Dipartimento di Fisica, Universit\`a di Milano-Bicocca and} \\
{\it INFN - Sezione di Milano Bicocca, Milano I-20126 - Italy} }

\maketitle

\begin{abstract}
After nearly 80 years since the first guess on its existence, neutrino still escapes our insight: the mass and the true nature (Majorana or Dirac) of this particle is still unknown. In the past ten years, neutrino oscillation experiments have finally provided the incontrovertible evidence that neutrinos mix and have finite masses. These results represent the strongest demonstration that the Standard Model of electroweak interactions is incomplete and that new Physics beyond it must exist. None of these experimental efforts could however shade light on some of the basic features of neutrinos. 
Indeed, absolute scale and ordering of the masses of the three generations as well as charge conjugation and lepton number conservation properties are still unknown. 
In this scenario, a unique role is played by the Neutrinoless Double Beta Decay searches: these experiments can probe lepton number conservation, investigate the Dirac/Majorana nature of the neutrinos and their absolute mass scale (hierarchy problem) with unprecedented sensitivity. 
Today Neutrinoless Double Beta Decay faces a new era where large scale experiments with a sensitivity approaching the so-called degenerate-hierarchy region are nearly ready to start and where the challenge for the next future is the construction of detectors characterized by a tonne-scale size and an incredibly low background, to fully probe the inverted-hierarchy region. 
A number of new proposed projects took up this challenge. These are based either on large expansions of the present experiments or on new ideas to improve the technical performance and/or reduce the background contributions. 
On the other hand, nuclear theorists are making remarkable progress in the calculation of the double beta decay nuclear matrix elements in order to eliminate the theoretical uncertainties affecting the particle physics interpretation of this process. 
In this paper, a review of the most relevant ongoing experiments is given. The most relevant parameters contributing to the experimental sensitivity are discussed and a critical comparison of the future projects is proposed. 

\end{abstract}

%%%%%%%%% Introduction %%%%%%%%%%%%%%%%%%%%%%%%%%
\section{Introduction}
\label{sec:intro}
First suggested by M. Goeppert-Mayer in 1935, Double Beta Decay (DBD or \BBn) is a rare spontaneous nuclear transition in which an initial nucleus (A,Z) decays to a member (A,Z+2) of the same isobaric multiplet with the simultaneous emission of two electrons. 
Unfortunately also the equivalent sequence of two single beta decays can produce the same result and -- in experimental investigations -- the choice of the parent nuclei is therefore generally restricted to the nuclei which are more bounded than the intermediate ones. Because of the pairing term, such a condition is fulfilled in Nature for a number of even-even nuclei. 
The decay can then proceed both to the ground state or to the first excited states of the daughter nucleus. 
Double beta transitions accompanied by positron emission or electron capture are also possible. However they are usually characterized by lower transition energies and poorer experimental sensitivities. 
Different \BB decay modes are possible. Among them, two are of particular interest: the 2$\nu$
\footnote{
The neutrinos emitted in all \BB decays are electron neutrinos. It is generally understood, that where not explicitly indicated, ``$\nu$'' indicates an electron neutrino. We will follow such a convention everywhere in the text.
}
 mode (\BBdn) 
\begin{equation}\label{eq:bb2nu}
^A_ZX \to ^A_{Z+2}X + 2e^- + 2\overline{\nu_e}
\end{equation}
which obeys lepton number conservation and is allowed in the framework of the Standard Model (SM) of electro-weak interactions, and the 0$\nu$ mode (\BBzn) 
\begin{equation}\label{eq:bb0nu}
^A_ZX \to ^A_{Z+2}X + 2e^-
\end{equation}
which violates the lepton number by two units and occurs if neutrinos are their own antiparticles (i.e. the neutrino is a Majorana particle).
A third decay mode (\BBmn) in which one or more neutral bosons $\chi$ (Majorons) are emitted
\begin{equation}\label{eq:bbchi}
^A_ZX \to ^A_{Z+2}X + 2e^- + N\chi
\end{equation}
is also often considered. The interest in this decay is mainly related to the existence of Majorons, massless Goldstone bosons that arise upon a global breakdown of B--L symmetry~\cite{mohapatra81}.

From the point of view of Particle Physics, \BBz is of course the most interesting of the \BB decay modes for its important theoretical implications. 
In fact, after 80 years from its introduction~\cite{racah37,furry39},  \BBz is still the only practical way to probe experimentally missing neutrino properties like mass and nature. Indeed, it can exist only if neutrinos are Majorana particles and it can provide unique constraints on the neutrino mass scale. 
Furthermore, \BBz observation would prove that total lepton number is not conserved in physical phenomena, an observation that could be linked to the cosmic asymmetry between matter and antimatter (baryogenesis via leptogenesis \cite{fukugita86,kuzmin85,giudice04,bari05}). 

In addition to a theoretical prejudice in favor of Majorana neutrinos, there are other reasons to hope that experimental observation of \BBz is at hand. In particular the results of oscillation experiments which have demonstrated that neutrinos are massive particles. 
Although these results cannot provide a firm prediction for \BBz rates, they suggest that favorable conditions for its observation may be realized in Nature and have enormously increased the interest toward the experimental search for this decay. 
It should also be stressed that \BBz could have been already observed. Indeed, an extremely intriguing and debated claim for \BBz observation in $^{76}$Ge is awaiting unambiguous confirmation by upcoming experiments.

The important implications of massive Majorana neutrinos and the possible experimental observation of \BBz have triggered a new generation of experiments spanning a variety of candidate isotopes with different experimental techniques, all aiming at reaching a sensitivity allowing to test the region of neutrino masses indicated by neutrino oscillation experiments. Experimental techniques range from the well-established germanium calorimeters, to xenon time projection chambers and low temperature calorimeters. Some of the experiments are already running or will run very soon. Others are still in their R\&D phase, trying to reach the limit of their experimental  technique. 

In all cases, the common claim is of being sensitive to very light neutrino masses by assuming an improvement of one to three orders of magnitude in term of background suppression, detector performance or increase of the target mass. 

In this paper we review the state-of-the-art of this rapidly changing field. 
In Section \ref{sec:neutrinos} we summarize the general status of neutrino phenomenology while in Section \ref{sec:dbd} we analyze the case of \BBzn. Section \ref{sec:nme} is devoted to the nuclear part of the problem, the calculation of the transition probabilities (or Nuclear Matrix Elements, NME). In Sections \ref{sec:expovv} and \ref{sec:expmet} the most important experimental aspects are described. 
In Section \ref{sec:exppas} we summarize the results of previous experiments. In Section \ref{sec:expfut} we introduce the challenging aspects of present and future projects while in the following Sections we review and compare them.
Our conclusions are summarized in Section \ref{sec:sum} 
%---------------------------------------------------------
\section{Neutrinos}
\label{sec:neutrinos}
Today, we know there are three generations of neutrinos, distinguished by their leptonic flavor. 
%These are the only not-sterile neutrinos with masses lower than the Z$^0$ mass. 
These are the only known neutrinos with mass lower than the Z$^0$ mass which interact with matter via the exchange of W$^{\pm}$ or Z$^0$ bosons (``active'' neutrinos).
A number of experiments in the past 20 years have monitored intense neutrino sources (solar, atmospheric, reactor and accelerator neutrinos) and have reported the observation of neutrino flavor conversion during propagation (neutrino oscillations and Mikheyev-Smirnov-Wolfenstein (MSW) effect), either in terms of neutrino disappearance or in terms of the appearance of a \emph{wrong} neutrino flavor. This phenomenon has its natural explanation when assuming that neutrinos are massive particles and mixing among mass eigenstates is assumed, which implies the need to modify or better \emph{extend} the Standard Electroweak Model to include massive neutrinos.

Massive neutrino phenomenology (see for example~\cite{zuber,giunti,strum08,petcov2013}) is described in the framework of three distinguishable particles provided with their own leptonic number, flavor and mass eigenvalue. As for the quark sector, a not diagonal matrix -- the Pontecorvo-Maki-Nakagawa-Sakata matrix (PMNS) -- describes the mixing of neutrinos. The PMNS matrix, in its most general case, is parametrized by 3 angles ($\theta_{12}$, $\theta_{23}$ and $\theta_{13}$) and 3 CP-violating phases ($\delta$, $\lambda_2$, and $\lambda_3$) for a total of 6 parameters to be added to the 3 unknown values of the neutrino masse eigenstates ($m_i$). The PMNS matrix can be expressed as:

\begin{center}
\begin{equation}
{U_{i,j}}=\left(
\begin{array}{ccc}
c_{12}c_{13}	& s_{12}c_{13} 	 & s_{13}e^{-i\delta} \\
-s_{12}c_{23}-c_{12}s_{23}s_{13}e^{i\delta} 	& c_{12}c_{23}-s_{12}s_{23}s_{13}e^{i\delta}	 & s_{23}c_{13} \\
s_{12}s_{23}-c_{12}c_{23}s_{13}e^{i\delta}		& -c_{12}s_{23}-s_{12}c_{23}s_{13}e^{i\delta}	 & c_{23}c_{13}
\end{array}
\right)
\times \rm{diag} (1, e^{i\alpha_1}, e^{i\alpha_2})
\label{eq:PMNS}
\end{equation}
\end{center} 

\noindent where $c_{ij} \equiv \cos\theta_{ij}$ and $s_{ij} \equiv \sin\theta_{ij}$. When neutrinos are Dirac particles, the two Majorana phases can be re-absorbed by a rephasing of the neutrino fields and the PNMS matrix has therefore only 4 free parameters.

Neutrino oscillation probabilities are described in terms of the PNMS angles and of the square mass differences $(m_i^2-m_j^2)$ of the three eigenstates.
The results from oscillation experiments (see for example~\cite{fogli2012} and references therein) constrain neutrino square mass differences and most of the PMNS mixing parameters within rather narrow bands (Table~\ref{tab:nmasses}).
In particular, the measured square mass differences prove that one neutrino state is much more split than the other two. This allows three different mass orderings: direct hierarchy ($m_1 \lesssim m_2 \ll m_3$, \dma$>$0), inverted hierarchy ($m_3\ll m_1\lesssim m_2$, \dma$<$0) and degenerate hierarchy ($m_1 \simeq m_2 \simeq m_3$)~\cite{vissani99,bilenky99,klapdor01,matsuda01,pascoli02,pascoli02a,feruglio02}. 

Only two of the three possible square mass differences are independent and presently constrained. 
These are \dms,~generally labeled as the \emph{solar} term and \dma, the \emph{atmospheric} one (see Table~\ref{tab:nmasses} for their definition). The only parameters irrelevant for oscillations are the Majorana phases $\alpha_1$ and $\alpha_2$ In fact, as pointed out above, they are strictly related to the possible Majorana nature of the neutrinos and appear only in phenomena where such a condition is essential.
Table~\ref{tab:nmasses} summarizes the present status of our knowledge about PNMS matrix elements and neutrino mass split. 

Few experimental results cannot be accommodated in this framework: the LSND \emph{anomaly} \cite{LSND} (further investigated by MiniBoone \cite{MiniBoone}) as well as a possible neutrino deficit observed in reactor~\cite{reactordeficit} and Gallium measurements with very intense (Mci) radioactive neutrino sources~\cite{sourcedeficit}. If confirmed, these could prove the existence of sterile neutrinos. These interact with ordinary matter only through gravitation and can be observed only indirectly in oscillation experiments if they mix with active neutrinos.

\begin{table}
\begin{center}
\label{tab:nmasses}
\caption{Summary table of $\nu$ properties from~\cite{fogli2012}. We use the convention where $m_2>m_1$ (therefore \dms$>$0 by construction) and $m_3$ is the most split state. We report here the 1$\sigma$ range for each parameter (note that in the case of $\theta_{23}$ we report a different range for normal  (NH) or inverted hierarchy (IH)).}
{\begin{tabular}{@{}lc@{}}
\hline
$sin^2\theta_{12} $ &   (0.29 - 0.32) \\
$sin^2\theta_{23} $ &  (0.037 - 0.0041) for NH or (0.037 - 0.0043) for IH \\
$sin^2\theta_{13} $ &   (0.022 - 0.027) \\
%mass eigenstates & m$_1$, m$_2$, m$_3$ \\
$\delta m^2 = (m_2^2 - m_1^2) $ & (7.3-7.8) $\times$10$^{-5}$ [eV$^2$]\\
$ |\Delta m^2|=|m_3^2 - (m_1^2+m_2^2)/2|$ &  (2.3-2.5)$\times$10$^{-3}$ [eV$^2$]\\
\hline
\end{tabular}}
\end{center}
\end{table}

The challenge of next generation oscillation experiments is to be able to measure the sign of \dma~ and therefore fix the neutrino mass hierarchy problem~\cite{measnuhierarchy}.

\begin{figure}
\begin{center}
\includegraphics[width=1\textwidth]{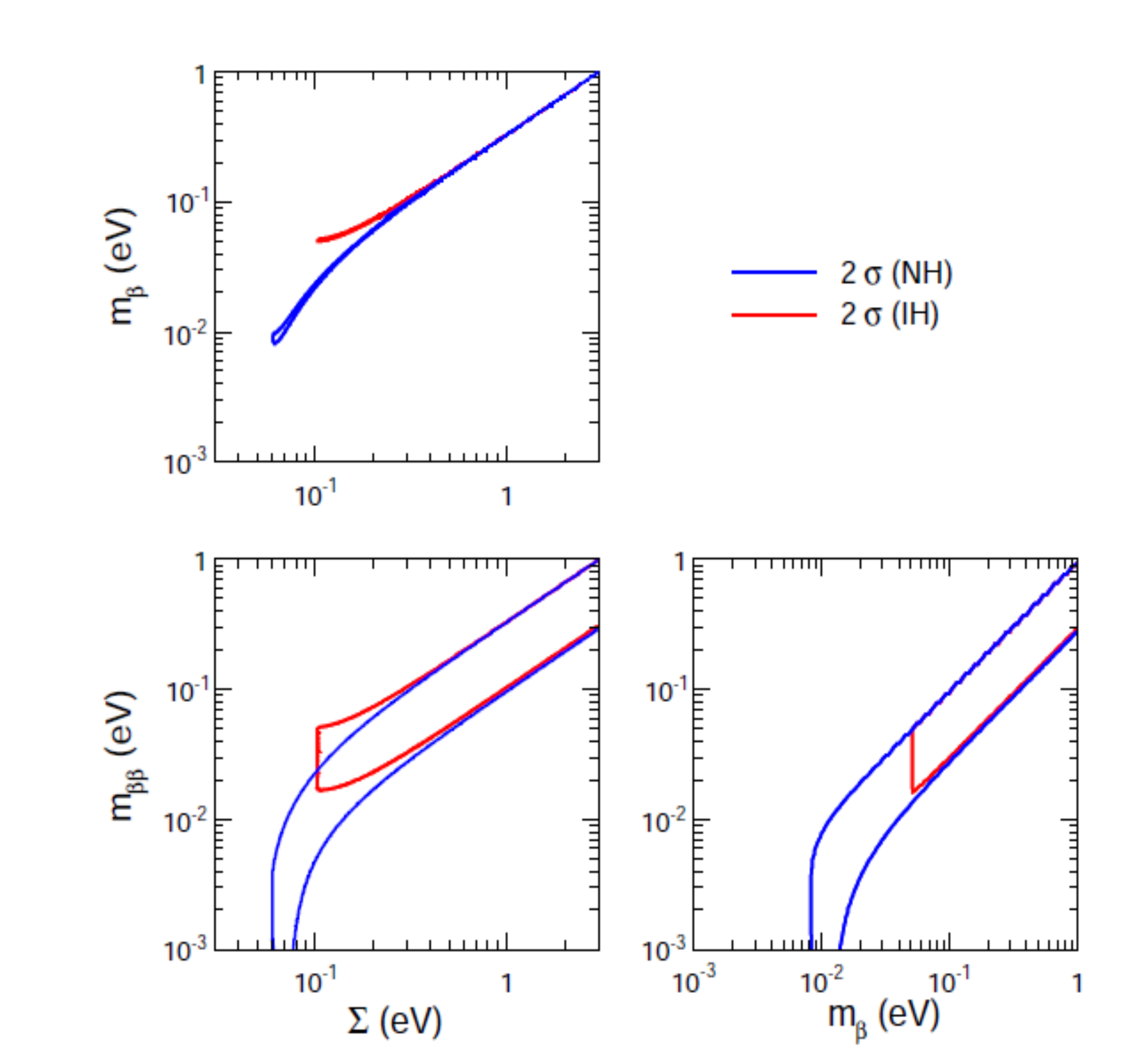}
\end{center}
\caption{Constraints induced by oscillation data (at 2$\sigma$ level) in the planes charted by any two among the absolute mass observables. Blue (red) bands refer to normal (inverted) hierarchy. Figure from G. L. Fogli et al.~\cite{fogli2012}. Here \amnu is indicated as $m_{\beta\beta}$.} 
\label{fig:allowedmass}
\end{figure}

Although the hierarchy can be accessible by oscillation experiments, nevertheless they will not be able to provide information on the absolute scale of neutrino masses which is presently only constrained by experimental measurements of the following three parameters: 
\begin{enumerate}
\item $\sum$ = $\sum m_i$ (Cosmology);
\item $m_{\beta}$ = $\sqrt{(\sum|U_{1i}|^2 m_i^2)}$ (Beta Decay);
\item \amnu $ = \vert \sum U_{1i}^2 m_i  \vert$  (Neutrinoless Double Beta Decay)
\end{enumerate}
These three parameters are strictly correlated among each other and bounded by oscillation results within well defined regions shown in Fig~\ref{fig:allowedmass}. In particular in the case of $\Sigma$ and of $m_{\beta}$ lower bounds of $\sim$0.04 and $\sim$0.008 eV, respectively, are obtained. In the case of \amnu (also called neutrino \emph{Majorana mass}) cancellations among the complex terms of the mass combination are always possible and consequently \amnu has no lower bound.

Upper limits on $\Sigma$ are derived from astronomical observations by fitting the experimental data to complex cosmological and astrophysical models.
Actually, cosmological neutrinos (i.e. neutrinos produced just after the Big Bang) influence the evolution of the Universe and the Large Scale Structures (LSS) formation in a way that is strictly dependent on the size of $\Sigma$, with effects on astrophysical observables such as the anisotropies of the Cosmic Microwave Background (CMB) or the power spectrum of mass-density fluctuations. Despite their increasing sensitivity, cosmological bounds on neutrino masses are considered with caution since they are (strongly) model dependent. The most recent result in this field comes from the Planck collaboration~\cite{Planck2013} and yields a 1$\sigma$ upper limit on $\Sigma$ ranging from about 1 eV to 0.23 eV depending on the set of data and models used in the computation. 

The study of the end point in the beta decay Kurie plot provides a straightforward and direct technique to measure $m_{\beta}$. Present experimental results come from Tritium experiments providing an upper bound on $m_{\beta}$ of 2 eV at 95\% C.L.~\cite{misureKurieplot}. This bound will be improved in the next future by the KATRIN spectrometer~\cite{KATRIN} that aims at reaching a sensitivity of the order of $\sim$0.2 eV. KATRIN is considered as the final step in the use of spectrometers for beta decay measurements, while new ideas and project are emerging in the case of calorimetric measurements of the beta spectrum~\cite{olmio}.

\section{Neutrinoless Double Beta Decay}
\label{sec:dbd}
The neutrinoless mode of nuclear double beta decay (Eq.~(\ref{eq:bb0nu})) is a hypothetical, very rare transition in which two neutrons undergo $\beta$-decay simultaneously without the emission of neutrinos. It was immediately recognized as a powerful method to test Majorana's theory with neutrinos. Indeed, it can be derived from the \BBd mode assuming a Racah sequence of two single beta decays in which the (anti) neutrino emitted at one vertex is absorbed at the other. This is only possible if neutrino and anti-neutrino coincide, i.e. are Majorana particles. In contrast to the two-neutrino mode, it violates total lepton number conservation and is therefore forbidden in the Standard Model. Its existence is linked to that of Majorana neutrinos even though a variety of exotic models can account for it. So far, no convincing experimental evidence of this decay has been found.

When mediated by the exchange of a light virtual neutrino, the \BBz rate is expressed as:

\begin{equation}\label{eq:bbrate}
[ T_{1/2}^{0\nu}]^{-1} = 
G^{0\nu}~|M^{0\nu}|^2~{\vert\langle m_{\nu} \rangle\vert^2}/m_{\mathrm{e}}^2
\end{equation}

\noindent where $G^{0\nu}$ is the phase space integral (exactly calculable, but affected by the uncertainties on the axial coupling constant, as discussed in the next section), $|M^{0\nu}|^2$ is the nuclear matrix element and \mel is the electron mass. Finally, \mnu -- introduced in the previous section -- is the so called  \emph{Majorana mass} of the neutrino that can be expressed in terms of the PNMS matrix elements as:
\begin{equation}
\langle m_{\nu} \rangle =
c_{12}^2 c_{13}^2 m_1+s_{12}^2 c_{13}^2 e^{i\alpha_1} m_2+s_{13}^2 e^{i\alpha_2}m_3
\end{equation}\label{eq:mnueff2}

As evident from Fig.~\ref{fig:allowedmass} oscillation results constrain \amnu to be between 20 and 50~meV in the case of inverted hierarchy (above \ca 50 meV the bands representing the two hierarchies merge in the same \emph{degenerate} band). This is more or less the sensitivity range of forthcoming \BBz experiments. If these would not observe any decay (and assuming that neutrinos are Majorana particles) the inverse ordering could finally be excluded thus fixing the problem of the neutrino absolute mass scale~\cite{strum08,petco05}. If, on the other hand, other experiments would demonstrate that neutrino mass ordering is inverted, then \BBz non observation would demonstrate that neutrinos are Dirac particles.

\amnu is the only experimental observable presently studied where Majorana phases appear explicitly, these phases measure CP violation for Majorana neutrinos (if CP is conserved they are integer multiples of $\pi$). Their presence implies that cancellations are possible (see Fig.~\ref{fig:allowedmass}).
In principle Majorana phases can have measurable consequences even if in practice their determination is very difficult. Many authors have examined the potential to combine \BBz measurements with single beta and cosmology results to determine their value~\cite{pascoli02,pascoli02a,sugiyama03,abada03}. The general conclusion is that at least two experiments that depend on the phases are required to unambiguously determine both. Moreover, a significant improvement in the precision of nuclear matrix elements (Sec.~\ref{sec:nme}) is also required.

\amnu is also the only \BBz measurable parameter containing direct information on the neutrino mass scale. Unfortunately its derivation from the experimental results on \BBz half-lifetimes requires a precise knowledge of the transition nuclear matrix elements M$^{0\nu}$ appearing in Eq.~(\ref{eq:bbrate}). 
Many evaluations are available in the literature, but they are often in considerable disagreement,
leading to large uncertainty ranges for \amnun. This has been recognized as a critical problem by the \BB community. 

Neutrinoless double beta decay is presently the only practical way to discover if the neutrino is its own anti-particle. Its observation would have dramatic consequences for nuclear and particle physics as well as for astrophysics and cosmology. Indeed, one of the most intriguing problem in accommodating massive neutrinos in a Standard Model extension is to be able to explain the smallness of neutrino masses. The see-saw mechanism -- which predicts the existence of Majorana neutrinos -- is a very attractive solution which could also provide an explanation for one of the biggest cosmological puzzle, that of the 
the observed matter-antimatter asymmetry of the Universe (via the leptogenesis mechanism~\cite{fukugita86,kuzmin85,giudice04,bari05}).

Lepton number violation and Majorana neutrinos are the distinctive features of \BBzn, and they represent the primary mission of upcoming \BBz experiments. However, the exchange of a light massive Majorana neutrino is not the only mechanism able to account for \BBzn. Actually many extensions of the standard model include mechanisms that can explain it.  This is the case, for example, of the L/R symmetric GUT's with the exchange of right-handed W-bosons, or of SUSY models with R-parity violation. In all the cases however, the observation of \BBz is irremediably linked to the Majorana nature of neutrinos~\cite{simkovic99}.

A possibility to distinguish between different mechanisms could consist in the analysis of the energy and angular distributions of the emitted electrons  and the study of the transitions to the ground and excited states. 
Unfortunately, the study of the single electron distributions is possible only for a very limited number of experimental techniques. Moreover, in most cases the decay is mediated by the exchange of heavy particles which give rise to similar terms and produce, in particular, the same single electron distributions.

The measurement of the transitions to different final states in the same nucleus seems then the only viable solution~\cite{simkovic02}, taking advantage of the different nuclear matrix elements that enter the decay amplitudes. This requires an accurate calculation of all the nuclear matrix elements, a goal still far from being reached.

Constraints coming from other experiments that study extensions of the Standard Model can of course provide some help. This is the case for example of the LHC measurements on supersymmetric particles which will limit the parameter space reducing the number of possible contributions.

% contains a short intro to bb decay. both 0nu and 2nu

%%%%%%%%% NME %%%%%%%%%%%%%%%%%%%%%%%%%%
\subsection{Nuclear Matrix Elements}
\label{sec:nme}
%--- Introductory notes
The most relevant parameter available from \BBz is the effective neutrino mass \amnun.
According to Eq.~(\ref{eq:bbrate}) it can be obtained from the measured half-lifetime once all the other terms appearing in the equation are known. 
This requires a precise knowledge of the phase space factor G$^{0\nu}$ and of the Nuclear Matrix Elements (NME) M$^{0\nu}$ which cannot be separately measured and therefore can only be evaluated theoretically. 

While precise calculations of the phase space factors have been carried out by many authors~\cite{simkovic99,pantis96,kotila12}, only approximate estimates of the NME's have been so far obtained, due to the many-body nature of the nuclear problem. 
NME's include all the nuclear structure effects of the decay and are indispensable not only to extract the value of \amnu but also to compare the sensitivities and the results of the experiments based on different nuclei.
 
In this respect, it should be stressed that uncertainties on NME's and on the experimental value of the decay half-lifetime concur in the same way to the uncertainty on \amnun. Comparable efforts should be therefore addressed for both aspects of the problem.

A lot of work has been actually devoted in the last decade to develop a proper many-body technique.
Indeed, the calculation of \BBz NME's has been carried out by many authors using different methods: the Quasi-particle Random Phase Approximation~\cite{kortelainen07,kortelainen07a,rodin07} (QPRPA, RQPRPA, pnQPRPA etc.), the Nuclear Shell Model~\cite{menendez09,menendez09a} (NSM), the Interacting Boson Model~\cite{ibm2} (IBM), the Generating Coordinate Method~\cite{rodriguez10} (GCM), and others. 
These models have complementary virtues and flaws. The true problem is that it is not always easy, if not impossible, to establish which is providing the correct answer so that the spread in the theoretical calculations is generally considered as an estimate of the uncertainty. 

At first, really large discrepancies (by orders of magnitude) were observed. After discarding some evident pathologic calculations, discrepancies shrank to about one order of magnitude. However, despite the significant improvements obtained in the past years, the QRPA matrix elements still exceed those of the shell model by factors of up to about two in the lighter isotopes (e.g. $^{76}$Ge and $^{82}$Se), and somewhat less in the heavier isotopes (see Table~\ref{tab:nme}). On the other hand, IBM results are in reasonable agreement with QRPA calculations~\cite{dueck11}.

The origin of the discrepancies is still unclear and attempts to constrain the models by referring to additional observables have been pursued. Actually, the more observables a calculation can reproduce, the more trustworthy it probably is. 
This is the case, for example, of Gamow-Teller distributions which enter indirectly into \BBzn, and can be measured through (p,n) reactions~\cite{aunola98}.
The nuclear process most close to \BBz is however \BBdn, which has now been measured in 10 different nuclei. \BBd results have been used to calibrate QRPA calculations~\cite{rodin03}. 
In particular, when renormalizing all QRPA strengths by the same amount, no dependence on model-space size, or on the form of the nucleon-nucleon interaction, or on the QRPA flavor is observed. 
This is an astonishing result which has been interpreted as an indication of the correctness of the method.

\begin{figure}
\begin{center}
\includegraphics[width=0.9\textwidth]{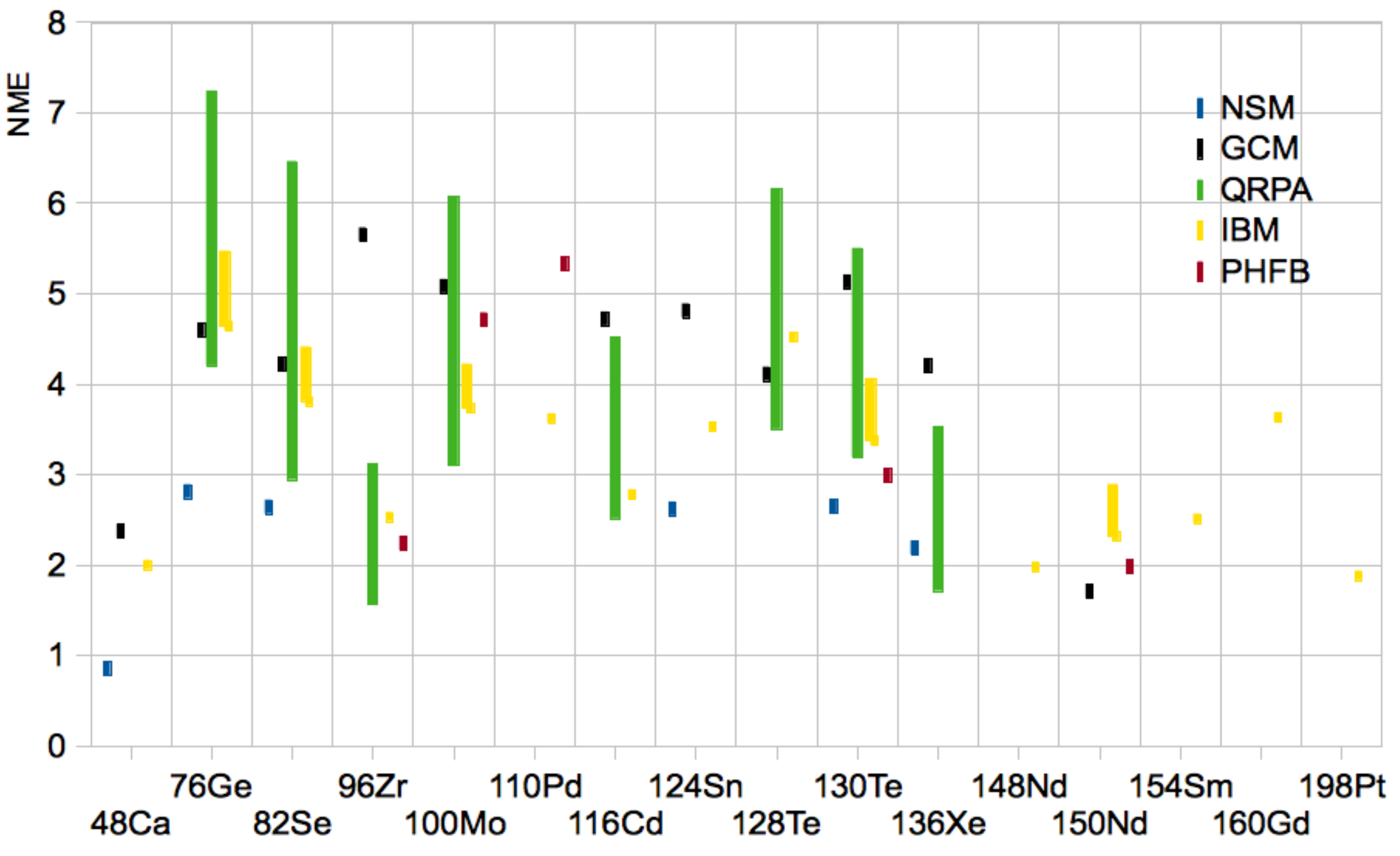}%
\caption{\BBz NME calculations as reported in Table~\ref{tab:nme}.}
\label{fig:nme}
\end{center}
\end{figure}

A number of common approximations characterize all the calculation methods while the most significant differences relate to the details of the nuclear part. 
In all cases, the reaction amplitude is factorized into the product of a leptonic and a hadronic part. 
As already mentioned above, in the case of a decay mediated by the exchange of a light neutrino, the leptonic part is proportional to the Majorana mass \amnu and to a potential N(r) describing the effects of the neutrino propagator. 
N(r) has two most relevant consequences in the calculation of the nuclear matrix elements: it introduces a dependence on the excitation energies E$_m$ of the virtual states in the odd-odd intermediate nucleus, as well as a dependence of the transition operator on the coordinates of the two nucleons.
Given the relatively high momentum of the exchanged virtual neutrino ($\langle p_{\nu}\rangle$ \ca 1/R \ca O(100 MeV), where R is the nuclear radius) a closure approximation is then applied when integrating over the virtual neutrino energies. This consists in neglecting the energy variation of the intermediate nuclear states and adding coherently the contributions of the two electrons.
The impulse and long wave approximations are then used to get rid of the hadronic current and lead to Eq.~(\ref{eq:bbrate}) for the decay rate.

The different nuclear models are then used to estimate the purely nuclear term M$^{0\nu}$. 
All models agree that only nucleons which are very close (d $\lesssim$ 2-3 fm) contribute (somehow justifying the closure approximation) although none of them takes care of the short range repulsive core (r $\lesssim$ 0.5 fm), introducing on the contrary further approximations to get rid of it.

The basic assumption of the Nuclear Shell Model is that the nucleons move independently in a proper mean field. A strongly attractive spin-orbit term is then introduced to describe the correct level separation and explain magic numbers.
As the number of protons and neutrons depart from the magic numbers, the introduction of a “residual” two-body nucleon interaction among nucleons is needed to move particles through orbits while respecting angular momentum conservation and the Pauli principle. The calculation problem consists then in the diagonalization of a matrix over a sufficiently large (valence) basis. The use of a limited valence space represents the most relevant limitation. On the other hand, all the configurations of valence nucleons are included and the NSM describes well properties of low-lying nuclear states.

%--- QRPA
In the Quasiparticle Random Phase Approximation the residual interaction among nucleons is dominated by the pairing force. As it is well known, this force accounts for the tendency of nucleons to couple pairwise to form particularly stable configurations in even-even nuclei.
As a result of the strong coupling between homologous nucleons, the orbital angular momentum and spin of each pair adds to zero with a J$^{\pi}$=0$^+$ for the nuclear ground states. The nucleon pairing is introduced via a BCS approach applied to a quasiparticle basis obtained after a unitary (Bogoliubov) transformation. Quasiparticles are thus generalized fermions with a finite probability of being either particles or holes and the net effect of the transformation is to smear out the nuclear Fermi surface for both protons and neutrons. Quasiparticles are, to a first order, independent nevertheless allowing a simple description of the pairing force between like nucleons. 
Once the vacuum of quasiparticles in the even-even nucleus has been fixed, the problem of QRPA consists in evaluating the transition amplitudes to arbitrary J$^{\pi}$ excited states in the neighboring odd-odd nuclei through a proper charge changing single-body operator.

The main advantage of QRPA is the inclusion of correlations in a ground state characterized by purely independent quasiparticles. 
As a consequence, the vacuum state can accommodate two-particle two-hole excitations so that new processes can be taken into account. 
The corresponding transition amplitudes can be written in terms of particle-hole (p-h) and particle-particle (p-p) matrix elements which are usually parametrized in terms of two adjustable coupling constants, g$_{ph}$ and g$_{pp}$ respectively. The realistic nucleon-nucleon interaction is then recovered for  g$_{ph}$ \ca g$_{pp}$ \ca 1, a condition which is unfortunately often unstable.
Many different variants of the QRPA method have been considered to get rid of this undesired behavior and produce to a more realistic description.

%--- GCM
The Generating Coordinate Method refers to the so-called aligned coupling scheme for describing the nucleon pairing and fix the equilibrium shape of a nucleus. In this scheme, each nucleon has the tendency to align its orbit with the average field produced by all other nucleons thus giving rise to nuclei with deformed equilibrium shapes and collective rotational motion. A common representation of the shape of these nuclei is that of an ellipsoid. 
A self-consistent field approach is then used to reduce the multi-body problem into one of non-interacting particles in a mean field (including deformation effects) to obtain a set of Hartree-Fock-Bogoliubov (HFB) wave functions, whose eigenstates can be found by projecting the components of well defined angular momentum, proton and neutron number (PHFB \cite{rath10}).

The Gogny interaction~\cite{gogny75} is then used as the underlying nucleon-nucleon interaction.
Different deformations are then allowed leading to a superposition of wavefunctions with coefficients which can be found by solving the so-called Hill-Wheeler-Griffin (HWG) equation \cite{rowe10}.

%--- IBM2
The Interacting Boson Model~\cite{iachello87, ibm2} can be considered somehow halfway between the “microscopic” view of NSM and the “collective” ones of QRPA and GCM.
The collective nuclear states of NSM are assumed while collective excitations are described by bosons.
However, as the number of valence nucleons increases, the direct application of the shell model becomes prohibitively difficult, and it is usually assumed that the closed shells are inert. Furthermore it is also assumed that the dominant configurations in even--even nuclei are those in which identical particles are paired together in states with total angular momentum and parity 0$^+$ or 2$^+$. Particle pairs are then treated as bosons, like Cooper pairs in a gas of electrons.
The result is a system of interacting bosons of two types, protons and neutrons.
The number of shells is reduced to the simple s-shell (J = 0) and d-shell (J = 2) and the number of proton and neutron bosons is counted from the nearest closed shell in terms of particles or holes depending if the current shell is less or more than half filled.
All fermionic operators are mapped into bosonic operators~\cite{otsuka78} 
and the matrix elements between fermionic states in the collective subspace are identical to the matrix elements in the bosonic space~\cite{ibm2}.
A realistic set of wavefunctions for even-even nuclei with mass A $\gtrsim$ 60 is provided by the IBM-2 extension \cite{iachello87} which provides an accurate description of many properties (energies, electromagnetic transition rates, quadrupole and magnetic moments, etc.) of the final and initial nuclei and allows to calculate \BBz NME through proper bosonic operators~\cite{ibm2}.
A peculiar feature of IBM-2 is its independence from nuclear deformation details which allows the calculation of NME also for heavily deformed nuclei (e.g.$^{150}$Nd) which is almost prohibitive with other methods.
 
The different methods provide an important cross check of the NME calculations  although the effect of the different approximations still needs to be explored. 
The clear advantage of the NSM calculations is the full treatment of the nuclear correlations. On the other hand, the limitations in the valence spaces can underestimate the NME's~\cite{blennow10}. 
On the contrary, all the other methods tend to underestimate the correlations thus overestimating the NME's~\cite{caurier08,menendez11}.

Unfortunately, as already mentioned above NME results are still into significant disagreement and despite a better relative agreement (Fig.~\ref{fig:nme} and Table~\ref{tab:nme}) they do not provide yet an answer to the question of which method is closer to the truth, nor to the origin of the observed disagreement. 

The careful check of the models in order to account for the omitted physics or the important missing information seems the only way out of the problem. A systematic analysis of the calculation methods and their basic hypotheses has been therefore started. 
However, the inclusion of the missing correlations into the QRPA looks a very difficult task (because of the several uncontrolled approximations of the method) while for the shell model, at least in principle, a systematic procedure for adding the effects of missing states exists.

The ultimate limitation of the QRPA method seems the perturbative approach which is implemented in a renormalized nuclear interaction and requires always some adjustment to the data. 
Reasonably good results are usually obtained by a proper parametrization of the short range correlations or the reduction of the axial-vector coupling constant g$_A$. This corresponds to a phenomenological correction of the \BBz operator whose reliability is not easy to assess.
A better approach could consist in obtaining an effective double-beta-decay operator~\cite{engel09}. 

A statistical analysis of the different NME calculation (comparison of different methods and model parameters) has also been recently considered~\cite{nmestat}. Besides providing useful recipes for the comparison of the experimental results on different isotopes this approach can help in identifying systematic effects in the different calculations.

\begin{table}[ht]
\caption{\label{tab:nme} Theoretically evaluated \BBz\ matrix elements M$^{0\nu}$ according to different authors and methods. Where needed, values have been scaled  to R$_0$ = 1.2 fm and g$_A$ = 1.25 (M'$^{0\nu}$=M$^{0\nu}$(1.25/g$_A$)$^2$) for a uniform comparison. Ranges refer to variation which arise due to model details.}
\centering
\begin{tabular}{@{}lcccccc@{}}
\hline
 Isotope & NSM\cite{menendez09}  & GCM\cite{rodriguez10} &QRPA\cite{simkovic09,fang10,civitarese12}  & IBM\cite{ibm2} & PHFB\cite{rath10}    \\
\hline
$^{48}$Ca&	 0.85&	2.37&			&		2.00&		&	\\
$^{76}$Ge&	 2.81&	4.60&	4.20-7.24&	4.64-5.47&		&	\\
$^{82}$Se&	 2.64&	4.22&	2.94-6.46&	3.81-4.41&		&	\\
$^{96}$Zr&	 	&	5.65&	1.56-3.12&		2.53&	2.24&	3.46\\
$^{100}$Mo&	 	&	5.08&	3.10-6.07&	3.73-4.22&	4.71&	7.77\\
$^{110}$Pd&	 	&		&			&		3.62&	5.33&	8.91\\
$^{116}$Cd&	 	&	4.72&	2.51-4.52&		2.78&		&	\\
$^{124}$Sn&	 2.62&	4.81&			&		3.53&		&	\\
$^{128}$Te&	 	&	4.11&	3.50-6.16&		4.52&		&	\\
$^{130}$Te&	 2.65&	5.13&	3.19-5.50&	3.37-4.06&	2.99&	5.12\\
$^{136}$Xe&	 2.19&	4.20&	1.71-3.53&		3.35&		&	\\
$^{148}$Nd&	 	&		&			&		1.98&		&	\\
$^{150}$Nd&	 	&	1.71&	3.45	&	2.32-2.89&	1.98&	3.70\\
$^{154}$Sm&	 	&		&			&		2.51&		&	\\
$^{160}$Gd&	 	&		&			&		3.63&		&	\\
$^{198}$Pt&	 	&		&			&		1.88&		&	\\
\hline
\end{tabular}
\end{table}

\begin{figure}
\begin{center}
\includegraphics[width=0.9\textwidth]{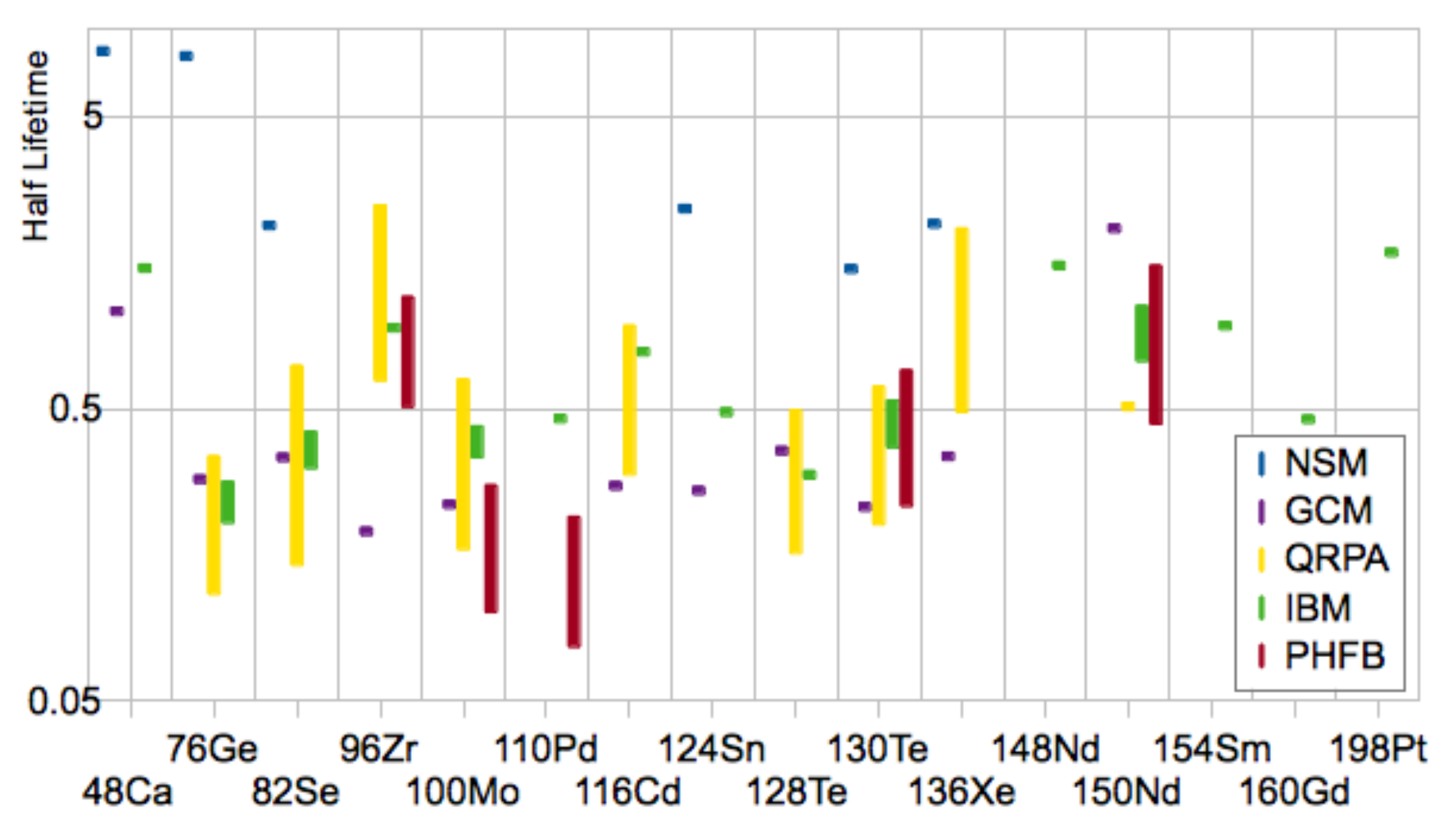}%
\caption{Ranges of theoretical \BBz half-lifetimes in unit of 10$^{26}$ yr, evaluated for \amnu = 50~meV and g$_A$=1. Discrepancies among different calculations are still of the order of a factor 2-3. Bars identify calculation ranges obtained within the same model, using different parameterizations. Dots refer to single calculations. }
\label{fig:tau}
\end{center}
\end{figure}

\begin{table}[h]
\caption{\label{tab:phsp} Phase-space factors $G^{0\nu}$ in units of $10^{-15}$ yr$^{-1}$~\cite{kotila12} and specific phase space $H^{0\nu}$~\cite{robertson13} in units of tonne$^{-1}$ y$^{-1}$  eV$^{-2}$ for \BBz candidate isotopes. Q-values and natural isotopic abundances are reported in the second and third columns.}
\centering
\begin{tabular}{@{}lrrrr@{}} 
\hline
 Isotope & \QBB (keV) &I.A.(\%)& $G^{0\nu}$ & $H^{0\nu}$    \\
\hline
$^{48}$Ca&	4272&0.187&  24.81&	826.2\\
$^{76}$Ge&	2039&7.8&  2.36&	 49.6\\
$^{82}$Se&	2995&8.73&  10.16&	198.1\\
$^{96}$Zr&	3350&2.8&  20.58&	342.7\\
$^{100}$Mo&	3034&9.63&  15.92&	254.5\\
$^{110}$Pd&	2018&11.72&  4.82&	 70.0\\
$^{116}$Cd&	2814&7.49&  16.70&	230.1\\
$^{124}$Sn&	2287&5.79&  9.04&	116.5\\
$^{128}$Te&	866&31.69&  0.59&	  7.4\\
$^{130}$Te&	2527&33.8&  14.22&	174.8\\
$^{136}$Xe&	2458&8.9&  14.58&	171.4\\
$^{148}$Nd&	1929&5.76&  10.10&	109.1\\
$^{150}$Nd&	3371&5.64&  63.03&	671.7\\
$^{154}$Sm&	1215&22.7&  3.02& 	 31.3\\
$^{160}$Gd&	1730&21.86&  9.56&	 95.5\\
$^{198}$Pt&	1047&7.2&  7.56&	 61.0\\
\hline
\end{tabular}
\end{table}

Particular attention deserves the attitude, adopted in many occasions in the past, to consider the disagreement between different calculations as a measure of the theoretical error. 
This is a very dangerous approach which creates a lot of confusion especially when comparing  the experimental sensitivities. Indeed, it does not take into account the above mentioned correlations between different calculations (for the same isotope) and suggests an improper use of the error intervals.
Although characterized by good common sense, the proposed \emph{Physics motivated intervals}~\cite{gomez10} or \emph{Educated ranges}~\cite{giuliani13} do not add any clarification and limit themselves to propose better intervals (uncertainties at the level of 20-30\%). 

A possible (and provocative) solution consists in the (arbitrary) choice of a single calculation~\cite{cremonesi10}. 
This could be, somehow, justified by the recently recognized trend of NME calculations which show only small differences among different nuclei (Fig.~\ref{fig:nme}), generally within the uncertainty interval. 
Known as the \emph{no super-element} conjecture, such an observation has been very recently strengthened by the astonishing discovery of a possible anti-correlation between phase space factors and NME's~\cite{robertson13}. 

\begin{figure}
      \centering\includegraphics[width=0.7\textwidth]{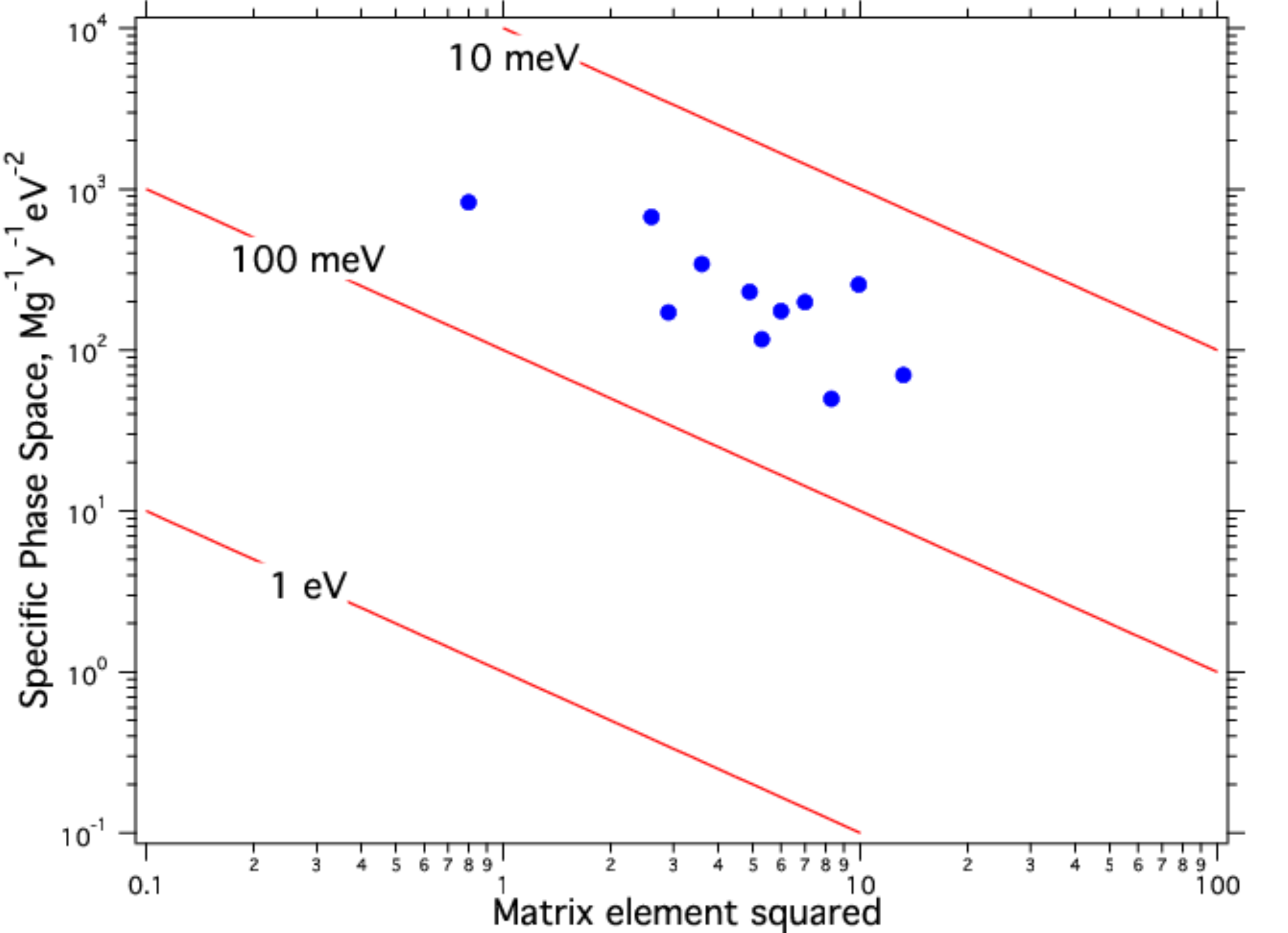}%
\caption{Specific phase space vs the geometric mean of the squared NME's reported in Table~\ref{tab:nme}, for (in increasing order of abscissa) $^{48}$Ca, $^{150}$Nd, $^{136}$Xe, $^{96}$Zr, $^{116}$Cd, $^{124}$Sn, $^{130}$Te, $^{82}$Se, $^{76}$Ge, $^{100}$Mo, and $^{110}$Pd. Phase-space factors are evaluated at g$_A$=1. From \cite{robertson13}.}
\label{fig:rob}
\end{figure}

This is easily realized when plotting (Fig.~\ref{fig:rob}), the available NME's versus the respective specific phase space (defined as $H^{0\nu}=ln(2)N_AG^{0\nu}/A$, where N$_A$ and A are the Avogadro and atomic number respectively, Table~\ref{tab:phsp}) for \BBz emitters with  Q-values larger than 2 MeV (the most relevant from the experimental point of view).

The general conclusion is that, within a factor of 2-3 (i.e. of the same order of the present NME discrepancies), the decay rate per unit (isotope) mass does not depend on the nucleus or, equivalently, that there are no especially favored or disfavored isotopes.
This also means that (within the same approximation) experimental sensitivities on the half-lifetime would translate directly (apart from a common scaling factor) in sensitivities on \mnu. 

Phase space factors reported in Table~\ref{tab:phsp} are taken from the recent extensive calculations of Kotila et al.~\cite{kotila12}. As recognized by the authors, uncertainties in G$^{0\nu}$ arise from the possible choices for the renormalization of the axial vector coupling g$_A$. 
In order to decouple this problem from other sources of uncertainty an explicit factor g$_A^4$ is suggested in the expression of G$^{0\nu}$.
Indeed, calculated phase-space factors for neutrinoless decay are generally presented for different free-nucleon g$_A$ values in the range 1-1.269.
The difference between these  values and the minimum reported value 0.6 (renormalized to fit \BBd experimental lifetimes) is significantly large in terms of rates (\ca 20). g$_A$ renormalization is therefore another critical item in neutrinoless double beta decay, and still a topic of debate among theorists.

% gives a summary of the status of NME calculations and related problems

%%%%%%%%% EXP %%%%%%%%%%%%%%%%%%%%%%%%%%
% gives general info on experiments
\section{Experimental overview}
\label{sec:expovv}

The observation of neutrinoless double beta decay would unambiguously prove that neutrinos are Majorana particles and lepton number is violated. 
This ambitious goal is challenging experimental physicists since about fifty years, justifying the enormous efforts in searching  for such an evanescent decay. 
The most suitable and best performing experimental techniques have been designed to build massive detectors operating in the most extreme conditions of low radioactivity.
However, the discovery of neutrino oscillations and the measurement of the oscillation parameters has dramatically changed the experimental situation, fixing a clear target for next generation experiments whose primary goal is to reach the needed sensitivity to study the inverted hierarchy of neutrino masses.
The intriguing claim of \BBz observation in $^{76}$Ge has further rocked the boat with a new unexpected milestone.

The size of the challenge is essentially the rarity of the decay which asks for increasingly larger masses while maintaining an excellent performance and ultra-low background environments. According to Fig.~\ref{fig:tau} a sensitivity to \BBz half-lifetimes in the range of 10$^{26-27}$ yr is required to enter the inverted hierarchy region,~{\amnu \ca ~50~meV}. This is equivalent to about a count per year in 10$^4$ moles of isotope, or in one tonne of isotopically enriched material on the average. Consequently, to record a sizable number of \BBz events over its operation time, an experiment needs to have a \Mbb~of at least 100 kg if \amnu \ca ~50~meV and few tonnes if \amnu is as low as the lower bound of the inverted hierarchy (i.e. 10~meV).

On the other hand, the decay signature exploited by most experiment is simply based on the monochromatic energy of the two emitted electrons (the sum kinetic energy of the electrons is equal to the transition energy since nuclear recoil is negligible). Unfortunately, as discussed later, there are several sources that can produce background counts in this same energy region. Their fluctuations can easily hide very faint peaks like the \BBz one, spoiling the effectiveness of the signature.
A better signature is often synonymous of a lower background and, definitely, of a better sensitivity. In principle the reconstruction of the single electron energies, the angular correlations and the identification and/or counting of the daughter nucleus could result in a large improvement of the signal to background ratio of an experiment. However, exploiting these complementary signatures is not simple and in general it has a price. All experiments tend therefore to find a compromise between the desire to collect the maximum information and the best way in which such a goal can be accomplished.

\subsection{The experimental sensitivity}
The performance of the different \BBz experiments is usually expressed in terms of an experimental {\em sensitivity} or detector {\em factor of merit}, defined as the process half-lifetime ($\tau^{Back.Fluct.}_{1/2}$) corresponding to the maximum signal $n_B$ that can be hidden by the background fluctuations at a given statistical Confidence Level (C.L.). 

The sensitivity expresses the capacity of a detector to maximize the \BBz signal while minimizing the background and is given, at 1$\sigma$ level by:
\begin{equation}\label{eq:Sensitivity}
F_{0\nu} = \tau^{Back.Fluct.}_{1/2} =
\ln 2~N_{\beta\beta}~\epsilon~\frac{T}{n_B} 
\end{equation}
where N$_{\beta\beta}$ is the number of \BB decaying nuclei under observation,
$\epsilon$ is the detection efficiency, 
$T$ is the measure time, and $n_B$ is the maximum number of counts hidden by fluctuations of the background. 

In the raw (but often well motivated) assumption that the background rate scales with the mass $M$ of the source, one can obtain the expected total number of background counts by integrating over a proper interval (customarily chosen equal to the FWHM resolution of the detector): $N_B=(b\cdot\Delta\cdot T\cdot M$), where $b$ is the specific background rate per unit mass, time and energy and $\Delta$ is the FWHM resolution. 
On the other hand $N_{\beta\beta}$ can  be rewritten as $N_{\beta\beta}=(x\cdot \eta \cdot N_A\cdot M/A$), where $x$ is the number of \BB atoms in the molecule, $\eta$ is the \BB isotopic abundance, $N_A$ is the Avogadro number and $A$ is the molecular weight.
Assuming then a Poisson statistics one gets (at 1$\sigma$) $n_B =\sqrt{N_B}=\sqrt{b\cdot\Delta\cdot T\cdot M}$ and the sensitivity formula can be rewritten as

\begin{equation}\label{eq:sensitivity1}
F_{0\nu} = \ln 2\times \frac{x ~ \eta ~ \epsilon ~ N_A}{A} 
\sqrt{ \frac{ M ~ T }{b ~ \Delta} } ~ (68\% CL)
\end{equation}  

A slightly different version of this formula can be obtained by introducing a new specific background rate $B$ normalized to the mass of the \BB isotope $M_{\beta\beta}=(M\cdot x\cdot\eta \cdot A_{\beta\beta}/A$), where \Abb~is the atomic weight of the \BB isotope. The new background rate $B$ is then related to $b$ by $B=b/(x\cdot\eta)$ while $N_B=(B\cdot\Delta\cdot T\cdot M_{\beta\beta}$). Then the sensitivity becomes:

\begin{equation}\label{eq:sensitivity}
F_{0\nu} = \ln 2\times \frac{\epsilon ~ N_A}{A_{\beta\beta}} 
\sqrt{ \frac{ M_{\beta\beta} ~ T }{B ~ \Delta} } ~ (68\% CL)
\end{equation}  

Despite their simplicity, Eqs.~(\ref{eq:sensitivity1})  and (\ref{eq:sensitivity}) have the unique advantage of emphasizing the role of the essential experimental parameters: mass, measuring time, isotopic abundance, background level and detection efficiency. 

Of particular interest is the case when the background rate $B$ is so low that the expected number of background events in the region of interest along the experiment life is close to zero. In such cases, one generally speaks of \emph{zero background} (ZB) experiments, a condition sought by a number of future projects. In such conditions Eq.~(\ref{eq:sensitivity}) is no more valid. Indeed $n_B$ is given by a constant term $n_L$ (the maximum number of counts compatible, at a given C.L. with no counts observed~\cite{pdg}) and the sensitivity reads:

\begin{equation}
\label{eq:0sensitivity}
F_{0\nu}^{ZB} = 
\ln 2~N_{\beta\beta}~\epsilon~\frac{T}{n_L} 
= \ln 2\times \frac{x ~\eta ~ \epsilon ~ N_A}{A} 
\frac{ M ~ T }{n_L} 
= \ln 2\times \frac{\epsilon ~ N_A}{A_{\beta\beta}} 
\frac{ M_{\beta\beta} ~ T }{n_L}
\end{equation}  

The most relevant feature of Eq.~(\ref{eq:0sensitivity}) is that it does not depend on the background level or the energy resolution and that it scales linearly with the sensitive mass~\Mbb and the measure time $T$. 
On the contrary, in the finite background case of Eq.~(\ref{eq:sensitivity}) the sensitivity depends only on the square root of \Mbb ~ and $T$. The dramatic effect of background is therefore not only to limit the sensitivity but even to change its dependence on the other experimental parameters.

The intermediate situation in which the expected number of counts is close to unity  marks the transition between the two regimes: ($B\cdot M_{\beta\beta}\cdot T\cdot \Delta) \simeq O(1)$. No equation exists that can properly describe this condition and one has to rely here on numerical estimates of the sensitivity.

Since $T$ is usually limited to a few years and $\Delta$ is usually fixed for a given experimental technique, there is little room to improve these terms and the transition to the ZB condition is ruled by the ($B$\dot \Mbb)~term only. 
This means that the ZB condition can be obtained because of a very good background level or of an insufficient mass of the source. 

On the other hand, Eq.~(\ref{eq:0sensitivity}) indicates that in the ZB regime, the sensitivity does not depend anymore on the background rate but only on $M_{\beta\beta}$ and further improvements in the background are useless without corresponding increases of the experimental mass. 

Similar considerations apply to the discovery potential usually defined in terms of the ratio of the observed effect and background events. Also in this case, in the ZB regime the background contribution is constant and the  discovery potential scales linearly with (\Mbb \dot $T$).

We conclude this section with the following note: there are sometimes ambiguities in the sensitivity numbers reported in literature, often because the parameters/confidence level/technique used for sensitivity computation are not clearly stated. In this article, we will adopt the following convention: provide our own evaluation of a 68\% C.L. sensitivity, that we will label as \SFz (when computed according to Eq.~(\ref{eq:sensitivity})) or \SFzz (when computed according to Eq.~(\FZ0)). We will use the latter whenever ($B\cdot M_{\beta\beta}\cdot T\cdot \Delta) < 1$ (making an approximation for the \emph{grey} zone where the background is only nearly zero). We will use \Sz to indicate sensitivities estimate provided by the authors, for which we will either specify the hypotheses under which they have been evaluated or we will report a reference where that sensitivity estimate is discussed.

\subsection{Experimental parameters}
Most of the criteria that need to be considered when optimizing the design of a new \BBz experiment follow directly from Eqs. (\ref{eq:sensitivity}) and (\ref{eq:0sensitivity}):
\begin{itemize}
\item a well performing detector (e.g. good energy resolution and time stability) giving the maximum information (e.g.
electron energies and event topology); 
\item a reliable and easy to operate detector technology requiring a minimum level of maintenance (long underground running times);
\item a very large (possibly isotopically enriched) mass, of the order of one tonne or larger;
\item an effective background suppression strategy.
\end{itemize}

Unfortunately, these simple criteria cannot be satisfied simultaneously and actual experiments have to find always, for any given technique, the best compromise between incompatible requests.

Among the experimental parameters entering Eq.~(\ref{eq:sensitivity}), the background rate $B$ is probably the one presently attracting most of the interest of the \BBz researchers. The main reason behind this is that $B$ and \Mbb ~are the only parameters on which improvement by orders of magnitude still looks possible. Moreover the possibility to reach the \emph{zero-background} region, with its linear dependence on \Mbb ~and $T$ is particularly appealing.

$B$ integrates the contributions from all the physical processes which produce measurable effects that are not distinguishable from a \BBz decay. Unfortunately they are many and only two approaches can be devised: identify their origin and eliminate their sources or find a recipe to recognize and separate each single event.

The natural radioactivity of detector components (bulk or surface) is often the main background source.  
Even traces of nuclides from the natural radioactive chains can become a significant background. 
A serious problem is becoming  the availability of a proper diagnostic technique with the required sensitivity to measure trace levels well below the capability of conventional techniques. 
The decays of $^{208}$Tl and $^{214}$Bi (due, respectively, to the \thdt and \udt chains) with their high Q-values, populate the region above 2 MeV and are therefore particularly pernicious.
In some specific case (e.g. bolometers), surface contaminations of alpha emitters have demonstrated a limiting problem.
In all cases, a careful selection of material and purification is mandatory and next-generation experiments are being built with extremely radio-pure components.
Radon isotopes, either $^{222}$Rn or $^{220}$Rn, are liberated in natural decay chains and can contaminate all materials with their progeny. 
Special care is usually requested for them.

External backgrounds originated outside the detector have also to be taken into account. 
Underground location is the usual (and fundamental) recipe to get rid of cosmic rays. Depth requirements vary from case to case and depend on the experimental technique. In many cases, well designed effective shields and/or additional detection signatures compensate the benefits of a very deep laboratory. 
Besides the depth, other important factors characterize the underground sites like the accessibility,  the size and the availability of services in the halls and, of course, a low environmental radioactivity~\cite{bettini11} (starting from the rock itself). 
%---
In the underground laboratories, muons and neutrinos are the only surviving radiation from cosmic rays. Even if muons can be easily eliminated with proper veto systems, their interactions can produce high-energy secondaries such as neutrons or electromagnetic showers (as well as nuclear activation) that can represent a more serious problem.
The effects of this secondary radiation can be particularly dangerous above ground (e.g. during detector components preparation) so that when material activation can be a concern (e.g. for germanium or copper), underground fabrication and/or storage of the detector components is essential. 
Electromagnetic showers and $\gamma$-rays from radioactive decays produced in the rock surrounding the underground halls can produce background. Detectors need therefore to be surrounded by heavy shields to reduce the effects of this radiation. To this end, layers of increasing radio-purity are used as the innermost parts of the detector are approached.
Shields against neutrons are also usually implemented with layers of a moderating (hydrogenous) material followed by materials with a high cross-section for neutron capture.
Finally, even solar neutrinos can be an irreducible source of background when very massive detectors (e.g. huge liquid-scintillator calorimeters) are used.

\begin{figure}
      \centering\includegraphics[width=0.7\textwidth]{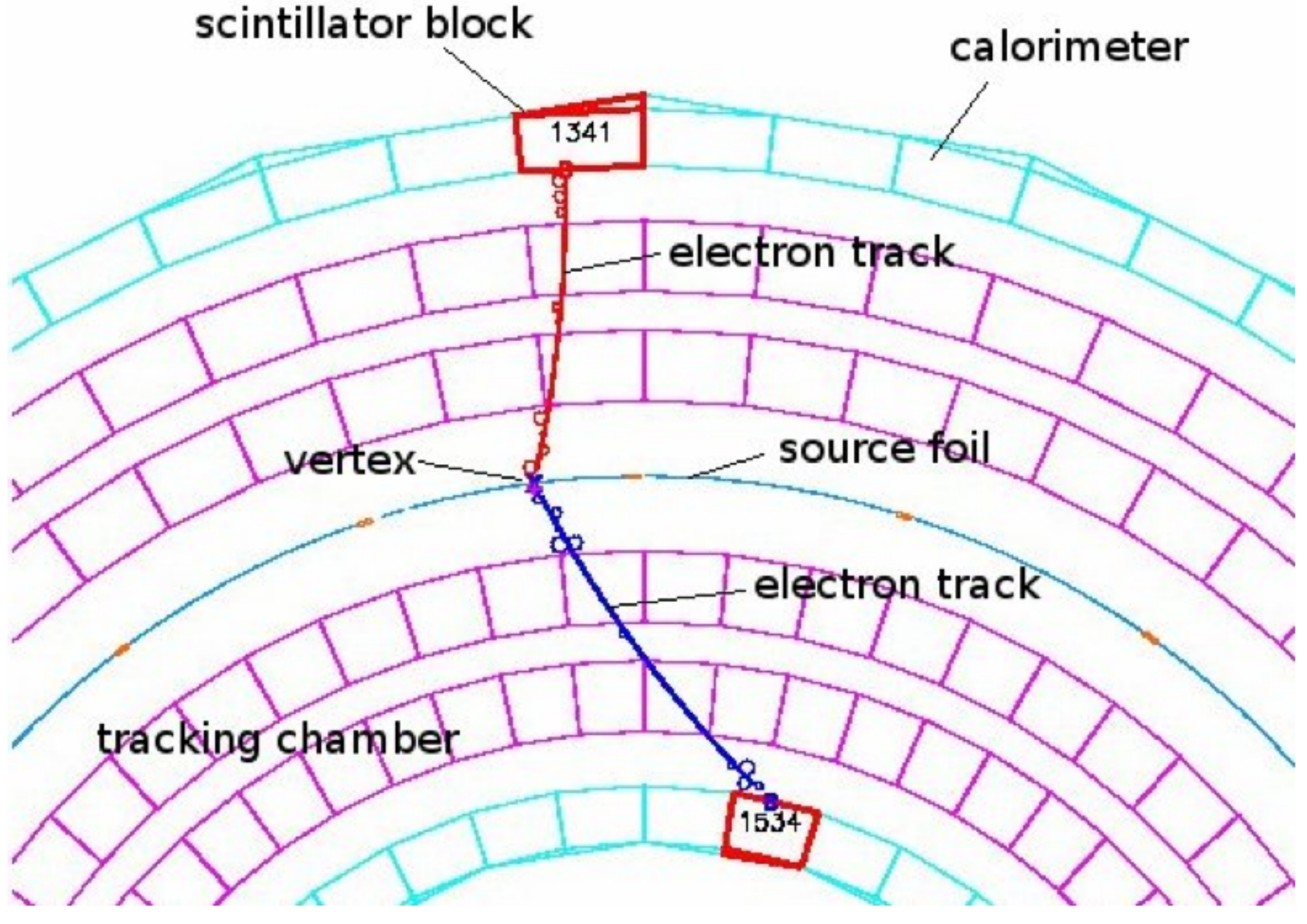}%
\caption{The figure shows how a \BBz decay candidate event would appear in the NEMO3 detector \cite{arnold05}.}
\label{fig:nemo3}
\end{figure}

In most cases, detectors are designed to measure only the total energy released in the \BBz decay (sum of the electron kinetic energies). Additional information (e.g. topological reconstruction) can be extremely helpful in identifying background contributions. Actually the lowest background rate so far  was achieved by the NEMO3 experiment~\cite{arnold05}, a calorimeter with tracking capabilities (Fig.~\ref{fig:nemo3}).

Given the rarity of \BBz decays, a high detection efficiency is another important requirement, as Eqs. (\ref{eq:sensitivity})  and (\ref{eq:0sensitivity})  clearly indicate. In general, simple calorimeters have the highest detection efficiency. 

Even if not appearing explicitly in Eq.~(\ref{eq:sensitivity}), the choice of the \BB isotope is particularly important since it influences all the relevant factors that characterize the design of an experiment:
\begin {itemize}
\item the isotopic abundance
\item the nuclear details of the decay (i.e. the nuclear factor of merit)
\item the Q-value (\QBB)
\item \BBd background
\item the choice of the experimental approach or technique
\end{itemize}
Of the 35 naturally-occurring isotopes that are \BB emitters none can match simultaneously all the requirements here listed. For each isotope a figure of merit can be drawn considering all the listed factors and this allows to identify the best candidates. 

As discussed in the introduction to this section, even in ideal conditions of efficiency and background, any experiment aiming at entering the inverted hierarchy region needs at least a mass of 100 kg of \BB isotope. 
Isotopic abundance is therefore a key ingredient in the choice of the \BB isotope.

The natural isotopic abundances of some of the most relevant \BB emitters are reported in Table~\ref{tab:phsp}. In most of the cases, the listed values are in the few \% range, with two significant exceptions: $^{130}$Te and $^{48}$Ca. 
With its 33.8\% $^{130}$Te is the only case in which a high sensitivity is possible even with natural samples. On the contrary, the natural abundance of $^{48}$Ca is well below 1\% and isotopic enrichment is indispensable. 
In order to limit the detector size and taking into account that the background level scales roughly with the total mass of the detector (and not simply the isotope fraction), it is evident that isotopic enrichment is a necessity for almost all next generation experiments. 

A further criterion can then affect the choice of the isotope: the availability and the cost of the enrichment techniques. In particular, $^{48}$Ca, $^{96}$Zr and $^{150}$Nd cannot be enriched with centrifuges and the cost becomes a limiting factor. 

The nuclear structure of each specific isotope can affect the value of the respective \BBz amplitude in a peculiar way. Indeed, a favorable value of the NME can identify some specific \emph{super-element}. This has been the case of $^{150}$Nd some year ago but, as discussed in Sec.~\ref{sec:nme}, present calculations seem to level the values of NME's which are becoming therefore a less relevant criterion. 

The Q-value is also particularly critical since it has a double effect on sensitivity, affecting both the phase space factor G$^{0\nu}$ (which varies as Q$^5$) and the background contributions (natural radioactivity populates the energy region below ∼3 MeV). Isotopes with large Q-values are therefore favored and the choice is usually restricted to \QBB \gt 2 MeV (the lowest of them is $^{76}$Ge). 
Only 9 \BB emitters survive this request.

From an experimental point of view, \BBd and \BBz decays can be distinguished from the shape of the two-electron sum energy spectrum which is a continuum between 0 and \QBB for \BBd and a sharp line at the transition energy \QBB for \BBzn. However, these distributions are smeared by the finite energy resolution of the detector and the tail of the \BBd distribution can overlap the \BBz peak. \BBd half-lifetime and energy resolution of the detector are the critical parameters, although for next generation experiments this is not a concern when the resolution is better than 1\% (Fig.~\ref{fig:bbdb}).

The relation between the choice of the \BBz isotope and the experimental approach will become more clear in the following when specific detection methods will be described. In practice, only two general experimental approaches have been so far devised: an external-source  (or {\it inhomogeneous}, or passive source) approach in which the electrons emitted by a very thin source sample (\ca60 mg/cm$^2$ in NEMO3) are observed by means of (usually very complex) external detectors, and a calorimetric (or {\it homogeneous}, or active source) approach in which the source sample is active and acts simultaneously as detector of the \BB decay.
Calorimetric detectors present serious limitations in the choice of the \BBz isotope  since only few materials can satisfy the request to be at the same time the active material of a detector. Few emblematic exceptions are $^{76}$Ge (germanium diodes), $^{136}$Xe (gas and liquid chambers) and $^{130}$Te (bolometers). On the other hand the calorimetric approach has provided so far the best sensitivities and this justifies the effort for the quest of a technology able to enlarge the list of isotopes that can be studied with a calorimetric approach. Bolometers have actually provided such an answer although few exceptions still exist (e.g. $^{150}$Nd).

% gives general info on experimental approach
\section{Experimental methods}
\label{sec:expmet}

Two main general approaches have been followed so far for \BB experimental investigation: i) indirect or inclusive methods, and  ii) direct or counter methods.
Inclusive methods are based on the measurement of anomalous concentrations of the daughter nuclei in properly selected samples, characterized by very long accumulation times. They include Geochemical and Radiochemical methods which, being completely insensitive to different \BB modes, can only give indirect evaluations of the \BBz and \BBd lifetimes. They have played a crucial role in \BB searches especially in the past.

Counter methods are based instead on the direct observation of the two electrons emitted in the decay. Different experimental parameters (energies, momenta, topology, etc.) can then be registered according to the different capabilities of the employed detectors. These methods are further classified in {\em inhomogeneous} (when the observed electrons originate in an external sample) and {\em homogeneous} experiments (when the source  of \BB's serves also as detector). 

Given the limited information coming from the decay, the experimental strategy generally adopted to investigate the \BBz decay consists in developing a proper detector to measure in real time the properties of the two emitted electrons.
The minimal request is to collect the sum energy spectrum of the electrons. However, when possible, additional pieces of information can be useful to lower background effects or constraining theoretical models. They consist usually of the single-electron energy and initial momentum, of the event topology and, in one specific case, of the species of the daughter nucleus. 
The next step consists then in the optimization of most of the experimental parameters addressed by the sensitivity equation (Eq.~(\ref{eq:sensitivity})):
\begin{itemize}

\item Energy resolution $\Delta$. A very good energy resolution is maybe the most relevant feature to identify the sharp \BBz peak over an almost flat background. It is however very useful also to keep under control the background induced by the unavoidable tail of the \BBd spectrum. Although almost negligible when the energy resolution is better than about 2\% (Fig.~\ref{fig:bbdb}), it represents a limiting factor in low resolving detectors. In these cases, candidates with a slow \BBd decay rate (e.g. $^{136}$Xe) are of course preferred.

\begin{figure}
      \centering\includegraphics[width=0.7\textwidth]{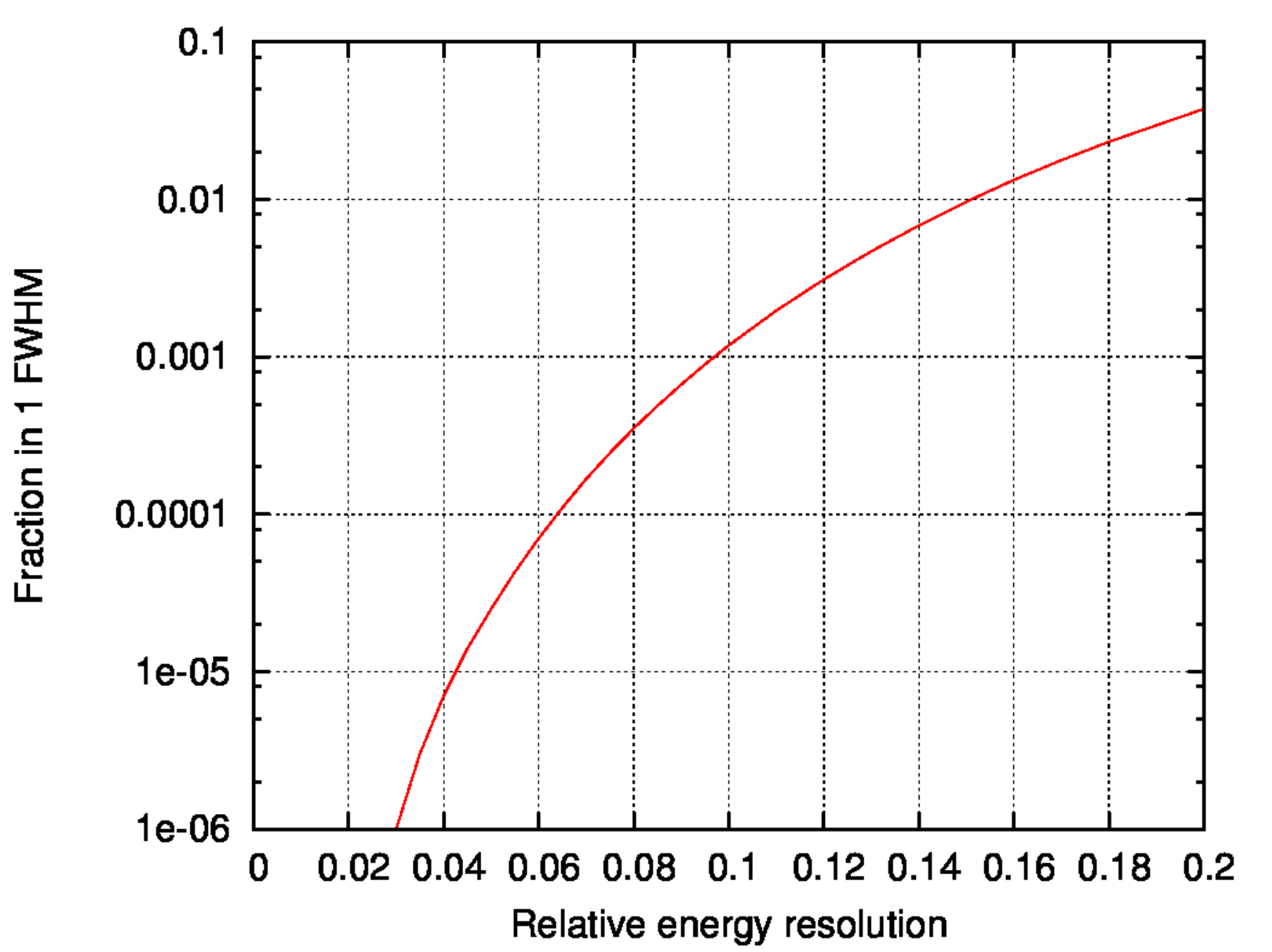}%
\caption{Fraction of the total \BBd counts expected in a window of total width equal to 1 FWHM as a function of the relative ($\sigma$/E) energy resolution.}
\label{fig:bbdb}
\end{figure}

\item Background rate $B$. As already discussed above, a very low background requires a proper  underground laboratory, extremely radio-pure materials and effective passive and/or active shields against environmental radioactivity.

\item Mass of the isotope M$_{\beta\beta}$. A large number of candidate nuclei is an inalienable constraint. Present experiments are characterized by masses of the order of few tens of kg (hundred in the most sensitive detectors) while experiments aiming at covering the inverted hierarchy region should reach the 100 -- 1000 kg scale.

\end{itemize}

Normally, these features cannot be met simultaneously in a single detection method and compromise solutions have to be worked out, privileging some properties with respect to others while having in mind of course the final sensitivity of the setup.
As already mentioned above, the searches for \BBz can be further classified into two main categories: calorimetric and external-source systems.

Originally proposed for Germanium diodes~\cite{fiorini67}, the calorimetric technique has been implemented with many types of detectors, such as scintillators, bolometers, solid-state devices, and gaseous chambers. 
Advantages and limitations of this technique can be summarized as follows:
\begin{description}
\item{$\uparrow$} The intrinsically high efficiency of the method allows large source masses. O(100 kg) has been already demonstrated and the tonne scale seems possible.
\item{$\uparrow$} With a proper choice of the detector type, a very high energy resolution is achievable (e.g. Ge-diodes and bolometers).
\item{$\downarrow$} Severe constraints arise from the request that the source material is embedded in the structure of the detector. These constraints have been however weakened by the use of liquid scintillator (e.g. Kamland-Zen and SNO+) and bolometers.
\item{$\downarrow$} Topology reconstruction is usually difficult. Also here, exceptions exist (liquid or gas Xe TPC).
\end{description}

Different detection techniques have been adopted also for the external-source approach, namely scintillators, solid state detectors and gas chambers.
Also here positive and negative aspects can be listed:
\begin{description}
\item{$\uparrow$} Reconstruction of the event topology is possible, making easier the achievement of the zero background condition. Such a beautiful feature is unfortunately masked by the negative effects of a bad energy resolution which mixes \BBz and \BBd events.
\item{$\downarrow$} Large masses of the isotope can be hardly gathered. Self-absorption in the source is the limiting factor and only masses of the order of 10 kg have been possible so far. The target of 100 kg seems possible even if at the cost of an extraordinary effort, while the tonne scale looks presently unreachable. 
\item{$\downarrow$} Typical energy resolution are of the order of 10\% mainly determined by source effects.
\item{$\downarrow$} Low detection efficiencies (of the order of 30\%) are another typically negative aspect of this approach.
\end{description}

Besides having provided so far the best experimental results on \BBzn, the calorimetric approach is still promising the best sensitivities and is therefore characterizing most of the future projects. 
Here, the well performing detectors seem limited by the scalability while the opposite holds for the very big liquid scintillation detectors. The quest for the \emph{zero background} condition is common to both, but let us remind that the golden rule is that the best sensitivity is achieved when $(M_{\beta\beta} \cdot T)\times(\Delta \cdot B ) \lesssim O(1)$. 
This is easily recognized when reworking Eq.~(\ref{eq:sensitivity}) as follows~\cite{biassoni13}:
\begin{equation}\label{eq:sensitivityM}
F_{0\nu} = 
\ln 2\times N_A \sqrt{ \frac{ n_{\beta\beta} ~ T }{B' ~ \Delta} } \equiv
\ln 2\times N_A \sqrt{ \frac{ Scale }{Performance} }
\end{equation} 
where n$_{\beta\beta}$=(\Mbb$\cdot x\cdot \epsilon$/A) is the number of moles of isotope rescaled for the efficiency while $B'$ is the background rate per unit of n$_{\beta\beta}$.
Equivalently for Eq.~(\ref{eq:0sensitivity})
\begin{equation}\label{eq:0sensitivityM}
F_{0\nu}^{0B} = 
\frac{ M ~ T }{n_L}
\ln 2\times N_A \frac{ n_{\beta\beta} ~ T }{n_L } \equiv
\ln 2\times N_A \frac{Scale }{n_L }
\end{equation}  
for the zero background regime.  
It is then apparent that $Performance=(B' \cdot \Delta)$ and $Scale=(n_{\beta\beta} \cdot T)$ must proceed hand in hand and that big efforts to reduce background without a corresponding increase in the source mass  risk to be a waste.

% gives the up to date limits and discusses Hans' claim
\subsection{Past experiments}
\label{sec:exppas}

Started in the 40's, with the first experimental work of E.~Fireman~\cite{fireman48} and soon after its theoretical proposal by W.H.Furry in 1939~\cite{furry39},
the research in double beta decay has been characterized for about half century by continuous attempts to improve the limits on lepton number conservation exploiting the improvements in the available technology.
The first direct measurement of \BBd dates back to 1987~\cite{elliott87} 
when Moe and collaborators observed the first tracks of the electrons emitted by a source of 14 g of 97\% enriched $^{82}$Se deposited on a thin mylar foil inside their Time Projection Chamber (TPC) at Irvine. 
Until that moment, the only evidence of the existence of double beta decay came from geochemical methods. 
Then, starting in the 80's, the scene was dominated for about 20 years by germanium diodes, which demonstrated an excellent technique to search for \BBz and established the superiority of the calorimetric approach. 
The discovery of neutrino oscillations at the end of the 90's has marked a true revolution in the field, providing for the first time a clear target for the \BBz experimental search. Since then, a rich and varied list of new experiments have been proposed. 

Next generation experiments will be reviewed in the next section while here we would like to summarize the most recent results.

Experimental evidence for several \BBd decays has been provided in recent years (see Table~\ref{tab:BBres}) mainly exploiting the external source approach to measure the \BBd two-electron sum energy spectra, the single electron energy distributions and the event topology. 
Impressive progress has been obtained in the same periods also in improving \BBzn ~half-life limits for a number of isotopes. The best results are still maintained by the use of isotopically enriched HPGe diodes for the experimental investigation of $^{76}$Ge (Heidelberg-Moscow~\cite{hmosc01} and IGEX~\cite{aalse02}) but two other experiments have reached comparable sensitivities: NEMO3~\cite{barab06,klang09} at Laboratoire Souterrain de Modane (LSM) and Cuoricino~\cite{arnab08} at Laboratori Nazionali del Gran Sasso (LNGS). 

NEMO3 was a large inhomogeneous detector aiming at overcoming the intrinsic limits of the technique (relatively small active masses) by expanding the setup dimensions. The big advantage of the NEMO3 technique was the possibility to access single electron information. This making possible to measure a variety of \BBd half-lives and to reach an excellent background rate. 
Cuoricino was, on the other hand, a TeO$_2$ granular calorimeter based on the bolometric technique. Its goal was to exploit the excellent performance of the bolometers (and the possibility they offer to be built with any material of practical interest~\cite{fiori84,alepe}) to scan the most interesting \BBz active isotopes. Apart from the relevant result on \BBzn, Cuoricino has the big merit of having demonstrated the scalability of the technique, paving the way to CUORE.
NEMO3 and Cuoricino were stopped in 2010 and 2008 respectively. \\

The evidence for a \BBz signal has also been claimed ~\cite{klapd04} (and confirmed later~\cite{klapd06,klapd08}) by a small subset (KHDK) of the HDM collaboration at LNGS. 
The latest reported result amounts to a 6$\sigma$ evidence with a \BBz half-life measurement of $T^{0\nu}_{1/2}=2.23^{+0.44}_{-0.31}\times~10^{25}$~yr. 
It corresponds to 11\pom 1.8 counts in the peak and agrees with the the previously quoted value within a 1.7$\sigma$ error~\cite{klapd06}.
The result is based on a complex re-analysis of the HDM data, leading to the observation of a \BBz peak in the sum energy spectrum at 2039~keV. 
This claim has triggered an intense debate in the community. No consensus still exists about its validity.
The only certain way to confirm or refute it is with additional sensitive experiments.  
Its verification is actually one of the goals of the next generation experiments. 
Preliminary results (Sec.~\ref{sec:expfut}) seem to exclude it according to most of the theoretical NME calculations.

\begin{table}[htb]
\caption{Best reported results on \BB processes. \BBz limits are at 90\% C.L. Where non explicitly referenced, the effective neutrino mass ranges are obtained according to the QRPA calculations reported in Table~\ref{tab:nme}.}
\label{tab:BBres}
\newcommand{\m}{\hphantom{$-$}}
\newcommand{\cc}[1]{\multicolumn{1}{c}{#1}}
\renewcommand{\tabcolsep}{0.9pc} 
\begin{center}
\begin{tabular}{@{}lllc}
\hline
Isotope 	& T$_{1/2}^{2\nu}$           & T$_{1/2}^{0\nu}$      &  \amnu \\ 
 	& (10$^{19}$ yr)           & (10$^{24}$ yr)     & (eV)  \\ 
\hline
$^{48}$Ca   & $(4.4^{+0.6}_{-0.5})$\cite{balish96,brudanin00,flack08}	 & $>0.058 $\cite{umehara08}	      	    & $<19-36$     \\
$^{76}$Ge   & $(150\pm 10)$\cite{avignone94,morales99,dorr03,agostini13}	 & $22.3^{+4.4}_{-3.1}$\cite{klapd06} 	    & $0.32^{+0.03}_{-0.03}$\cite{klapd06}\\  
            &           		 & $>19$\cite{hmosc01}  	      	    & $<0.17-0.29 $\\
            &				 & $>15.7$\cite{aalse02} 	      	    & $<0.19-0.32 $\\  
$^{82}$Se   & $(9.2 \pm 0.7)$\cite{arnold05b,elliott92}		 & $>0.36$ \cite{barabash11} 		    & $<1.23-1.88  $\\
$^{96}$Zr   & $(2.3\pm0.2)$\cite{arnold10,argyriades10}		 & $>0.0092$\cite{argyriades10}		    & $<5.24-10.83 $\\  
$^{100}$Mo  & $(0.71\pm0.04)$\cite{dassie95,desilva97}		 & $>1.1$\cite{barabash11}  	      	    & $<0.71-1.05  $\\ 
$^{116}$Cd  & $(2.8\pm 0.2)$\cite{flack08,ejiri95,danevich03,arnold96}		 & $>0.17$\cite{danevich03} 	      	    & $<1.64-2.69  $\\ 
$^{130}$Te  & $(70^{+9}_{-11})$\cite{arnaboldi03,arnold11}		  &$>2.8$\cite{andreotti11}    				    & $<0.45-0.70  $\\ 
$^{136}$Xe  & $(217\pm 6)$\cite{EXO-2n-2013}	 & $>1.6$\cite{EXO-0n} 	      	    & $<2.10-3.37  $\\ 
$^{150}$Nd  & $(0.82\pm 0.09)$\cite{desilva97,argyriades09}  	 & $>0.018$\cite{argyriades09}	      	    & $<9.01-16.07 $\\
\hline
\end{tabular}
\end{center}
\end{table}

% sets the scale for the next experiments and for one generation beyond
\section{Goals and methods of the Next Generation Experiments}
\label{sec:expfut}

The conclusion of Cuoricino and NEMO3 mark in some way the transition toward a new generation of experiments characterized by bigger detectors (100 - 1000 kg of isotope), designed and constructed by wide international collaborations sharing work and costs. The ultimate goal of these \emph{next generation} projects would be to explore the inverted-hierarchy region of neutrino masses, a very ambitious objective which requires the realization of experiments at the multi-tonne scale with background levels of the order of 1~\cpkty. The cost, the risk profile and the time scale (of the order of ten or more years) that characterize the preparation phase of these big experiment motivates the adoption of a cautious strategy, generally based on the construction of a 100~kg scale experiment that can be expanded at a later time to 1 or more tonnes. Scalability as well as performance are therefore the key issues on which next generation experiments will select the future technique.

Some of the parameters appearing in Eq.~(\ref{eq:sensitivity}) (e.g. the energy resolution) only depend on the experimental technique and cannot be improved at will. On the other hand, sizable improvements of the sensitivity can be obtained acting on:
\begin{enumerate}
\item background level;
\item isotopic enrichment;
\item active mass.
\end{enumerate} 

Next generation experiments are therefore facing the challenge of developing detectors characterized by masses of isotopically enriched materials of the order of \ca 1 tonne, operating underground in conditions of extremely low radioactivity. In this game, a further, certainly not naive and not always properly mentioned, difficulty is the unavailability of proper diagnostic methods to certify the assessment of a given level of background. In these conditions detector prototypes characterized by intermediate masses (the mentioned 100 kg scale phase) are the only possibility.

So far, the best results have been pursued exploiting the calorimetric approach which characterizes therefore most of the future proposed projects. 
They can be classified in three broad classes:
\begin{enumerate}
\item Dedicated experiments using a conventional detector technology with improved background suppression methods (e.g. GERDA and MAJORANA).
\item Experiments using unconventional detector (e.g. CUORE) or background suppression (e.g. EXO and SuperNEMO) technologies.
\item Experiments based on suitable modifications of an existing setup aiming at a different search (e.g. SNO+, KAMLAND)
\end{enumerate}

Experimental methods and expected sensitivities of the proposed projects are compared in Tables~\ref{tab:BBfutexp} and \ref{tab:BBfutprm}. 
As discussed above, technical feasibility tests are requested in some cases, but the crucial issue will be the capability of each project to pursue the expected background suppression.

Calorimetric detectors are usually preferred for future experiments since they have produced so far the best results. The calorimetric approach suffered for years from a strong limitation: it was possible only for a  small number of \BBz isotopes (e.g. $^{76}$Ge, $^{136}$Xe, $^{48}$Ca), thus limiting the number of experimentally accessible isotopes. Today, the multiple choices offered by new detectors and techniques (e.g. bolometers) show that a possible way out exists. 

\begin{sidewaystable}
\begin{center}
\caption{A selected list of the next generation \BBz experiments. Important features that characterize each experiment are indicated in the last column: $\Delta$E refers to a good energy resolution while PSA (Pulse shape analysis) stands for the capability to discriminate events topologies. The isotope mass, M$_{\beta\beta}^{fid}$, here reported takes into account isotopic abundance and fiducial volume as described in the following sections. }
\label{tab:BBfutexp}
\begin{tabular}{@{}lcccccc} \\ 
\hline
Experiment & Isotope & M$_{\beta\beta}^{fid}$ & Technique & Location & Start date \\
 &  & (kg) &  &  &  &\\
\hline
$^{130}$Te  & CUORE0/CUORE  &  11/206  &  Bolometric  &  LNGS  &  2012/2014  &   $\Delta E$, Cu+Pb shield  \\
$^{76}$Ge  &  GERDA I/II  &  11/30  &  Ionization  &  LNGS  &  2012/2014  &  $\Delta E$, PSA, LAr shield \\
$^{82}$Se  & LUCIFER  &  9  &  Bolometric  &  LNGS  &  2014  &   $\Delta E$, scintillation, Cu+Pb shield  \\
& MJD  &  26  &  Ionization  &  SUSEL  &  2014  &  $\Delta E$, PSA, Cu+Pb shield \\
$^{130}$Te  &  SNO+  &  163  &  Scintillation  &  SNOlab  &  2014  &  Size/Shielding \\
$^{82}$Se or $^{150}$Nd  &  SND/SuperNEMO  &  6/100  &  Tracko-calo  &  LSM  &  2014/2015  &  Tracking \\
$^{136}$Xe  &  EXO-200  &  79  &  Liquid TPC  &  WIPP  &  2012  &  Ionization + Scintillation \\
$^{136}$Xe & KamLAND-ZEN& 179  &  Scintillation  &  Kamioka  &  2012  &  Low Background Environment \\
$^{136}$Xe & NEXT-100 &  90 & Gas TPC &  Canfranc & 2014 & Tracking\\
\hline
\end{tabular}
\end{center}
\end{sidewaystable}

%%%%%%%%% ISOTOPES %%%%%%%%%%%%%%%%%%%%%%%%%%
% lists isotopes with natural concentration, Qvalue and maybe comments
%\subsubsection{Double Beta Decay Isotopes For Large Scale Experiments}
%\subsubsection{Enrichment Techniques For Large Scale Isotope Production}
%\input 04-enr.tex

%%%%%%%%% EXP %%%%%%%%%%%%%%%%%%%%%%%%%%
%\input 05-bkg.tex

%%%%%%%%%%%%% TPC's and tracking detectors %%%%%%%%%%%%%%%
\section{Time Projection Chambers}
\label{sec:tpc}

Particle tracking is a powerful technique to distinguish a \BBz signal from a background signal. A \BBz event is characterized by a pair of very short tracks originated at the source position if compared with background events with the same energy (most of the studied isotopes have Q-values of 2-3 MeV) that are usually characterized by much longer tracks (as in the case of cosmic ray muons) and/or by multi-site energy depositions (as in the case of $\gamma$ or $\gamma$+$\beta$ emissions).

Tracking is accomplished by the use of gas counters or Time Projection Chambers (TPC's)  where the \BB source is introduced in the form of thin foils or -- in the special case of \xe \BBz decay -- as the TPC filling gas/liquid. A magnetic field can be used to improve particle identification capability (which is the case for NEMO3 and also for the Moe pioneering experiment).
A segmented detector is used to reconstruct the spatial distribution of the ionization cloud, deriving event topology with a resolution that strongly depends on details of detector implementation: vertex position, number of interactions and track length are among the information that can be obtained. These are used for background rejection and background identification, the latter is of primary importance for background modeling. In the case of a high spatial resolution it becomes also feasible to disentangle \BBz from \BBd and to study the different \BBz decay mechanisms (see SuperNEMO description). 
Whatever the choice done for the tracking read-out, the energy is measured through a scintillation signal that in the case of xenon TPC's is produced by Xe itself, while in the other cases it is obtained by the introduction of an array of scintillators in the TPC. Energy resolution is often much worse than in pure calorimetric approaches such as those involving HPGe diodes or bolometers with two consequences: the increase of the number of sources able to mimic \BBz events and the need of a background reconstruction to disentangle the \BBz signal. 

Tracking can provide multiple techniques for background rejection, varying according to the specific characteristics of the detector. For example, a powerful and simple way to get rid of some radioactive source (in particular those emitting short range particles) is the definition of a fiducial volume. Requiring that the interaction vertex has to be within a volume that is sufficiently far from important sources as the TPC vessel, most of $\beta$+$\gamma$ and $\alpha$ events from natural chains (or other $\beta$ or $\alpha$ decaying isotopes) are rejected. Obviously a compromise has to be reached between the benefit -- in terms of background rate -- of a small fiducial volume and the corresponding reduction of the \BB active mass, this compromise can change in time according to the changes in intensities and locations of the background sources.

\subsection{$^{136}$Xe TPCs }

\xe is an attractive \BB candidate for various reasons: 
\begin{itemize}
\item it has a high \QBB (2457~keV), therefore the \BBz signal grows in a region that is less contaminated by radioactive background events;
\item its \BBd mode is slow (even slower than expected, as proved by EXO-200 and later confirmed by KamLAND-ZEN) and hence its contribution in the \BBz decay Region Of Interest (ROI) is irrelevant even when the energy resolution is poor;
\item xenon can be used for the realization of a \emph{homogeneous} detector since it provides both scintillation and ionization signals;
\item is a gas and can be easily and cheaply enriched (its natural isotopic abundance (i.a.) is 8.86\%) and purified. 
\end{itemize}

The running experiment EXO-200 and the projected NEXT-100 use xenon in an active source approach, while in the KamLAND-ZEN experiment the \xe \BB passive source is dispersed in a liquid scintillator (see Sec.~\ref{sec:scint}). 

At 2457~keV, multiple sources can mimic a \BBz decay. The dominant background comes from the high energy $\gamma$ lines due to isotopes in the \udt and \thdt natural chains: the 2448~keV $\gamma$ from $^{214}$Bi ($^{222}$Rn progeny) and the 2615~keV $\gamma$ from $^{208}$Tl. The former is certainly the most threatening one since it is less than 10~keV apart from the \BBz signal. The implementation of radon suppression techniques is a mandatory requirement for these experiment, while mitigation of radon-induced background can be obtained by improving the energy resolution of the calorimeter, the accuracy of energy calibration and the ability to identify and subtract $^{214}$Bi contributions from the measured spectrum. 
In particular cases, short-living nuclei produced by cosmic ray activation or by fallout, can be important background contributors as proved by the KamLAND-ZEN experience (see Sec.~\ref{sec:scint}). Finally, cosmic rays -- although potentially dangerous -- can be easily suppressed through the use of optimized veto systems and underground deep locations.

\subsection{EXO}

\begin{small}
\begin{tabular}{|p{2.5cm}|p{12cm}|}
\hline 
\textbf{EXO-200} & \xe (\QBB = 2457~keV) running experiment \\
\hline
\textbf{FWHM} & 96 keV \\
\textbf{EXPOSURE} & 32.5 kg(\xen) \per~yr \\
\textbf{MASS} & 79 kg of \xe (LXe fiducial volume of 98.5 kg)\\
\textbf{BKG} & \\
~~rejection & single-site vs. multi-site events separation + fiducial volume + \al~rejection through light/charge ratio\\
~~rate & (1.1$\pm$0.1)\per 10$^{-3}$~\cpkky \\
~~sources & $^{214}$Bi (localized outside the TPC, therefore contributes mainly through its 2448~keV $\gamma$), \thdt and \udt in the TPC vessel (2448~keV from $^{214}$Bi and 2615~keV from $^{208}$Tl \\
\textbf{\Tz} & $<$ 1.6\per 15$^{25}$ yr at 90\% C.L.~\cite{EXO-0n}\\
\textbf{\SFz} & 1.2\per 10$^{26}$ yr in 5 years (with the same background, detection efficiency (82.5\%) and active mass of~\cite{EXO-0n})\\
\hline
\end{tabular}
\end{small}
\newline
\newline

The Enriched Xenon Observatory (EXO) Collaboration is planning a series of experiments to search for \BBz decay of \xe with progressively higher sensitivity using liquid xenon (LXe) TPC's. Within this program, EXO-200 is a 200 kg-scale experiment designed to achieve a 2 year \BBz sensitivity of 6.5\per 10$^{25}$ yr. However, this was computed assuming a fiducial mass of 140 kg of Xe, namely higher than in the actual case, meaning that the same sensitivity will be reached in a longer time.  
The experiment is located at a depth of 1585 m water equivalent in the Waste Isolation Pilot Plant (WIPP) near Carlsbad, New Mexico. 
The advantage of LXe over a gaseous xenon TPC lies mainly in the reduced volume where the same mass can be concentrated, at the price of a worse energy resolution. EXO-200 exploits both the scintillation and ionization signals produced by particle interactions in xenon, while the future plans of the collaboration include the implementation of a \emph{Ba tagging} technique. This aims at the identification (through laser excitation) of the \xe \BB decay daughter ($^{136}$Ba$^{++}$) as a further and unambiguous signature of a \BB decay. If successful, this technique would impressively improve background discrimination.

\begin{figure}
\begin{center}
\includegraphics[width=1\linewidth]{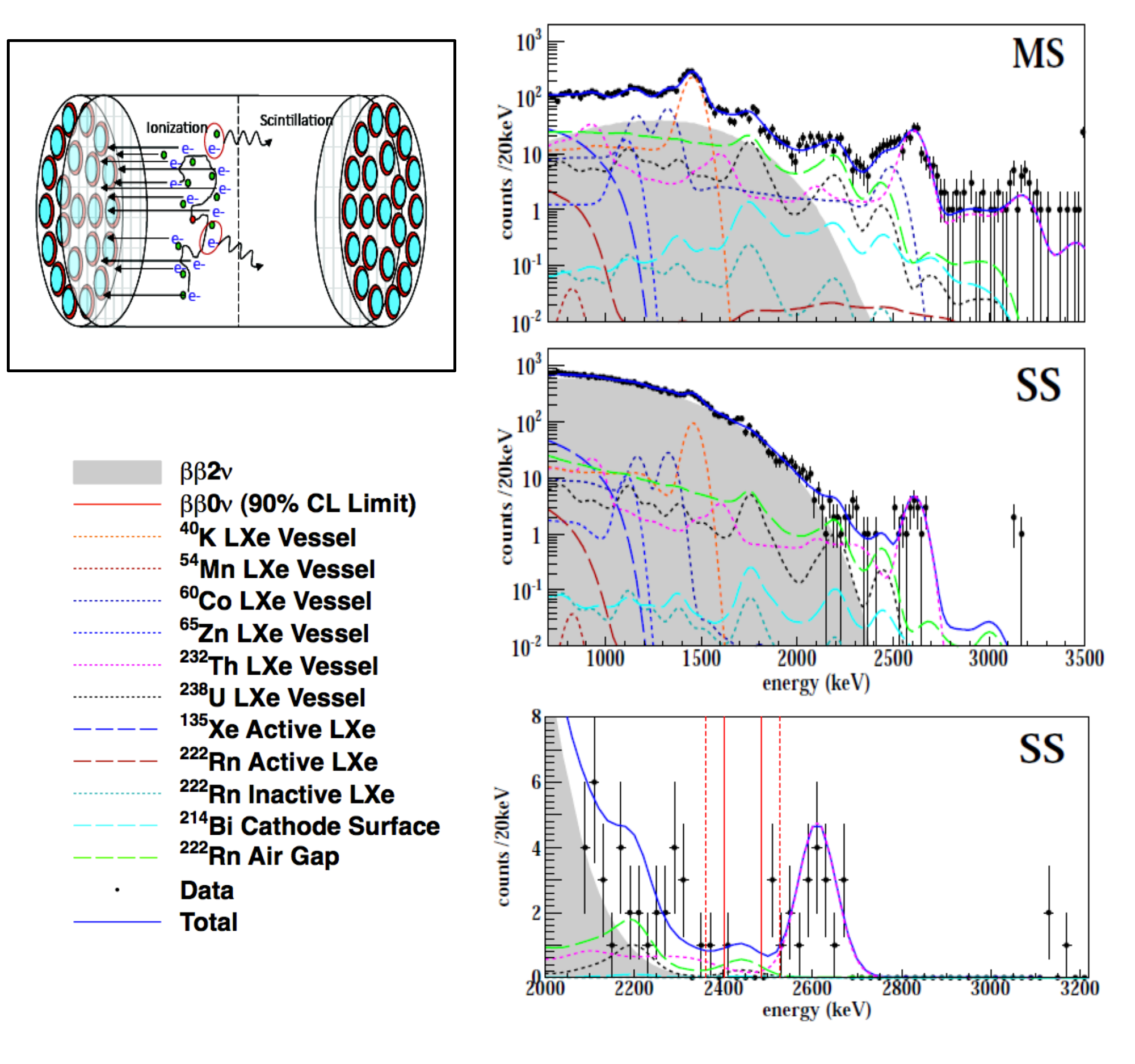}
\end{center}
\caption{Top left: EXO detector concept (figure from~D.~Auty presentation at 48th Rencontres de Moriond, year 2013). Right: EXO-200 results, SS and MS fits with a zoom (bottom right) of the SS fit in the \BBz region (figure from~\cite{EXO-0n})}
\label{fig:EXO}
\end{figure}

The EXO-200 detector consists in a cylindrical TPC filled with LXe (see Fig.~\ref{fig:EXO}) mounted inside a cryostat and externally shielded from cosmic rays and radioactivity by 25~cm of lead. A further thickness of 5~cm in copper and of 50~cm in the liquid refrigerator are provided by the cryostat itself. All components used for the construction of the detector were carefully selected for low radioactive content. The clean room module -- housing the cryostat and the TPC -- is surrounded on four sides by an array of plastic scintillators acting as cosmic rays veto. At WIPP, the muon rate is of about (3.10$\pm$0.07)\per 10$^{-7}$ $\mu$/(s$\cdot$cm$^2\cdot$sr) (\ca 10 times higher than at LNGS), while $\mu$'s traversing the TPC are easily rejected any $\mu$'s traversing the experimental apparatus but not tracked in the TPC can produce dangerous background events via bremsstrahlung or spallation. A cosmic-ray induced background rate 10 times higher than the EXO-200 goal (3 events/year in the \BBz ROI) was estimated in absence of the veto. This rate is reduced to negligible levels by the veto \cite{EXO-dete}.

EXO-200 uses about 200 kg of xenon, enriched to (80.6\pom0.1)\% in the isotope \xen. Xenon is continuously recirculated, therefore only a fraction of it (110~kg) is in the liquid phase inside the detector chamber. The cylindrical TPC (44~cm in length and 40~cm in diameter) is divided into two identical volumes (two halves) by a cathode grid held at negative high voltage, located in the mid-plane of the cylinder. The ionization signal is read-out at the two ends of the cylinder by two wire planes held at virtual ground potential (charge collection U-wires). A further plane of wires (induction V-wires) oriented at 60 degree with respect to U-wires, is positioned at each end of the TPC, at a distance of 6 mm from each U-wire plane. The electrically induced signal is used to have a second coordinate that allows two-dimensional localization of the ionization cloud. 

In order to improve the energy resolution of the detector, also the scintillation signal produced by particle interactions in LXe is read-out using two arrays of large area avalanche photo-diodes (preferred to phototubes mainly for the lower radioactivity), one behind each of the two charge collection planes. The scintillation signal provides a complementary energy information used to improve the energy resolution, to reject events corresponding to incomplete charge collection or alpha particles (that are characterized by a different charge-to-light ratio with respect to $\beta/\gamma$'s) and to achieve a three-dimensional position sensitivity: the z-coordinate is indeed obtained by using the difference in the arrival time between the ionization and scintillation signals (electron drift time).

The spatial information allows to reject events coming from the chamber walls (by the definition of a fiducial volume) and to classify signals in single-site (SS) and multi-site (MS). The majority (about 82.5\%) of \BB events is SS (a fraction of events is MS because of bremsstrahlung). MS events are mainly used to constrain background components.

Periodic calibrations of the apparatus are necessary in order to monitor continuously the free electron life-time and the overall charge-to-energy conversion. Source measurements are also used to verify the SS and MS reconstruction efficiencies through comparison with Monte Carlo simulations.

Data collected between May 21, 2011 and July 9, 2011  were used for \BBd analysis ~\cite{EXO-2n} with the discovery (later confirmed by KamLAND-ZEN) that \Td was shorter than what previously reported in the literature ~\cite{bernabei02} . 
 
In June 2012 the first result on \BBz was published, using a detector exposure of 32.5 kg(\xe)\per yr (corresponding to a fiducial volume of 
98.5 kg of LXe). 
Here the combination of the charge and light signals is used for the first time  to improve the energy resolution, with a gain of about a factor 2 with respect to the use of the ionization signal alone. The resolution at \QBB is 1.67\% for SS events and 1.84\% for MS events (i.e. the FWHM at the \BB transition energy is 96~keV in SS events and 106~keV in MS events).
The calibration error is lower than 1\%.
The \BBd and \BBz signals are extracted by a simultaneous fit of SS and MS spectra (the fitting region covers the range from 700 keV to 3.5 MeV, see Fig.~\ref{fig:EXO}) with the spectral shapes predicted by the Monte Carlo simulation for \BBd and \BBz decays and for the main radioactive sources responsible of the background counting rate. While the SS spectrum is dominated by \BBd events (according to the best fit, the ratio of \BBd events to background ones is 9.4 to 1 ~\cite{EXO-2n}) only a small fraction of them contributes to the MS spectrum which on the other hand is dominated by background sources (in the \BBz region the MS counting rate is about 10 times higher than SS one). The contamination levels yielded by the fit for the different background sources are consistent with the material screening measurements, which in some way proves the reliability of the background model.
Indeed, the consistency between contaminations extrapolated from the data and those measured for the single detector parts before assembly is not trivial: in many cases only upper limits on contaminant concentrations are available and moreover new contributions are often introduced by components handling, machining and assembly.

The \Td already measured in~\cite{EXO-2n} has been recently updated to:
\newline ~~~~ \Td =2.172$\pm $0.0017(stat)$\pm$0.060(syst)\per 10$^{21}$ yr~\cite{EXO-2n-2013}.

No peak is observed in the \BBz ROI. The fit yields a background rate in the (1$\sigma$ region) of (1.1$\pm$0.1)\per 10$^{-3}$~\cpkky~due to external background sources (i.e. not coming from \xe itself). 
The main contributors are identified in the 2448~keV $\gamma$ line of $^{214}$Bi, ascribed to   $^{222}$Rn in the cryostat-lead air-gap, \thdt (contributing through Compton scattering of the 2615~keV $\gamma$ line) and \udt (again the 2448~keV peak) in the TPC vessel. 
Actually, the spectral shape of a $^{222}$Rn contamination in the air gap cannot be distinguished from that of a \udt contamination in materials outside the cryostat but $^{222}$Rn measurements confirm the assumed hypothesis and allow for the possibility of a background improvement in the near future.

A lower limit on \BBz half-life is evaluated corresponding to 1.6\per 10$^{25}$ yr at 90\% C.L. 
The future evolution of EXO will go in the direction of a tonne scale experiment that aims at an active mass of 4 tonnes of \xen, a slightly improved energy resolution (1.4\% at 1$\sigma$) and a background reduction obtained through an improved radon suppression and the different surface/volume ratio.

\subsection{NEXT}

\begin{small}
\begin{tabular}{|p{2.5cm}|p{12cm}|}
\hline 
\textbf{NEXT-100} & \xe (\QBB = 2457~keV) under construction \\
\hline
\textbf{FWHM} & 12.5 keV\\
\textbf{MASS} & 90 kg of \xe (100~kg of enriched Xe)\\
\textbf{BKG} &  \\
~~rejection & event topology (predicted background rejection ratios are of \ca 2\per 10$^{-7}$, detection efficiency of 25\%)\\
~~goal & 8\per 10$^{-4}$ \cpkky (evaluation done on the basis of the background budget and rejection factors) \\
\textbf{\SFz} & 1.6\per 10$^{26}$ yr in 5 years\\ 
\hline
\end{tabular}
\end{small}
\newline
\newline

The concept of the NEXT project is very similar to the one of EXO: use ionization and scintillation signals in a xenon TPC. However, in NEXT xenon is in its gaseous phase where energy and tracking resolutions are better, an advantage whose price is the larger volume needed for the same xenon mass: LXe has a density of 3 g/cm$^3$ while in NEXT (that plans to work at a pressure of \ca 15 bar)  density is 0.075 g/cm$^3$.

\begin{figure}
\begin{center}
\includegraphics[width=1\linewidth]{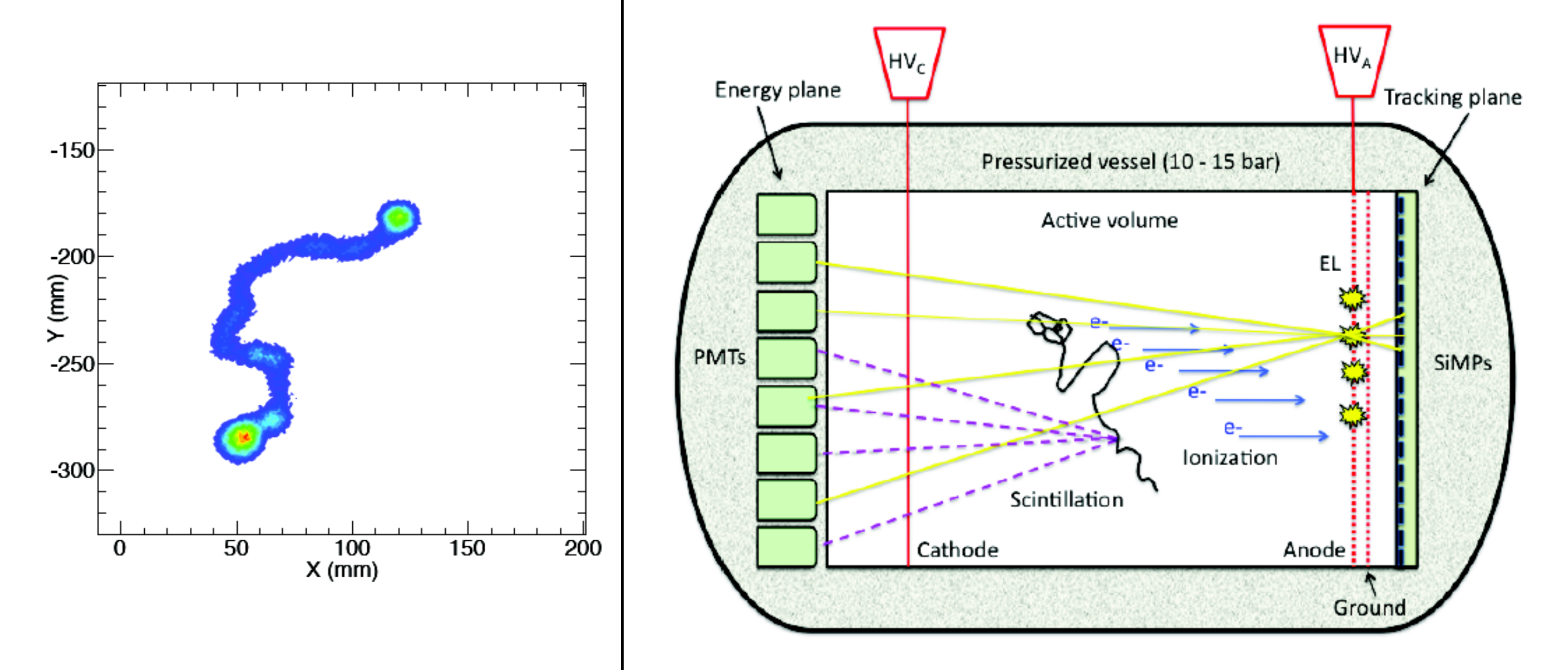}
\end{center}
\caption{Left: a Monte Carlo simulation of the \BBz ionization track in a 10 bar Xe chamber. Right: the NEXT detector (pictures from~\cite{NEXT-100})}
\label{fig:NEXT}
\end{figure}

In NEXT-100, scintillation and ionization are read-out as a light signals, with a solution that aims at reaching the best energy resolution (down to about 12~keV FWHM) and high resolution tracking: in a high pressure Xe chamber the two electrons emitted in a \BB decay produce a characteristic track \ca 30~cm long (see Fig.~\ref{fig:NEXT}), easily distinguished from most radioactive-induced events.
The detection principle is the following: a particle interacting in the chamber produces excitation and ionization of Xe atoms. The former mechanism gives rise to the prompt emission of scintillation light (this is the start of the event) while the latter produces charges (distributed along the particle track) that are drifted on a long length (of the order of 1 m) in an electric field of relatively low intensity. At the end of the drifting region, between the gate and the anode, a much more intense electrical field induces electro-luminescence (EL): drifted electrons acquire so much energy that scattering on Xe atoms produce their excitation followed by scintillation. In this way, the ionization signal is converted into scintillation light which is used for both energy measurement and tracking.

The NEXT-100 detector is a cylindrical, stainless steel pressure vessel containing a polyethylene field cage (see Fig.~\ref{fig:NEXT}). A 12~cm thick copper shield separates the cage from the vessel and is used to mitigate the possible effect of vessel radioactivity.

Three wire-meshes, cathode, gate (ground) and anode separate the two electric  field regions of the detector. The drift region, between cathode and gate, is a cylinder of 107~cm diameter and 130 cm length. The EL region, between gate and anode, is 0.5 cm long. The tracking function is provided by a plane of multi-pixel photon counters placed behind the anode plane that measure EL signal. An array of PM is located behind the transparent cathode and is used to read-out the scintillation light in order to provide a precise measurement of the energy released by the interacting particle. The solution of using two different arrays of optical devices, one dedicated to tracking and the other to energy measurements, allows to optimize separately the two measurements. 

Tests on small scale prototypes have proved an energy resolution of 1\% FWHM at 662~keV which scales to 0.5\% at \QBB (namely 12.5~keV) and a track reconstruction with an uncertainty of the order of 5-10 mm \cite{NEXT-100}.

According to Monte Carlo simulations the background rejection efficiency obtained through the combination of cuts based on tracking and energy is impressive: ranging from 3 to 7 orders of magnitude. The latter is obtained exploiting full event topology and corresponds to a \BBz detection efficiency (i.e. the fraction of \BBz events that survives topology cut) of 25\%. A background level of 8\per 10$^{−4}$ \cpkky~is predicted for the energy region of interest on the basis of the background budget of the experiment (material radioactive screening) and of the efficiency of topology cut. The 5 years sensitivity, in these hypotheses, is \SFz =~1.6\per 10$^{26}$ yr.
NEXT-100 is approved for operation in the Laboratorio Subterra\'neo de Canfranc (LSC), in Spain, at a depth of 2450 m.w.e. The assembly and commissioning of the detector is planned for early 2014.

\section{Inhomogeneous tracking detectors}

A completely different approach to \BBz searches separates the \BB source from the detection device. In this case, the source is a thin foil made of the \BB candidate, while the detector consists in a tracker combined with a calorimeter. This technique was successfully employed for example by Elegans V whose planned prosecution is MOON~\cite{MOON}. 
However, the best example of passive-source tracking detectors is certainly the NEMO3~\cite{NEMO3} experiment where tracking was associated with particle charge identification (thanks to the presence of a magnetic field) allowing not only an efficient background rejection but also a precise measurement of the different background sources producing the experimental counting rate. 

\subsection{SuperNEMO}

\begin{small}
\begin{tabular}{|p{2.5cm}|p{12cm}|}
\hline 
\textbf{SND} & $^{82}$Se (\QBB= 2997~keV) under construction\\
\hline
\textbf{FWHM} & 120 keV\\
\textbf{MASS} & 6.3 kg of $^{82}$Se (7~kg of enriched Se)\\
\textbf{BKG} &  \\
~~rejection & particle charge identification + track \\
~~goal & 5\per 10$^{-4}$ \cpkky \\
\textbf{\SFz} & 3.3\per 10$^{25}$ yr in 5 years (detection efficiency 30\%)\\ 
%~***l'efficienza e' in 105.1241 ma non trovo il fondo
\hline 
\textbf{SuperNEMO} & $^{82}$Se (\QBB= 2997~keV)\\
\hline
\textbf{FWHM} & 120 keV\\
\textbf{MASS} & 100 kg of $^{82}$Se (110~kg of enriched Se)\\
\textbf{BKG} &  \\
~~rejection & particle charge identification + track \\
~~goal & 5\per 10$^{-4}$ \cpkky \\
\textbf{\SFz} & 1.3\per 10$^{26}$ yr in 5 years (detection efficiency 30\%)\\

\hline
\end{tabular}
\end{small}
\newline
\newline
The SuperNEMO project is an extension of the NEMO3 technique toward the realization of a new apparatus able to overcome NEMO3 limitations. The increase in sensitivity will be based on a larger isotope mass (i.e. a larger experimental apparatus) and on the reduction of background. A clear idea of the background sources that need to be controlled in SuperNEMO comes from the NEMO3 experience.
NEMO3 was a cylindrical detector combining gas tracking counters and calorimeters. It was divided in 8 sectors, each one dedicated to the specific study of a \BB isotope ($^{100}$Mo, $^{82}$Se, \tectn, $^{116}$Cd, $^{96}$Zr, $^{48}$Ca, $^{150}$Nd). The best \BBz results were obtained for the two isotopes present with the highest masses, $^{100}$Mo and $^{82}$Se, both having a \QBB at about 3~MeV. The latest NEMO3 results are~\cite{barabash11}:
\begin{itemize}
\item $^{100}$Mo: \Tz$>$ 1.1 10$^{24}$ years at 90\% C.L 
\item $^{82}$Se: \Tz$>$ 3.6 10$^{23}$ years at 90\% C.L. 
\end{itemize}
with a background counting rate as low as 0.003 \cpkky.
A \BB decay was identified as two electrons emitted from the \BB source foil.
Background sources that can mimic this kind of events are:
\begin{itemize}
\item the two electrons emitted by \BBd (i.e. the tail of the \BBd decay spectrum that is comprised in the ROI);
\item high energy $\gamma$'s impinging on the foil and producing two electrons, double Compton or Compton+Moller scatterings or also pair production (in the case of mis-identification of the positron charge). The highest contribution here comes from $^{214}$Bi due to $^{222}$Rn contamination in the gas counters;
\item internal contaminations of the source foils with $\beta$ decaying isotopes accompanied by internal conversion (IC), Moller or Compton scattering. Radioisotopes with a high enough energy to produce such kind of events in the \BBz ROI are $^{214}$Bi (Q=3.3 MeV) and \tld (Q=5 MeV), respectively from \udt and \thdt chains.
\end{itemize}

SuperNEMO will have to reach a much better radio-purity in the \BB source foils as well as a stronger Rn suppression. However, this will not be enough  to get rid of background due to \BBd  events and a reduction of the FWHM is also compulsory. 
SuperNEMO plans to improve the energy resolution by about a factor of 2 and to choose a \BB candidate with a sufficiently long \Td with respect to the expected \Tz. This excludes the already studied $^{100}$Mo. Favorite isotopes are therefore $^{82}$Se, $^{150}$Nd and $^{48}$Ca although the possibility of enriching the latter two isotopes is still under study.

\begin{figure}
\begin{center}
\includegraphics[width=1\linewidth]{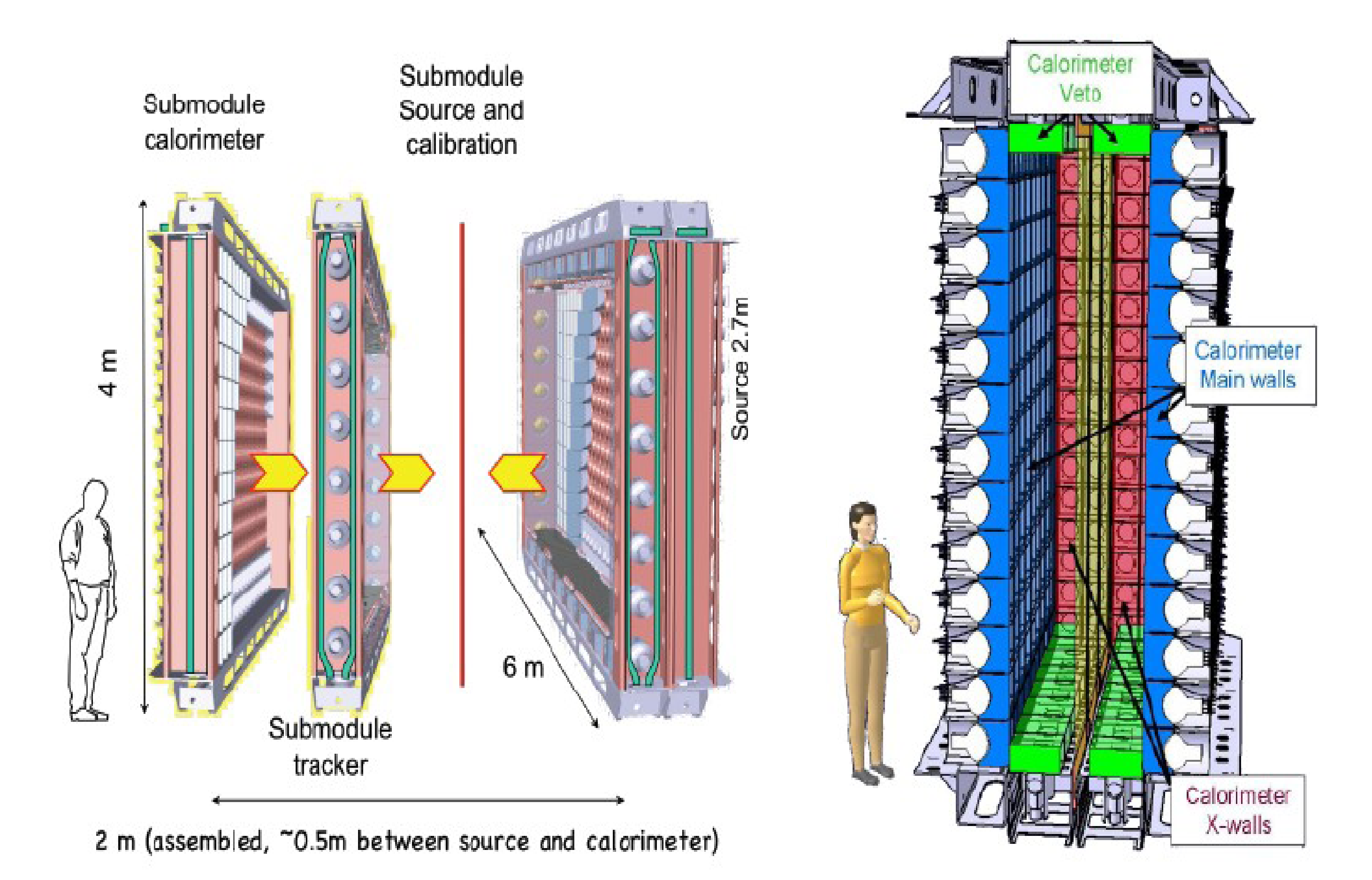}
\end{center}
\caption{The SuperNEMO single module (picture from X. Sarasin, arXiv:1210.7666v1)}
\label{fig:NEMO}
\end{figure}

SuperNEMO~\cite{SNEMO} is designed as an experiment made of 20 modules (Fig.~\ref{fig:NEMO}), each containing 5-7 kg of \BB emitter in the form of a thin foil of enriched material. The single module has a planar design (i.e. different from the NEMO3 cylindrical symmetry). The source is a thin (40 mg/cm$^2$) foil (3\per 4.5 m) mounted in the middle plane of a gas tracking chamber, the 6 walls of the chamber are covered by plastic scintillator blocks (500 to 700 depending on the design which is not yet fixed) to realize the calorimeter.
The tracking volume contains 2000 wire drift cells operated in Geiger mode in a magnetic field of 25 Gauss. These are arranged in nine layers parallel to the foil and will be able to provide particle identification, vertex reconstruction and angular correlation between the two electrons emitted in \BB decay. The expected spatial resolution is 0.7 mm in the direction perpendicular to the \BB source foils and 1 cm in the parallel one. The scintillators provide a calorimetric measurement of particle energy with an expected energy resolution of 7\% FWHM at 1~MeV (i.e. 120~keV at 3~MeV). The angular correlation between the two electrons emitted in the \BB decay can be used to study the \BBz decay mechanism~\cite{SNEMO-angular}.

The first module, the SuperNEMO Demonstrator (SND), containing 7 kg of $^{82}$Se (i.e. more than 7 times the isotope contained in NEMO3),  is presently under construction and will be installed in the Laboratoire Souterrain de Modane (LSM) within the year 2014. No background count is expected for the demonstrator in 2.5 years, corresponding to a sensitivity of 6.5\per 10$^{24}$ yr at 90\% C.L~\cite{NEMOtaup}. This is equivalent to a background counting rate of about 5$\times$10$^{-4}$ \cpkky, therefore the 
5 year sensitivity evaluated with our criteria is \textbf{\SFz}= 3.3\per 10$^{24}$ yr  (we assume a signal efficiency of 30\% as quoted in~\cite{SNEMO-angular}). SuperNEMO, which will require a much larger space, will be installed in the planned extension of the Modane laboratory, the 
5 year sensitivity evaluated on the basis of 100 kg~\cite{NEMOtaup} of $^{82}$Se is \textbf{\SFz} = 1.3\per 10$^{26}$ yr).

%%%%%%%%%%%%%%% Cryogenic detectors %%%%%%%%%%%%%%%%%%%%%
\section{Bolometric detectors}
\label{sec:bol}

A thermal detector is a sensitive calorimeter which measures the energy deposited by a single interacting particle through the corresponding temperature rise.
This is accomplished by using suitable materials (dielectric crystals, superconductors below the phase transition, etc.) and by running the detector at very low temperatures (usually below 100~mK) in a suitable cryostat (e.g. dilution refrigerators).
Indeed, according to the Debye law, the heat capacity of a single dielectric and diamagnetic crystal at low temperature is proportional to the ratio (T/T$_D$)$^3$ (T$_D$ is the Debye temperature) so that for extremely low temperatures it can become sufficiently small.
Of course, the measurement of the temperature change requires also a proper thermal sensor. 
A low temperature detector (LTD or “bolometer”) consists of three main components: 
i)  a particle absorber (the sensitive mass of the device where the particles deposit their energy); ii) a temperature sensor (or transducer) and iii) a thermal link to the heat sink.

The absorber material can be chosen quite freely, the only requirements being, in fact, a low heat capacity and the capability to stand the cooling in vacuum. The absorber can therefore be easily realized with materials containing any kind of unstable isotopes and many interesting searches are therefore possible (e.g. $\beta$ decay spectroscopy, neutrinoless double beta decay and dark matter). So far, absorbers with masses in the range from few micrograms to almost one kilogram have been developed.

In principle, the intrinsic energy resolution of a bolometer is limited only by the thermodynamical fluctuations of thermal phonons through the thermal link and it 
can be as small as few tens of eV even in the case of \ca kg bolometers. Besides the exceptionally low value, the intrinsic energy resolution does not depend on the deposited energy E.
In practical cases, $\Delta E$ is dominated by other noise contributions. A dedicated low-noise front-end electronics is therefore usually required in order not to spoil such a wonderful feature of these devices. However, important contribution to the detector noise come from vibrations (through the induced thermal dissipations) and are ofter referred to as \emph{microphonic} noise. In \teod bolometers (Cuoricino and CUORE \BBz experiments) energy resolutions lower than 1~keV at 10 keV (dominated by noise)~\cite{CUORE-lowene} and of \ca 5~keV at 2.6 MeV have been demonstrated (at the latter energy an additional contribution to the resolution is observed, in particular for $\gamma$'s, and is ascribed to an incomplete thermalization of the particle energy deposition).

The material choice flexibility together with the excellent energy resolution and the sensitivity to low or non-ionizing events are certainly the best features that make bolometers an excellent opportunity for rare events searches. On the other hand, the response slowness is an unavoidable limitation. Even if not actually a problem for the present generation of \BBz experiments, signal velocity could become important in approaching the Inverted Hierarchy region of neutrino masses, due to the unavoidable pile-up of \BBd events~\cite{beeman12,chernyak12,IHE}. 
One of the worst effects of the long thermal integration times is that they tend to wash out any possible difference in the time development of the signals (e.g. those arising from the interaction details of different particles). This is actually an undesired feature in the critical process of background abatement although hybrid techniques (e.g. the simultaneous detection of scintillation) can represent a practical solution. Very interesting results have already been obtained for a number of different absorbing materials as discussed in the following.

\subsection{Specific Backgrounds in Bolometers}
Bolometers can measure with high resolution the total energy deposited by any type of particle interaction. They rely on the observation of excess events above background in the region of the expected \BB signal as the primary (or unique) signature for neutrinoless double beta decay. 

The candidates that are presently used or proposed for a bolometric \BBz experiment are \tect (Cuoricino and CUORE), $^{82}$Se (LUCIFER), $^{100}$Mo, and $^{116}$Cd, selected according to their \QBB  and to the feasibility of a bolometric detector (with energy resolution of the order of 10~keV at \QBB) based on one of their compounds.
In the energy region where the \BBz line of these isotopes should appear (between 2.5 and 3 MeV) a number of sources contribute to background formation. Besides the usual sources, such as environmental and cosmogenic radioactivity, neutron and cosmic muon background (for which the already discussed mitigation solutions are generally adopted), bolometers are particularly sensitive also to an usually minor source of background signals: surface contaminations. While most of the other kind of detectors can rely on the use of topological information to reject surface events or -- in other cases - can be completely insensitive to them thanks to the existence of a surface dead layer protecting the sensitive volume, in bolometers this is not the case.
Surface contaminations can be therefore considered a specific background to bolometers, whose effects represent today the worst limitation to \BBz sensitivity.

Most of the information on the nature and effects of background sources for bolometric detectors come from the Cuoricino~\cite{Qino} experiment (the CUORE prototype which collected data at LNGS from January 2003 until June 2008) and a series of dedicated measurements carried out in the past years at LNGS, on smaller arrays of bolometers prepared under different conditions and with different materials ~\cite{paperiEPJ}.
All these measurements confirm a background model according to which the dominant sources in the ROI (\tect \QBB \ca2527 keV) are (with different weights)\cite{qprop05}: i) unshielded $^{208}Tl$ $\gamma$'s from the environment and the setup materials; ii) U and Th surface contaminations of the detector crystals and iii) U and Th surface contaminations of the Copper used for the detector supporting structure.

Concerning source i), it is important to recall that the \tld 2.6 MeV line is the highest natural \gm~line due to environmental contamination having a branching ratio $>$ 1\% . It appears as the dominant \gm~contribution in \tect ROI (through Compton events). 
In the case of $^{82}$Se, $^{100}$Mo, and $^{116}$Cd whose \QBB is $>$2.8~MeV, pure \gm~ contributions of natural radioactivity comes only from the low branching ratio $\gamma$ lines of $^{214}$Bi. 

The background measured above the \tld line in Cuoricino is ascribed mainly to degraded \al's coming from U and Th radioactive chains and due to surface contamination of the bolometric crystals (absorbers) or of (inert) detector elements directly facing the bolometers (the copper of the assembly structure, the PTFE stands that are used to secure the crystals in the copper structure ...). This continuum clearly extends below the \tld line thus participating to the background counting rate at lower energies (these are the contributions listed above as ii) and iii)). Besides degraded \al's,  surface contaminations produce also $\beta$+$\gamma$ events of the few isotopes belonging to U and Th chains that can produce a signal in the \BBz region when their Q value is greater than the isotope \QBB (e.g. \tld and $^{214}$Bi). This is generally a smaller contribution with respect to degraded \al's, that however becomes the only contribution from surface contamination in the case of scintillating bolometers where \al~events are rejected on the basis of their different scintillation yield.

While well designed heavy shields can ensure a strong reduction of the \gm~background, for \al~(and $\beta$+$\gamma$) background (that come only from the very inner part of the detector, i.e. the crystal themselves and the material directly facing the crystals) only a severe control of bulk and surface contaminations of the detector materials can guarantee the fulfillment of the sensitivity requirements. To this end, a correct identification and localization of the sources is mandatory, which requires a powerful diagnostic method able to detect and identify very small surface contaminations. For the same reasons for which surface \al~background is their worst enemy, bolometers are the best tools to study surface contaminations but measurements are long, difficult and very expensive. Diagnostic programs including analyses at different levels of sensitivities (with different techniques) are therefore the best choice~\cite{paperiEPJ}.

From the picture above, it is evident that surface contaminations are the worst background contribution in bolometers. Two main approaches can be adopted to mitigate their effects:

\begin{itemize}
\item {reduction of surface contamination;}
\item {identification and rejection of the events originated at the detector surface.}
\end{itemize}

The former implies the development of effective techniques for the cleaning of all the surfaces faced to the bolometer crystals, the latter the development of bolometers able to identify surface events or to identify particle type. Very promising results have been obtained -- in this framework -- with hybrid detectors exploiting the different scintillation properties of \al's and \gm's. Unfortunately they apply only to bolometers built with scintillating materials.
It should be finally pointed out that the two approaches are not mutually exclusive, and their development should run in parallel together with further checks of the radioactive contamination of all the detector parts and a complete scan of all the possible background sources.

\subsection{CUORE}

\begin{small}
\begin{tabular}{|p{2.5cm}|p{12cm}|}
\hline 
\textbf{CUORE-0} & $^{130}$Te (\QBB= 2527~keV) running \\
\hline
\textbf{FWHM} & (5.6$\pm$2.1) keV \\
\textbf{MASS} & 11 kg of $^{130}$Te (39 kg of natural \teod)\\
\textbf{BKG} & \\
~~rate & (0.074$\pm$0.012)~\cpkky\\
~~sources & still to be studied, from previous work: \tld in cryostat + degraded \al's\\
\textbf{\SFz} & 1.5\per 10$^{25}$ yr in 5 years (detection efficiency is 78\%)\\
\hline
\textbf{CUORE} & $^{130}$Te (\QBB=2527~keV) under construction \\
\hline
\textbf{FWHM} & 5 keV (predicted) \\
\textbf{MASS} & 206 kg of $^{130}$Te (741 kg of natural \teod) \\
\textbf{BKG} & \\
~~goal & 1 \per 10$^{-2}$ \cpkky\\
~~sources & degraded \al's from surface contaminations \\
\textbf{\SFz} &  2.1\per 10$^{26}$ years in 5 yr (detection efficiency is 86\%)\\
\hline
\end{tabular}
\end{small}
\newline
\newline

\begin{figure}
\begin{center}
\includegraphics[width=0.9\linewidth]{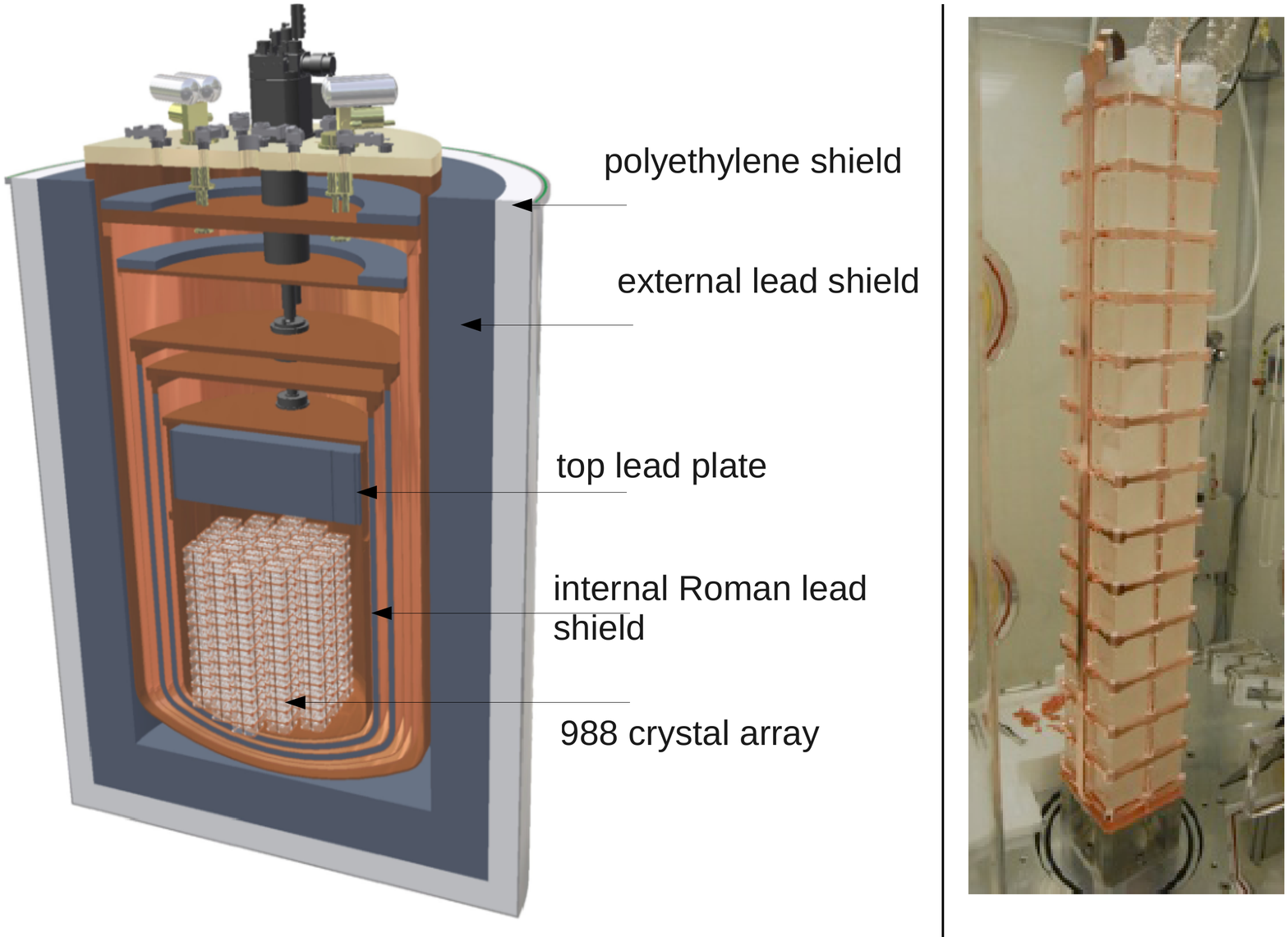}
\end{center}
\caption{Left: the CUORE set-up. Right: CUORE-0 tower.}
\label{fig:CUORE}
\end{figure}

CUORE~\cite{qprop05} ({\em Cryogenic Underground Detector for Rare Events}) is a next-generation experiment for the search of \BBz of \tect, which brings the concept of large mass bolometric detectors to the extreme. Its design is based on the successful and demonstrated technology of the pilot experiment Cuoricino.
It consists of an array of 988 (dielectric and diamagnetic) natural \teod cubic crystals grouped in 19 separated towers (13 planes of 4 crystals each) arranged in a rather compact cylindrical structure (Fig.~\ref{fig:CUORE}) designed in order to reduce to a minimum the distance among the crystals and the amount of inert material interposed (mainly copper from the mechanical support structure). 
Each crystal is 5~cm in side, with a mass of 750~g and is expected to operate at a temperature of ~10~mK. Neutron Transmutation Doped (NTD) Ge thermistors are used to detect the small temperature rise resulting from single nuclear decay events. 

\begin{figure}
\begin{center}
\includegraphics[width=0.7\linewidth]{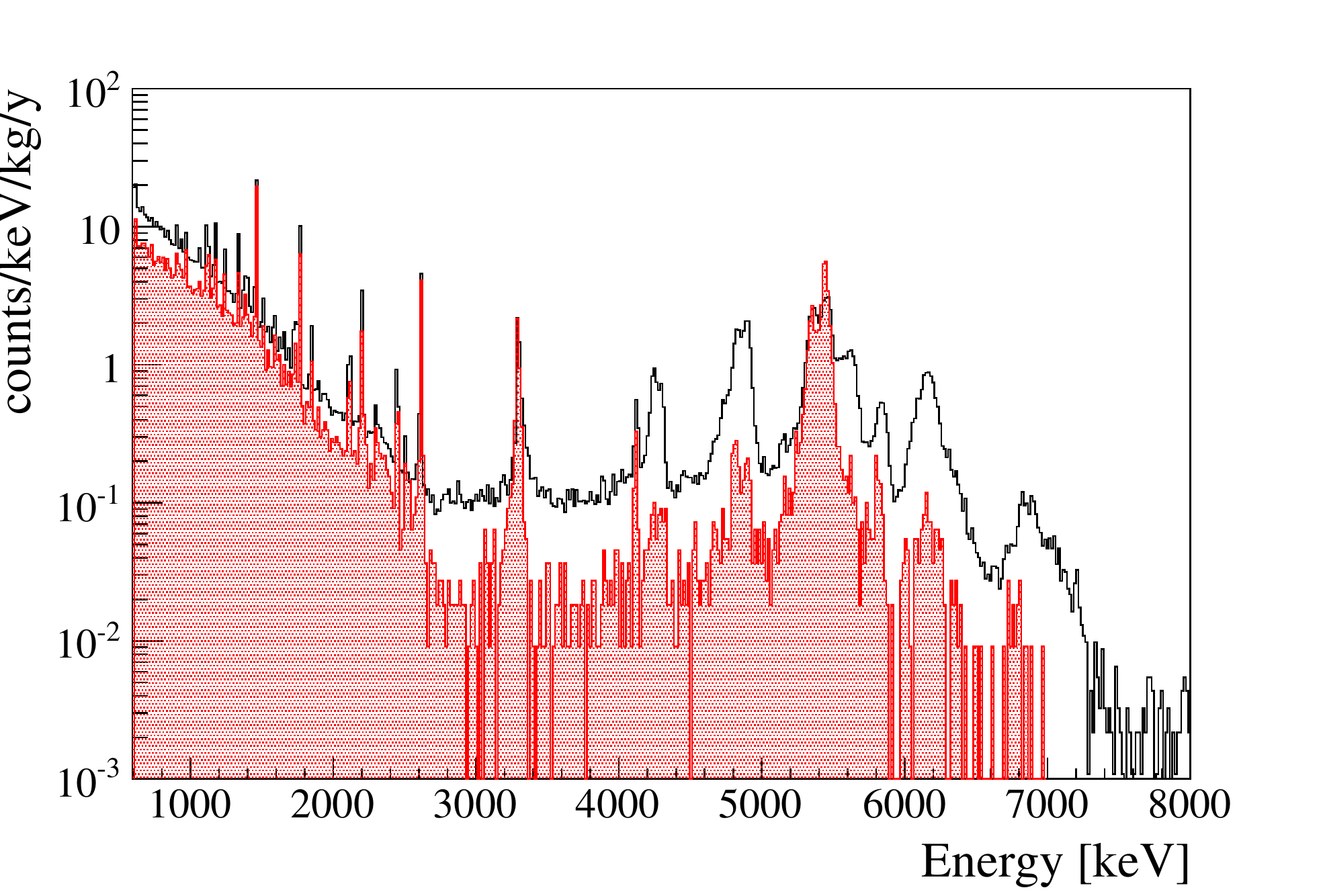}
\end{center}
\caption{Comparison of Cuoricino (black histogram) and CUORE-0 (red filled histogram) background spectra. Preliminary result from~\cite{cuore0}.}
\label{fig:CUORE0}
\end{figure}

The array, surrounded by a 6 cm thick lead shield (built with low activity lead from a sunk Roman ship), will be operated at about 10 mK in a He$^3$/He$^4$ dilution refrigerator (see Fig.~\ref{fig:CUORE}). A further thickness of 30 cm of low activity lead will be used to shield the array from the dilution unit of the refrigerator and from the environmental activity. A borated polyethylene shield and an air-tight cage will surround externally the cryostat. The experiment will be installed underground at LNGS, in the same experimental hall where Cuoricino was operated.
The design and construction of the cryostat that will be used to maintain the detectors at the necessary cryogenic temperatures is a rather unique undertaking. It is based on the comparatively recently developed technology of the cryogen-free dilution refrigerators, which utilizes pulse tube (PT) pre-cooling instead of a liquid helium bath; this should allow improved stability of the base temperature of the detectors as compared to the traditional He$^3$/He$^4$ refrigerator (used for Cuoricino). It will be the first cryostat of its kind big enough to house and cool the large detector mass represented by the CUORE array (\ca1 tonne) and its copper/lead shields.

For the point of view of the \BB candidate, Tellurium offers the advantage of a high natural abundance (33.8\%) of the \BBz candidate isotope, which means that enrichment is not necessary to achieve a reasonably large active mass. Also, the Q-value of the decay (2527 keV~\cite{qte09}) falls between the peak and the Compton edge of the 2615 keV gamma line of \tld; this leaves a relatively clean window in which to look for the signal. 

In addition, to the increase in scale from Cuoricino to CUORE, in order for CUORE to reach its anticipated sensitivity, improvement are required in two crucial aspects of detector performance: resolution and background. 

The resolution is expected to improve from the 6.3 $\pm$ 2.5 keV FWHM measured by Cuoricino (the error measuring the spread over the detectors) to about 5 keV FWHM which is the goal resolution for CUORE. This will be achieved both by the minimization of vibrational noise in the new cryostat and by progress (already achieved) in the crystal quality control, detector mounting structure design, and in the reproducibility of the thermistor-crystal couplings. 

Concerning background, an improvement of a factor \ca~20 with respect to Cuoricino is necessary to reach CUORE goal: from 0.18 \cpkky, as measured by Cuoricino, to 0.01~\cpkky that is the \emph{conservative} target for CUORE. As previously discussed, in Cuoricino an important contribution to the background counting rate in the ROI is ascribed to irreducible contaminations of the set-up that will be overcome in CUORE thanks to the new cryostat+shield system built with selected ultra-low radioactivity materials. On the side of the detector, large efforts have been spent to carefully select low background materials (starting from crystal production) and to clean their surfaces (focusing on crystals and copper that represent the largest area of the detector array). Finally, to prevent any re-contamination of surfaces after their cleaning the CUORE assembly-line allows the construction of the array without exposure of the detector parts to air (which prevents possible radon contamination) and minimizing the contact (in space and time) with other materials. 

Projections of CUORE sensitivity generally assume a 5~keV FWHM resolution and 0.01 \cpkky~background and results in \SFz = 2.1\per 10$^{26}$ years in 5 years exposure. Tests of the first batches of crystals produced for CUORE~\cite{CCVR} and of copper parts which have undergone special surface treatments prove that a background rate of the order of 0.01 \cpkky~is feasible~\cite{TTT}, but an important answer from this point of view will come from CUORE-0. 

CUORE-0 is the first CUORE tower that is now installed, as a stand alone experiment, in the Cuoricino cryostat and is taking data. Besides being a very important step in CUORE construction, CUORE-0 will be able to produce a meaningful improvement in the \tect \BBd results of Cuoricino. While CUORE-0 background rate in the ROI will be most probably dominated by cryostat contaminations (therefore will not be able to provide a direct check of CUORE background since the cryostat will be different), the information about degraded \al's contribution will be extracted from the counting rate recorded in the 3-4 MeV region, with the same technique discussed in~\cite{TTT}.
The total \teod mass is  39~kg, the expected background in the \BB region is higher than 0.06~\cpkky, being this the irreducible contribution evaluated for the cryostat contamination (the actual background rate of CUORE-0 will depend mainly on the success of the surface background control). 
Preliminary CUORE-0 data~\cite{cuore0} (see Fig.~\ref{fig:CUORE0}) prove the achievement of a relevant reduction of the 3-4 MeV counting rate with respect to Cuoricino (by a factor \ca 6), while -- as expected -- the counting rate in the \BBz regio is only a factor \ca 2 better than in Cuoricino.  With an energy resolution of 5.6 keV FWHM and a background counting rate of (0.074$\pm$0.012) \cpkky~\cite{cuore0} the 5 years sensitivity of CUORE-0 is \SFz = 1.5\per 10$^{25}$ years. Most probably CUORE-0 exposure will be of about 2 years since the experiment will close as soon as CUORE will start taking data, in this case the 2 years sensitivity is \SFz = 9.7\per 10$^{24}$ years

\subsection{R\&D programs and LUCIFER} 

\begin{small}
\begin{tabular}{|p{2.5cm}|p{12cm}|}
\hline 
\textbf{LUCIFER} & $^{82}$Se (\QBB= 2995~keV) under developement \\
\hline
\textbf{FWHM} & 13 keV \\
\textbf{MASS} & 9.3 kg of $^{82}$Se (15 kg of enriched materials, enrichment fraction 95\%, production yield \ca 65\%)\\
\textbf{BKG} & \\
~~goal & 1 \per 10$^{-3}$ \cpkky\\
\textbf{\SFzz} & 1.6\per 10$^{26}$ yr in 5 years (detection efficiency 76\%)\\
\hline
\end{tabular}
\end{small}
\newline
\newline

A very promising development of low-temperature calorimeters consists in the simultaneous detection of light and heat, i.e. in the construction of hybrid scintillating bolometers.
Pioneered by the Milano group with CaF$_2$ \cite{caf2scint} in the 90's, 
this approach  \cite{pirro05, IHE}
represents the basic idea behind the LUCIFER\cite{giuliani10}, LUMINEU\cite{beeman12}, and AMoRE\cite{amore12} projects. 
The detector in this case is made of a scintillating crystal containing the \BBz candidate. The read-out of the scintillating light escaping the crystal is done with an unconventional technique since both photomultipliers and photodiodes (commonly used for this purpose) are unsuited to the use in vacuum  and at very low temperature. The light is detected by a second bolometer, a Si or Ge undoped wafer provided with a temperature sensor. Thanks to the small volume of the wafer, the heat capacity of this bolometer is so low that even optical photons give rise to a sizable temperature increase. 

The simultaneous detection of the heat and scintillation components of an event allows to identify and reject $\alpha$ particles with very high efficiency (close to 100\%). The concept is very simple: the ratio between the light and phonon yield is different for $\alpha$ and for $\gamma$/$\beta$ interactions.
In addition, it has been shown that $\alpha$/$\gamma$ discrimination by pulse shape analysis is also possible in some crystals, both in the heat and light channel~\cite{arnaboldi11}. 
The $\alpha$ rejection capability is particularly appealing when applied to candidates with a large \QBB. In fact, above 2.6 MeV the natural $\gamma$/$\beta$ contributions from environmental and material radioactivity tend to vanish and $\alpha$’s are the only really disturbing background source. 
R\&D measurements carried out in the past decade have identified a full list of candidates (e.g. $^{48}$Ca, $^{100}$Mo, $^{116}$Cd, and $^{82}$Se) which are characterized by scintillating compounds such as PbMoO$_4$, CdWO$_4$, CaMoO$_4$, SrMoO$_4$, ZnMoO$_4$, CaF$_2$, and ZnSe~\cite{arnaboldi11,arnaboldi11a,arnaboldi10,beeman12a,beeman12b}. 
In particular, $^{82}$Se and  $^{100}$Mo look the most promising ones. 
Scintillating bolometers based on their compounds have been operated successfully and the complete elimination of $\alpha$ events is expected to lead to specific background levels of the order of 10$^{-4}$ \cpkky~\cite{beeman12}.
Therefore they have been selected as the basic ingredients of the above mentioned projects.

The choice of LUCIFER has fallen on ZnSe, because of the favorable mass fraction of the candidate, the availability of large radio-pure crystals, and the well-established enrichment/purification technology for Se. 

LUCIFER~\cite{Lucifer} will consists of an array of ZnSe crystals similar to the Cuoricino one and designed to fit exactly the experimental volume of the Cuoricino cryostat (since the baseline for the LUCIFER program is to use this cryostat). The array will be realized with ZnSe crystals grown from enriched material. About 15 kg of metallic Se (enriched to 95\% in $^{82}$Se) will be purchased and used to grow ZnSe crystals. The chemical process used to produce the ZnSe compound from the enriched material and the following crystal grow procedure imply -- as usual -- a material loss that in the case of ZnSe is quite relevant. The goal of the LUCIFER collaboration is to be able to achieve a production yield of about 65\% (still to be demonstrated), this will result in about 17 kg of ZnSe crystals corresponding to 9.3 kg of $^{82}$Se. Assuming an energy resolution of 13~keV FHWM~\cite{Lucifer} and a background rate of 10$^{-3}$~\cpkky~the experiment will work in nearly zero background condition. The sensitivity estimate yields \SFzz = 1.6\per 10$^{26}$ yr in 5 years.

The compounds ZnMoO$_4$ and CaMoO$_4$ are equally promising, and have been selected for the LUMINEU and AMoRE experiments. 
For other very interesting isotopes, like \tect employed in CUORE, scintillating materials have not yet been identified.
However, lso in this case the $\alpha$ rejection could be achieved by exploiting a similar approach based on the much weaker Cerenkov signal~\cite{tabarelli,miscerenkov}.
Indeed, the two electrons emitted in the \BBz decay are above threshold and can produce a flash of light with a total energy of approximately 140 eV. This is not the case however for $\alpha$ particles which are by far below threshold. 
The detection of the Cerenkov light would improve dramatically the sensitivity of CUORE, providing the possibility to reduce the present specific background (10$^{-2}$~\cpkky) by an order of magnitude. However, the detection of the Cerenkov light in bolometers, with the proper sensitivity to  discriminate events from natural radioactivity, still requires an intense R\&D program aiming at exceptionally sensitive light detectors.

%%%%%%%%%%%%%%%%% Semiconductor detectors %%%%%%%%%%%%%%%%%
\section{High Purity Germanium Detectors Enriched in $^{76}$Ge}
\label{sec:semi}

The use of Germanium diodes to search for \BBz decay dates back to 1967~\cite{fiorini67} when it was realized that the decay of $^{76}$Ge could be investigated with a calorimetric approach, using what was at the time -- and is still today -- the best detector for gamma spectroscopy in the MeV range. 

Today, standard HPGe diodes reach energy resolutions of the order of 0.2\% FWHM at 2 MeV and masses as high asfew kg.  To be efficiently used in \BBz searches, the germanium crystals have to be grown starting from isotopically enriched material since the natural isotopic abundance is low (Table~\ref{tab:phsp}). 
This has been done by the HDM \cite{hmosc01} and the IGEX \cite{aalse02} collaborations who carried out the reference experiments in the field. Using respectively 11 kg and 8 kg of isotopically enriched (86\%) germanium, these two experiments were located in deep underground laboratories (respectively, LNGS and LSC). In both experiments, the set-up consisted in HPGe diodes operated in a low contamination copper cryostat, surrounded by lead and/or copper thick shields. A Pulse Shape Analysis (PSA) technique was used to reject multi-site events (typical of non-\BB interactions). However, in both experiment this was possible only on a sub-set of the total exposure. The two experiments concluded their operation with two of the most sensitive \BBz result ever reached: a 90\% C.L. limit on $^{76}$Ge \Tz of 1.9 10$^{25}$ yr \cite{hmosc01} (HDM, exposure=35.5 kg\per yr) and 1.57 10$^{25}$ yr \cite{aalse02} (IGEX, exposure=8.9 kg\per yr). 

Today two large scale projects benefit of the heritage of HDM and IGEX for their ambitious program: GERDA, a mainly European collaboration,  and MAJORANA, mainly US collaboration. Both experiments have phased programs with time schedules dictated by funding, isotope production and a continuous update of the project on the basis of the knowledge acquired along the path. The ultimate goal is to merge the two experiments in a single one-tonne, zero background \BBz project.

\subsection{Specific Backgrounds in Germanium Experiments}

The transition energy of  $^{76}$Ge is considerably lower (\QBB=2039~keV) than that of most of the isotopes discussed so far. This implies that -- in spite of their high resolution --  experiments using Ge diodes fight against an unusually large number of dangerous background sources. 
Both \udt and \thdt can contribute to the \BBz ROI through their major $\gamma$ emissions while the short-range $\beta$ and $\alpha$ particles emitted by the same chains can mimic a \BB event only in the case of contaminations sufficiently close to the detectors. Furthermore, sizable background contributions can be due to a number of long-lived cosmogenically produced isotopes (e.g. $^{68}$Ge  with \TH=271 d, $^{60}$Co with  \TH=1925 d, $^{56}$Co with \TH =78 d), characteristic of copper and germanium activation, as well as a number of anthropogenic radioisotopes (i.e. artificially produced radioisotopes as $^{207}$Bi with \TH=31.5 yr). Thanks to the high energy resolution, \BBd yields a completely negligible background.

GERDA and MAJORANA aim at a background reduction of more than one order of magnitude with respect to HDM and IGEX. While both experiments are based on the same technology, the way they plan to achieve their background goal is influenced by the different conclusions of the respective precursors  concerning the most relevant background sources.
 
In HDM the main background sources were identified in radioactive natural/cosmogenic contaminations of the experimental apparatus (in the lead and copper shields and in the copper of the cryostat), with a negligible contribution coming from Ge diodes themselves (this contribution was excluded on the basis of the absence of \udt and \thdt $\alpha$'s peaks). This has biased the unconventional design of the GERDA project aimed at surrounding the detectors only with an ultra-pure material acting as passive or (better) active shield. 

In IGEX, on the contrary, the background counting rate was ascribed to radioisotopes produced by cosmic-ray neutron spallation reactions, that occurred in the detector and cryostat components while they were above ground. The major contributions were identified in $^{68}$Ge,  $^{56}$Co and  $^{60}$Co. This has influenced the choices of the MAJORANA collaboration that has focussed the attention on the control and reduction of cosmogenically generated isotopes through material preparation completely carried out underground.

As a concluding remark, in this section it is worth to underline how impressive are the background achievements -- already obtained by the past generation Ge experiments -- in spite of the low $^{76}$Ge transition energy. The extremely background counting rates characterizing these experiments have been obtained through a careful choice of the setup materials. Indeed, what is today the standard procedure in the field was just pioneered by germanium experiments.  

At the present stage of the realization of the next generation experiments a new ingredient has to be added to maintain the competitiveness of this technology: an active background reduction based on a new detector design. This represents the new frontier and is presently addressing large experimental efforts.  

\subsection{Pulse Shape Discrimination in HPGe Diodes}

Most of the background sources listed in the previous section produce events in the ROI through multiple Compton scattering of higher energy $\gamma$'s. This is the only possible contribution coming from radioactive contaminations far from the detectors, while for contaminations in  close proximity of the diodes (or in the HPGe itself) also $\beta$'s and $\alpha$'s can produce relevant energy depositions.
The HPGe used both by GERDA and MAJORANA are of p-type, with a large and 
thick n+ electrode which effectively shields the sensitive volume from impinging $\beta$'s or $\alpha$'s, and a thin p+ electrode that is the only entrance window for these particles, after an almost negligible energy degradation. 
In the case of $\gamma$ or $\beta$+$\gamma$ energy depositions in the sensitive volume, the topology of the event is characterized by multiple interaction sites inside the crystal (MSE), extending over several centimeters. 
Single site events (SSE) extend over volumes of few mm cube and originate from single Compton scattering, from photoelectric or multi-site interactions very close to each other. The latter category includes electron induced interactions and double-escape events. Double beta events are SSE.

As discussed below, in germanium diodes SSE and MSE have a different pulse shape which allows to implement background rejection techniques that can be highly efficient. As an example, the HDM experiment measured a background counting rate of about 0.19~\cpkky~in the region from 2000 to 2080 keV, and -- using a PSA technique based on neural network computations -- managed to reduce it  by a factor of 3 down to 0.06 counts/(keV kg yr). 

The reason for a different pulse shape is the lack of uniformity of the electric field over the detector sensitive volume. Indeed, the time structure of the charge signal changes according to the topology of the initial energy deposition: the current pulse is higher when charges drift through the volume of a large weighting potential gradient~\cite{GERDA-dete}. This implies that the number of sites where primary ionization occurs and the differences in charge trajectories and drifting times induce a shaping of the signal that can be used to distinguish single sites event (SSE) from multiple site events (MSE). 

The rejection capability can be optimized with a proper design of the detector. In p-type point contact detectors (PTPCGe) the signal electrode is very small if compared to standard coaxial HPGe detector, this results in a completely different field distribution capable of enhancing the differences between SSE and MSE pulses.  Examples of this technology are the commercially available Broad Energy Germanium detectors (BEGe) produced by Canberra Company and used in GERDA. Practically the same design is  used in the MAJORANA Demonstrator (MJD). These are p-type HPGe diodes with a point-like p+ electrode for induced charge collection and a Li-drifted n+ contact (0.5 mm thickness) covering the whole outer surface, including most of the bottom part. Due to their peculiar electric field configuration and limited size of the collection implant they exhibit a superior pulse shape discrimination performance: SSE and MSE can be easily distinguished simply on the basis of the ratio A/E with A being the pulse amplitude (measured as the maximum of the pulse current) and E being the energy~\cite{GERDA-PSA}.

On the contrary, in coaxial HPGe (namely the kind of detectors employed in the past generation experiments, like HDM and IGEX) the difference in shape is less pronounced and more varied, requiring sophisticate algorithms (as neural network systems) for event classification. 

Finally, alternative detector technologies aiming at very efficient background rejection capabilities have been also proposed (Canberra SEGA, ~\cite{MJD-SEG}). Based on n-type segmented diodes they are able to achieve remarkable event discrimination but have been so far superseded by the more practical PTCPGe design.

\subsection{The GERDA Experiment}

\def\coaxRate{1.75$^{+0.26}_{-0.24}$ \per 10$^{-2}$~}
\def\BEGeRate{3.6$^{+1.3}_{-1.0}$\per 10$^{-2}$~}
\begin{small}
\begin{tabular}{|p{2.5cm}|p{12cm}|}
\hline 
\textbf{GERDA-I} & $^{76}$Ge (\QBB=2039~keV) running \\
\hline
\textbf{FWHM} & \ca 4 keV \\
\textbf{EXPOSURE} & 21.6 kg\per yr \\
\textbf{MASS} & 13.5 kg of $^{76}$Ge, only 10.9 used for this result (18 kg coaxial HPGe + 3.6 kg BEGe) \\
\textbf{BKG} & \\
~~rejection & single-site vs. multi-site events separation \\
~~rate & \coaxRate \cpkky~on coaxial HPGe (no PSA) \\
& \BEGeRate \cpkky~on BEGe (no PSA) \\
~~sources & $^{42}$K and $^{222}$Rn in LAr + $^{214}$Bi and $^{228}$Th in detector assembly + surface $\alpha$'s \\
\textbf{\Tz} &  $>$ 2.1 \per 10$^{25}$ yr at 90\% C.L. (in the Bayesian case $>$1.9\per 10$^{25}$ yr)\\
\hline
\textbf{GERDA-II} & $^{76}$Ge (\QBB=2039~keV) upgrade of GERDA-I, under construction\\
\hline
\textbf{FWHM} & 3 - 4 keV \\
\textbf{MASS} & 30  kg of $^{76}$Ge (18 kg coaxial HPGe + 21 kg BEGe)\\
\textbf{BKG} & \\
~~rejection & single-site vs. multi-site events separation \\
~~goal & 10$^{-3}$ \cpkky (design value with PSA) \\
\textbf{\SFzz} & 6.0 \per 10$^{26}$ yr in 5 years (optimistic: assuming zero-background condition reached on both coaxial and BEGe detectors, detection efficiency of 66\%)\\
\textbf{\SFz} & 2.2 \per 10$^{26}$ yr in 5 years (conservative approach: assuming the best background counting rate measured by GERDA-I, detection efficiency of 83\%)\\
\hline
\end{tabular}
\end{small}
\newline
\newline

\begin{figure}
\begin{center}
	\includegraphics[width=1\linewidth]{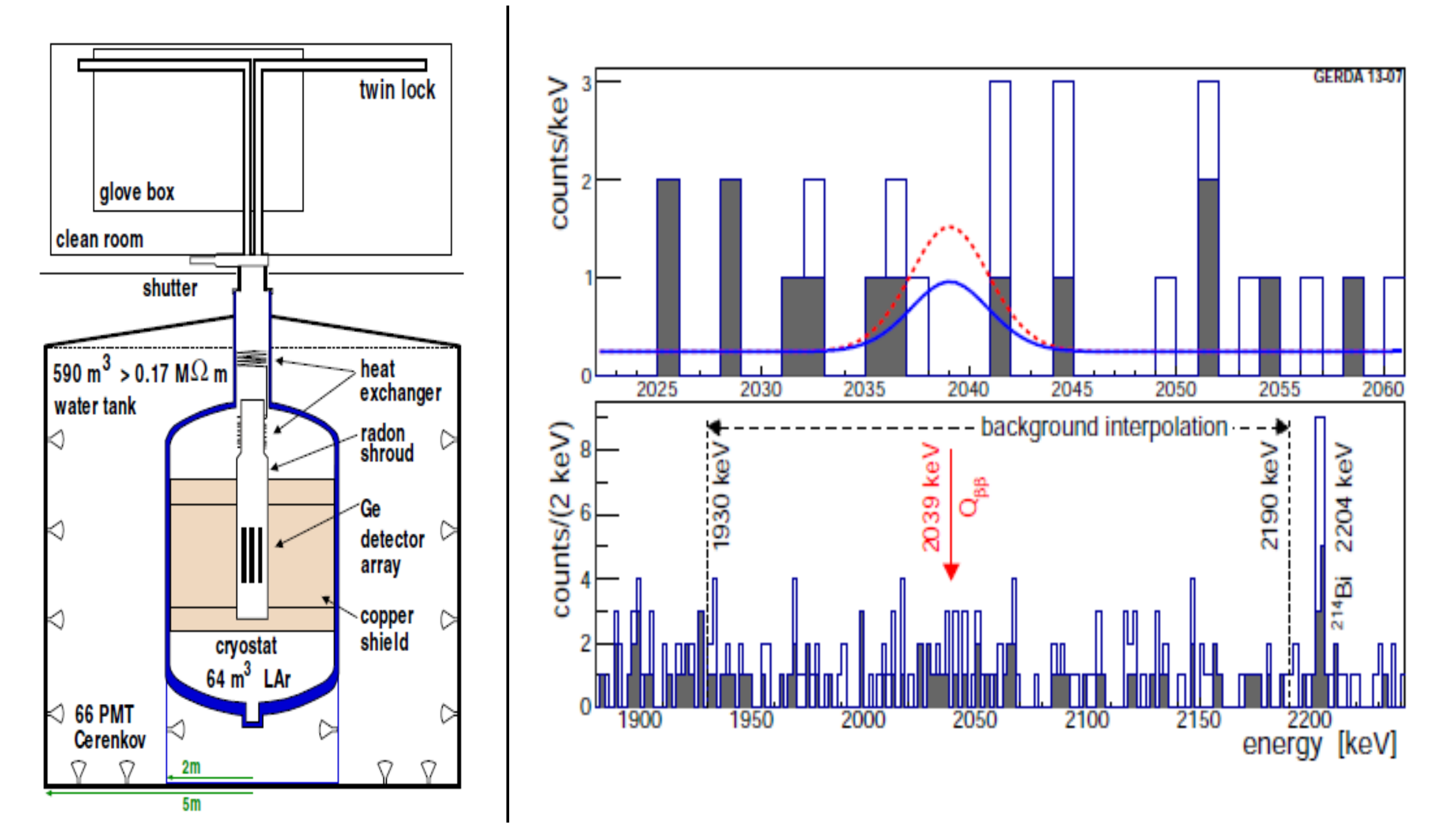}
\end{center}
\caption{Left: schematic drawing of the main components of the GERDA experiment. Right: GERDA phase I results. In the top panel is shown the combined energy spectrum from all enriched Ge detectors without (open histogram) or with (filled histogram) PSA. The lower panel shows the region used for the background interpolation. In the upper panel, the spectrum zoomed to \QBB is superimposed with the expectations (with PSD selection) based on the central value of KDHK~\cite{klapd04}, \Tz = 1.19 $\times$ 10$^{25}$ yr (red dashed line) and with the Gerda phase I result  \Tz $>$ 2.1$\times$ 10$^{25}$ yr at 90\% C.L. (blue solid line). (figure and description from~\cite{GERDA-0n})}
\label{fig:GERDA}
\end{figure}

Evolved from the HDM experiment, GERDA\cite{GERDA-dete} implements the concept of Ge diodes immersed in a liquid argon (LAr) bath~\cite{klapd01} for a radical background suppression. The experiment, installed in LNGS, looks today as shown in Fig.~\ref{fig:GERDA}. A stainless-steel cryostat filled with liquid argon (\ca 100 tonnes) is surrounded by a water Cherenkov detector.
\ca 86\% isotopically enriched HPGe detectors are mounted in strings (each of about 3-5 detectors) which are suspended from the top in the center of the cryostat. 
The water tank shields the inner part of the set-up from $\gamma$ radiation due to rock radioactivity and serves as muon veto (being completed  -- at the top of the cryostat -- by plastic scintillator panels, realizing a complementary muon coverage where the water Cerenkov detector is thinner). The cryostat has an internal lining of ultra-pure copper, used primarily to reduce the $\gamma$ radiation from the steel vessel itself (as a rule of thumb copper is less radioactive than most materials, including steel which however is preferred for its mechanical qualities and costs).
LAr serves both as a passive shield and as a refrigerant for the HPGe diodes. The motivation for this shielding configuration are various. With respect to conventional set-up, the naked diodes are far from any cladding materials (with their radioactive contaminations) and a liquid can be easily purified to extremely low levels of contaminants (the main worries in the case of LAr are radon and $^{42}$Ar, discussed below). Moreover, in LAr the $\gamma$ production from muons interaction is much lower -- thanks to its low Z -- than in the traditionally used high Z shielding materials (as copper and lead). Finally, LAr offers the future possibility of reading out the Ar scintillation light for additional background rejection.

The pre-operation phase of GERDA allowed to highlight two weak points in the project: a different behavior of HPGe in LAr with respect to liquid nitrogen (that was the refrigerant considered in the early phase of the project and the one where the naked HPGe were tested) and an unexpected high contribution from $^{42}$Ar. The former problem consisted in an excess leakage current appearing upon $\gamma$ irradiation of the detectors, it was solved by changing the passivation layer on the HPGe surface. The latter was ascribed to an anomalous concentration (20 times higher then expected) of $^{42}$K close to the detectors.
$^{42}$K is the progeny of $^{42}$Ar, a known radioactive contaminant of argon. It $\beta$ decays with a Q=3525 keV and \TH = 12.36 h, with a most intense $\gamma$ line at at 1524.7 keV (B.R.=18.1\%). When close to the detectors, the emitted $\gamma$ and $\gamma$+$\beta$ particles can produce events in the ROI. The reason of this surprise was at the end clarified: $^{42}$K is produced, after $^{42}$Ar decay, as a positively charged ion which migrates toward the diodes attracted by their externally-extended weak electrical field. The solution of the problem was obtained by the installation of a thin (60 $\mu$m) copper electrostatic shield (called mini-shroud) surrounding the detectors array at a very close distance (few mm) and permeable to LAr. A further thin copper shield protects the detector from radon emanation (radon shroud).

GERDA is designed to proceed in two phases:
\begin{itemize}
\item GERDA-I (presently taking data) is going to verify the KHDK claim~\cite{klapd06} using the coaxial HPGe enriched detectors inherited from the HDM and IGEX experiments (\ca18~kg of \ca86\% enriched Ge) and few new detectors (enriched BEGe diodes, deployed only in June 2012, having a total mass of \ca3.6 kg of \ca88\% enriched Ge).
\item GERDA-II will see the deployment of additional detector strings to achieve (21+18)~kg of germanium isotopically enriched in $^{76}$Ge to 86\% (for the old coaxial HPGe's) and 88\% (for the BEGe's), aiming at a 5 years sensitivity of $1.1\times10^{26}$yr.
\end{itemize}

Depending on the actual physics results of the two experimental phases, a third phase using 500 to 1000 kg of enriched germanium detectors is planned, merging GERDA (this is phase III) with MAJORANA.

The first result released by the collaboration was the \BBd one, confirming previous measurements~\cite{agostini13}, and a detailed background study~\cite{GERDA-bkg}.  While this review was written, the unblinding of the phase I \BBz data was presented with a paper dedicated to background modeling~\cite{GERDA-bkg}  (with the identification of major background)  and the paper reporting the \BBz result.
Phase I results are here summarized. 

The average energy resolution at \QBB is 4.8 keV and 3.2 keV respectively for the coaxial and the BEGe detectors.  The total exposure and the \BBz counting rate are:

\begin{itemize} 
\item 17.9 kg\per y (gold-data) plus 1.3 kg\per y (silver-data) collected with 6 of the 8 coaxial HPGe diodes. Two coaxial diodes had to be switch-off due to excess leakage current (one of them after having collected a fraction of data). The silver-data corresponds the deployment of the BEGe detectors, for a short period a slightly higher than usual counting rate was observed. 
The corresponding \BBz rate is \coaxRate \cpkky (with PSA cuts).
\item 2.4 kg\per yr collected with 4 of the 5 BEGe diode. One BEGe is not used in the analysis because of instabilities. The \BBz rate is \BEGeRate \cpkky (with PSA cuts).
\end{itemize}

No excess of signal counts is observed over the background in the ROI (\QBB$\pm$5 keV) where the observed events are 6 for the coaxial HPGe and 2 for the BEGe, reduced respectively to 2 and 1 with the application of PSA cuts (see right panel of Fig.~\ref{fig:GERDA}). This is translated into a 90\% C.L. lower limit of \Tz $>$ 2.1 \per 10$^{25}$ yr using a frequentist approach to be compared with a median sensitivity of 2.4\per 10$^{25}$ yr at 90\% C.L. (similar results are obtained with a Bayesian analysis). The compatibility of the result with the KDHK claim is studied comparing the probability of two models describing collected data: $H_0$ is the model of background without \BBz signal and $H_1$ is the model with background plus the same \BBz signal found by KDHK in~\cite{klapd04}. The Bayes factor P($H_1$)/P($H_0$) is found to be 0.024. Assuming model $H_1$ to be true, the probability of observing 0 \BBz events in GERDA, namely the Bayes factor, is P(N$_{0\nu}$=0|$H_1$)=0.01~\cite{GERDA-0n}. Extending the GERDA profile likelihood to include HDM and IGEX spectra (i.e. using the sum exposure or the three experiment) the Bayes factor is further reduced to 2$\times$10$^{-4}$, i.e. model $H_1$ is strongly disfavored.
It is worth to note that the GERDA collaboration decided to take into account only the 2004 KDHK publication~\cite{klapd04} where a 4.2$\sigma$ \BBz evidence was reported with a half-life of  \Tz = 1.19 $\times$ 10$^{25}$ yr. Indeed later papers, again based on re-analysis of the same data, are characterized by an improved statistical significance. For example, the latest reported result~\cite{klapd08})  amounts to a 6$\sigma$ evidence with \Tz = $2.23^{+0.44}_{-0.31}\times~10^{25}$~yr. 

The major background sources contributing in the ROI are identified in $^{42}$K and $^{222}$Rn in the LAr plus $^{214}$Bi and $^{228}$Th in detector assembly and a contribution from $\alpha$'s in the p+ electrode surface (i.e. the only portion of the diode surface where the dead layer is so thin that $\alpha$'s can enter the active volume without being too degraded in energy).

GERDA will conclude phase I (the target exposure is already reached) as soon as ready to start with the upgrades required for phase II. 25 new BEGe detectors have been prepared by Canberra, totaling -- with the five already installed -- 30 BEGe (20.8 kg of Ge) that once added to the old coaxial HPGe will reach the phase II goal of about 21+18 kg of enriched Ge detectors. 

The background goal of this latter phase is 10$^{-3}$~\cpkky, i.e. more than one order of magnitude lower than the BEGe counting rate recorded in phase I. With such a low counting rate, achieved on both coaxial and BEGe detectors, the experiment would reach a nearly zero background condition, corresponding to a 5 years sensitivity \SFzz = 6.0 \per 10$^{26}$ yr. This is probably a very optimistic case, since at least for coaxial detectors the achievement of this low background condition looks very difficult (for example, background rejection through PSA is more than 2 times better in BEGe than in coaxial diodes). A more conservative hypothesis is to assume a counting rate of 1.7$\times$10$^{-3}$~\cpkky (the best recorded in phase I) for all the detectors, in that case the 5 years sensitivity is reduced to \SFz = 2.2 \per 10$^{26}$ yr.

The upgrades foreseen for phase II include various modification of the apparatus to host an increased number of detectors, with improvements on both radioactivity and electronics. 
The efforts to get rid of $^{42}$K background will focus on detector performance: with a lower noise the $\beta$+$\gamma$ events induced by $^{42}$K can be rejected using PSA. The instrumentation of LAr (i.e. read-out of the scintillation light of Ar) is on the other hand the way to mitigate $^{214}$Bi background.

\subsection{The MAJORANA Experiment}

\begin{small}
\begin{tabular}{|p{2.5cm}|p{12cm}|}
\hline 
\textbf{MJD} & $^{76}$Ge (\QBB=2039~keV) under construction\\
\hline
\textbf{FWHM} & 4 keV\\
\textbf{MASS} & 26 kg of $^{76}$Ge (30~kg of germanium enriched to 86\%) \\
\textbf{BKG} &\\
~~rejection & PSA: single-site vs. multi-site events separation \\
~~goal & 7.5 \per 10$^{-4}$ \cpkky (with PSA) \\
\textbf{\SFzz} & 4.4\per 10$^{26}$ yr in 5 yr (detection efficiency 70\%)\\
\hline
\end{tabular}
\end{small}
\newline
\newline

\begin{figure}
\begin{center}
\includegraphics[width=1\linewidth]{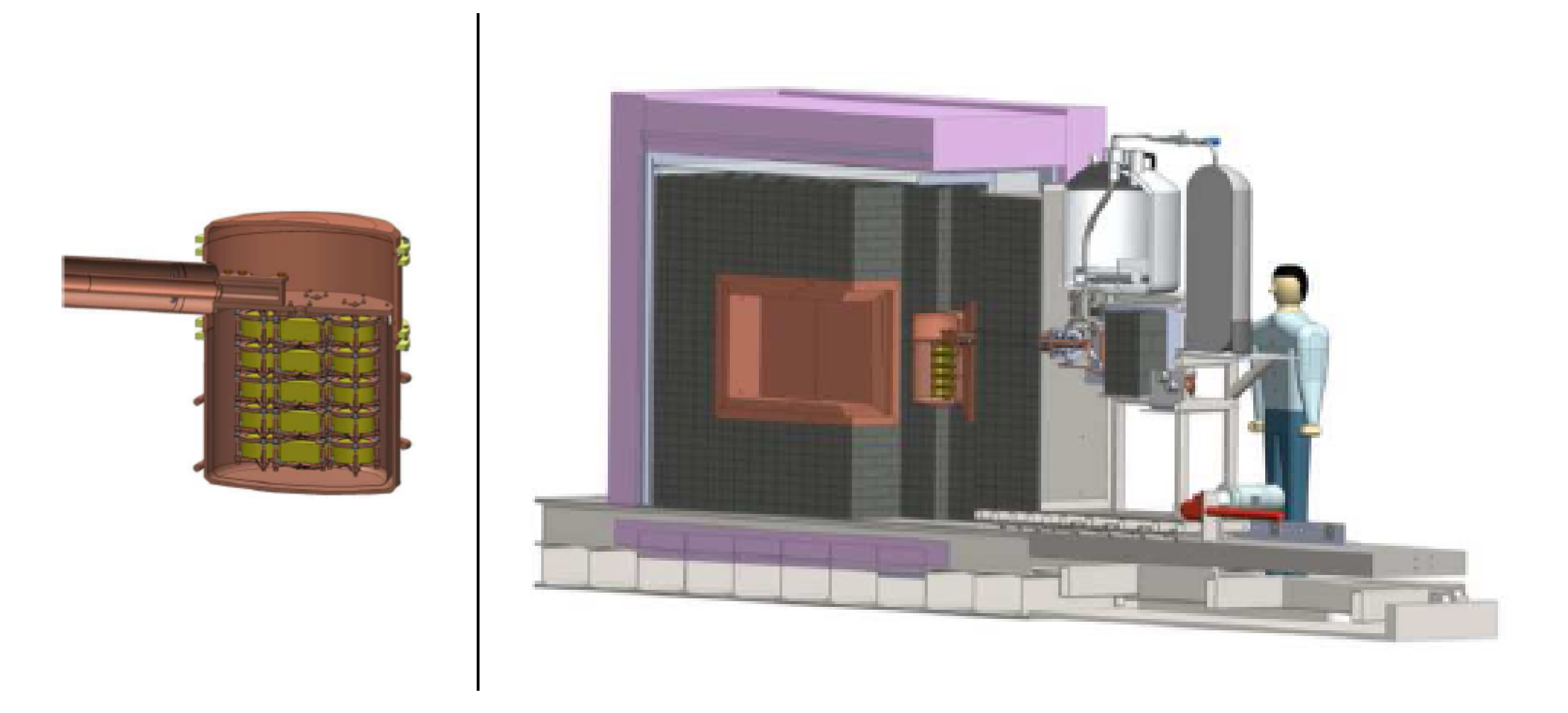}
\end{center} 
\caption{The MAJORANA Demonstrator: (left) cross sectional view of one cryostat and (right) of the entire apparatus showing one cryostat while being inserted in the shield (figure from~\cite{MajD2011}).}
\label{fig:MAJORANA}
\end{figure}

MAJORANA is an evolution of the IGEX experiment. The basic ideas behind the project are summarized in the year 2003 White Paper~\cite{Maj-white}: 

\begin{itemize} 
\item realize a large mass Ge experiment (final goal is a \Tz sensitivity of the order of 10$^{27}$~yr) based on a well known technology and design, i.e. using an array of hundreds of HPGe detector operated in a conventional configuration;
\item focus the main effort on two goals: the improvement of HPGe technology )aiming at the use of segmented HPGe with highly improved pulse shape capabilities) and the selection and/or custom production of high radio-purity materials. 
\end{itemize}
The proposed configuration~\cite{MajD2011} is based on an evolution of the traditional HPGe set-up: close-packed arrays of HPGe diodes (57 crystals each) are mounted inside ultra-clean electro-formed conventional cryostats, minimizing in this way the amount of structural materials in-between the diodes (see left panel of Fig.~\ref{fig:MAJORANA}). A number of these 57-crystal arrays is installed in a low-background passive shield provided with a muon active veto. The entire apparatus is installed in a deep underground laboratory. The ultimate goal of the project is the realization of a tonne scale experiment with a counting rate lower than 1 counts/(tonne $\cdot$ yr) in the ROI, i.e. nearly zero background condition. In addition to the extremely difficult challenge from the point of view of background rate achievement, both time and cost of this project are very high, in particular for what concerns germanium enrichment.
The present program of the MAJORANA collaboration is to realize a small-scale prototype to demonstrate the viability of the technique (the MAJORANA Demonstrator~\cite{MajD2011,TAUPmajorana}) and to define a one tonne scale project in collaboration with GERDA, aiming at a sharing of costs and of knowledge, having therefore the opportunity to benefit of the experience and skills deriving by the two initial stages of both experiments. 

The MAJORANA Demonstrator (MJD) will use about 40 kg of germanium diodes (\ca 30~kg will be of enriched $^{76}$Ge). The detector performance is comparable to GERDA's (the baseline for the MAJORANA demonstrator are the same PTPC Ge diodes used by GERDA) and the target background rate is i.e. $0.75\times10^{-3}$  \cpkky~ in the 4 keV ROI (with PSA)~\cite{TAUPmajorana}, nearly identical to that of GERDA-II. 
Screening and selection of commercially available materials may not allow to fulfill the background requirements, therefore special techniques have been developed not only for the custom production of the MJD enriched detectors (which is quite common in this field) but also for the custom production of the inner shielding material (which today is a standard procedure only for experiments using liquid detectors or shields, but not for solids). The cryostat enclosing the HPGe array and the inner shielding layer of the MJD are made of copper. The MJD radioactivity requirement for this material are extreme: \udt and \thdt contaminations below 1 $\mu$Bq/kg (a contamination level that -- by itself -- is very hard to measure) and a negligible cosmic ray activation. 
The solution was identified in the underground electro-forming of copper. The collaboration has realized a facility at 1500~m depth (SUSEL, Stanford Underground Science and Engineering Laboratory, South Dakota, USA) where electro-forming of copper is done in underground clean rooms, purifying in this way the copper from \udt and \thdt as well as from cosmogenically generated radionuclides (\cs is an example) that will not be regenerated thanks to the reduced cosmic ray flux. 

The same facility will host the MJD operation. This will consist (Fig.~\ref{fig:MAJORANA}) of two electro-formed cryostats, the first will be ready in 2013 and will contain both natural and enriched HPGe, surrounded by a onion-like shield made of 5~cm of electro-formed copper, 5~cm of Oxygen Free High Conductivity (OFHC) copper (the procedure used for the production of this special kind of copper ensures very high radiopurity levels), 45~cm of lead and 30 of polyethylene with embedded a plastic scintillator used as cosmic rays veto. The completion of this phase is expected in 2014.
The one-year sensitivity for the MJD is (according to Eq.(\FZ0)) \SFzz \ca 1.7\per 10$^{25}$ yr scaling to 4.4\per 10$^{26}$ yr in five years.

%%%%%%%%%%%%%%%%%% Liquid scintillator based experiments %%%%%%%%%%%%%
\section{Loaded Organic Liquid Scintillators}
\label{sec:scint}

In the last decade a new class of \BB experiments started occupying the international scenario. These are based on the conversion to \BBz decay searches of huge liquid-scintillator or water-Cherenkov detectors that were first designed and employed for neutrino oscillation measurements. Indeed, the need of a low background counting rate (low intrinsic radioactivity, shielding, underground location), of a high detector efficiency and of an optimized energy resolution is common to the two research fields. 
Once their campaign of measurements with solar/reactor neutrino is completed these detector can be dedicated -- with minor modifications and therefore at limited expenses -- to DBD searches. This is what happened with Kamland-ZEN and what is in progress with SNO+, although the original idea dates back to 2001 with the proposal of dissolving Xe in Borexino~\cite{Caccianiga} or of placing an array of CdWO$_4$ crystals inside its core (CAMEO proposal~\cite{CAMEO}).

These experiments are characterized by the capability of reconstructing the interaction vertex that allows to define a fiducial volume where the \BBz events have to be located in order to be accepted.
This allows to reduce the number of background sources that can mimic a \BBz decay. On the other hand, the poor energy resolution achievable in liquid scintillators imply first that \BBd is an irreducible background (i.e. the choice of the \BBd candidate has to take into account \BBd rate) and second that the \BBz result can be extracted only after a careful background reconstruction (similarly to what happens in the case of most experiments based on tracking detectors). 

\subsection{KamLAND-ZEN}

\begin{small}
\begin{tabular}{|p{2.5cm}|p{12cm}|}
\hline 
\textbf{K-ZEN} & \xe (\QBB= 2457~keV) stopped for upgrade\\
\hline
\textbf{FWHM} & 240 keV \\
\textbf{EXPOSURE} & 89.5 kg(\xen)~\per~yr\\
\textbf{MASS} & 179 kg (first data set) 125 kg (second data-set) of \xe\\
\textbf{BKG} & \\
~~rejection & prompt and delayed coincidences + fiducial volume (FV)\\
~~rate & \ca 1.5\per 10$^{-4}$ \cpkky \\
~~sources & $^{110m}$Ag (probably Fukushima fallout, produces a peak near \BBzn)\\
\textbf{\Tz} & $>$ 1.9\per 10$^{25}$ yr at 90\% C.L. \\
\hline
\end{tabular}
\end{small}
\newline
\newline

The KamLAND-Zen \cite{KamZen2012} experiment is based on a modification of the existing KamLAND detector carried out in summer of 2011: a mini-balloon filled with a Xe-loaded liquid scintillator have been added in the very core of the apparatus to search for \xe \BBz decay (for a discussion of \xe as a \BBz source see Sec.~\ref{sec:tpc}). 
KamLAND is located in the site of the earlier Kamiokande at a depth of 2700 m.w.e. and is used since 2002 for neutrino oscillation measurements.

\begin{figure}
\begin{center}
\includegraphics[width=1\linewidth]{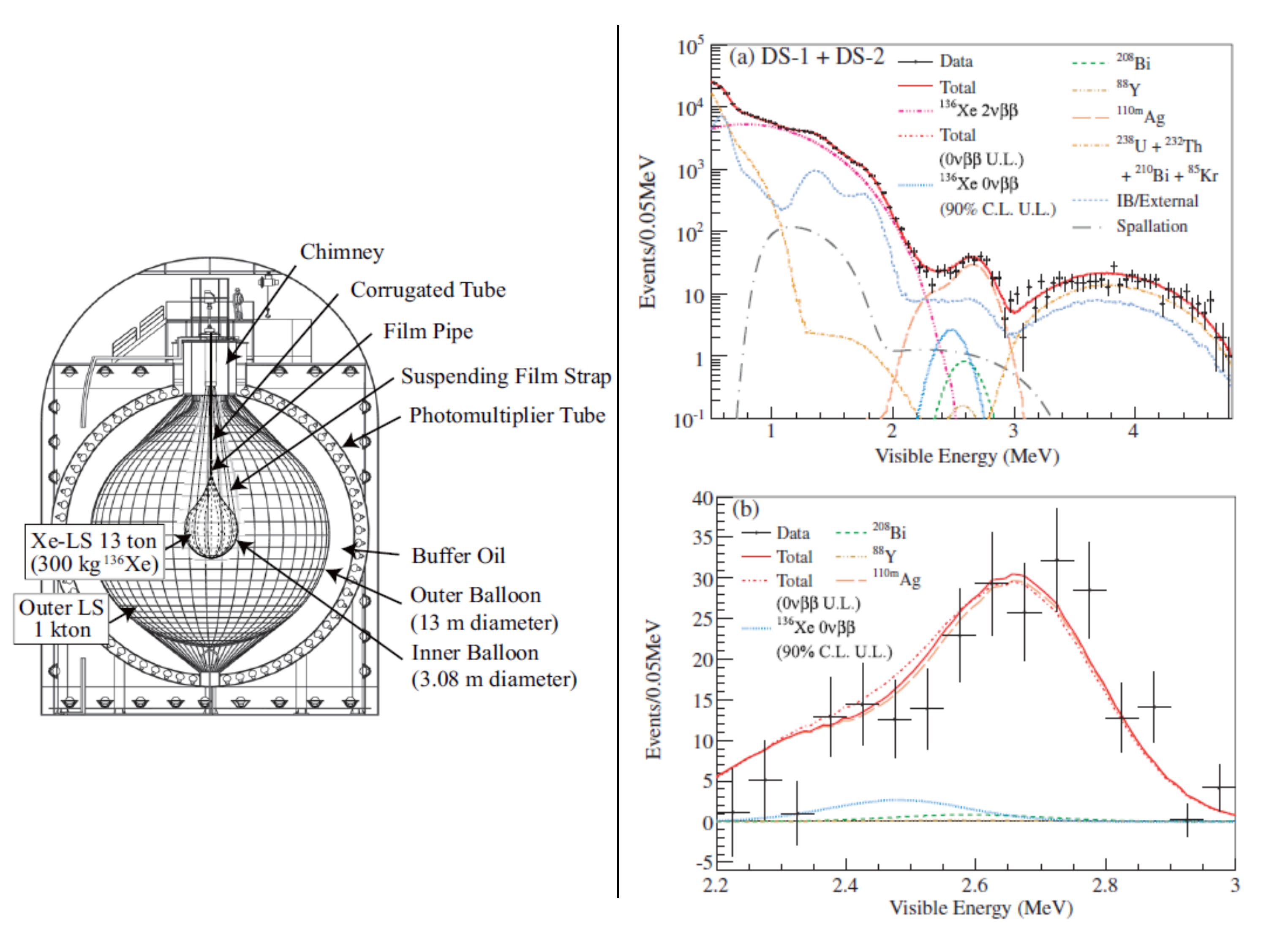}
\end{center}
\caption{Left: KamLAND-ZEN detector, see text for a description (figure from~\cite{KamZen2012}). Right: KamLAND-ZEN \BBz result (figure from~\cite{KamZen2013}). In the top panel (a) the energy spectrum of selected candidate events together with the best-fit backgrounds and \BBd decays, and the 90\% C.L. lower limit for \BBz decays. In the bottom panel (b) closeup of (a) in the \BBz region after subtracting known background contributions.}
\label{fig:KamZen}
\end{figure}

The detector today looks as in Fig.~\ref{fig:KamZen} (left panel). It comprises:
\begin{itemize}
\item the Inner Balloon (IB) (made of a 25-${\mu}$m-thick transparent nylon film) suspended at the center of the detector, this contains the \BB source in the form of 13 tons of Xe-loaded liquid scintillator (Xe-LS);
\item the Outer Balloon (OB) (135-${\mu}$m-thick nylon/EVOH film) filled with 1 ktonne of liquid scintillator (LS), this is the detector used for neutrino oscillation measurement in KamLAND while in KamLAND-ZEN acts as an active shield for external gamma’s;
\item the Stainless Steel Tank (SST) that is the containment vessel for the two balloons. The gap between the SST and the OB is filled with a buffer of mineral oil that passively shields the LS from external radiation. The inner surface of the SST is covered by an array of 1879 photomultiplier tubes (PMTs) that read-out the scintillation signal produced either in the IB (\BBz decay candidate events) or in the OB (background events).
\item a 3.2 ktonne water-Cherenkov detector -- read out by 225 PMTs -- that surrounds the whole structure. This Outer Detector (OD) absorbs gamma-rays and neutrons from the surrounding rock and provides a tag for cosmic-ray muons.
\end{itemize}

The LS is a mixture of ~80\% dodecane, ~20\% pseudocumene plus PPO. The Xe-LS has a similar composition to which is added a (2.52 \pom 0.07)\% in weight of enriched xenon gas (\ca 300 kg) with isotopic abundances (90.93 \pom 0.05)\%  for \xe and (8.89 \pom 0.01)\% for $^{134}$Xe. 

A \BBz decay is observed through the detection of the scintillation light from the two coincident electrons emitted in the transition. The two particles cannot be separately identified and only their summed energy at 2.458 MeV can be measured. Various background sources can hide this signal due to the poor  energy resolution of the detector. Indeed, $(6.6\pm0.3)\% / \sqrt{E[MeV]}$ is the 1$\sigma$ resolution estimated with multi-gamma calibration, which means a FWHM resolution of \ca 240 keV at the \BBz energy.

Data acquisition and analysis aims at the reconstruction of the background spectrum on a wide region (from \ca 0.5 to 5 MeV) besides using multiple cuts to select candidate \BB events. These are:

\begin{itemize}
\item a fiducial volume (FV) cut to select only events originating inside the IB (the \BB source);
\item a cut that removes both muon and muon induced events (i.e. events occurring within 2 ms after a muon);
\item a delayed coincidence cut, applied to remove events from the \bidq - $^{214}$Po cascade;
\item a delayed coincidence cut that removes antineutrino induced events (mainly from reactors); 
\item a cut based on the time-charge distribution in the vertex recorded by the photomultiplier array, that removes poorly reconstructed events.
\end{itemize}

The FV cut is designed in order to mitigate background coming from the radioactivity of the mini-balloon. Indeed the study of the vertex distributions of candidate \BBd and \BBz events shows an increase near the IB boundary that is ascribed to $^{134}$Cs in the case of the \BBd region and \bidq in the case of the \BBz region. The FV is therefore smaller than the IB volume, thus reducing the active mass of \xe. The presence of $^{134}$Cs and $^{137}$Cs and the ratio of their activities is compatible with a contamination of the IB balloon related to the Fukushima accident. Other fallout isotopes might therefore be present (although not directly observed).

Background events surviving cuts are ascribed to three categories: external to the Xe-LS (mainly from IB material), from the Xe-LS and induced by spallation. A careful study is performed to identify and disentangle the various background sources. \BBd and \BBz half-lives are estimated as the result of a best-fit spectral decomposition, MC simulations are used to represent the spectral shape of the different sources whose weight in the fit is in some case constrained by independent measurements of the source intensity.
The result is shown in Fig.\ref{fig:KamZen} (right panel): the spectrum shows a peak structure centered slightly above the \BBz region. To account for this peak all the isotopes in ENSDF database~\cite{ensdf} have been analyzed and few candidates (with the correct spectral shape and an ancestor live time greater than 30 days) have been identified. These are  $^{110m}$Ag (Q=3 MeV, $\tau$=360~day),  $^{88}$Y, \cs and  $^{208}$Bi that can be either Fukushima fallout products or (except $^{208}$Bi) the result of cosmogenic activation of Xe. These isotopes are therefore included in the likelihood function, with unconstrained weights. The peak structure is found to be compatible with a $^{110m}$Ag dominant contamination. 
The results reported in the more recent paper ~\cite{KamZen2013} refer to two data-sets collected before and after an attempt of Xe-LS purification. The second data-set has a smaller FV (125 kg instead of 179 kg of \xe) due to additional fiducial volume cuts made around the siphoning hardware left in place after the filtration. 
Unfortunately the filtration did not have the wanted effect: in the \BB window (the interval 2.2-3 MeV) the background counting rate due to $^{110m}$Ag is 0.19$\pm$0.02 counts/(tonne$\cdot$ day) in first data-set and 0.14$\pm$0.03 counts/(tonne$\cdot$ day) in the second data-set. 

\BBz and \BBd results reported so far are: 

\begin{itemize}
\item \Tz $>$ 1.9 \per 10$^{25}$ yr at 90\% C.L.  with an exposure of 89.5 kg\per yr of \xe (about 210 days) \cite{KamZen2013}
\item $T_{1/2}^{2\nu} = 2.38 \pm 0.02(stat) \pm 0.4(syst) \times 10^{21}$ yr for an exposure of 30.8~kg\per yr of \xe (77.6 days) \cite{KamZen2012}, compatible with the EXO~\cite{EXO-2n,EXO-2n-2013}.
\end{itemize}

The first phase of the experiment was terminated in order to start a purification campaign to remove the $^{110m}$Ag isotope. This is done by removing the Xe from the LS and distilling the LS to purify it, meanwhile considering the possibility of a substitution of the mini-balloon.

\subsection{SNO+}

\begin{small}
\begin{tabular}{|p{2.5cm}|p{12cm}|}
\hline 
\textbf{SNO+} & \tect (\QBB=2527~keV) under construction \\
\hline
\textbf{FWHM} & \ca 240 keV (evaluation done on the basis of photon statistic in the scintillator~\cite{SNO-2008}) \\
\textbf{MASS} & 163 kg of \tect (with 0.3\% of natural Te in the liquid scintillator and a fiducial volume of \ca 20\%~\cite{TAUPsno+}) \\
\textbf{BKG} & \\
rejection & prompt and delayed coincidences + fiducial volume (FV) \\
goal & \ca 3\per 10$^{-4}$ \cpkky \\
\textbf{ \SFz} & 2.0\per 10$^{26}$ yr\\
\hline
\end{tabular}
\end{small}
\newline
\newline

\begin{figure}
\begin{center}
\includegraphics[width=0.8\linewidth]{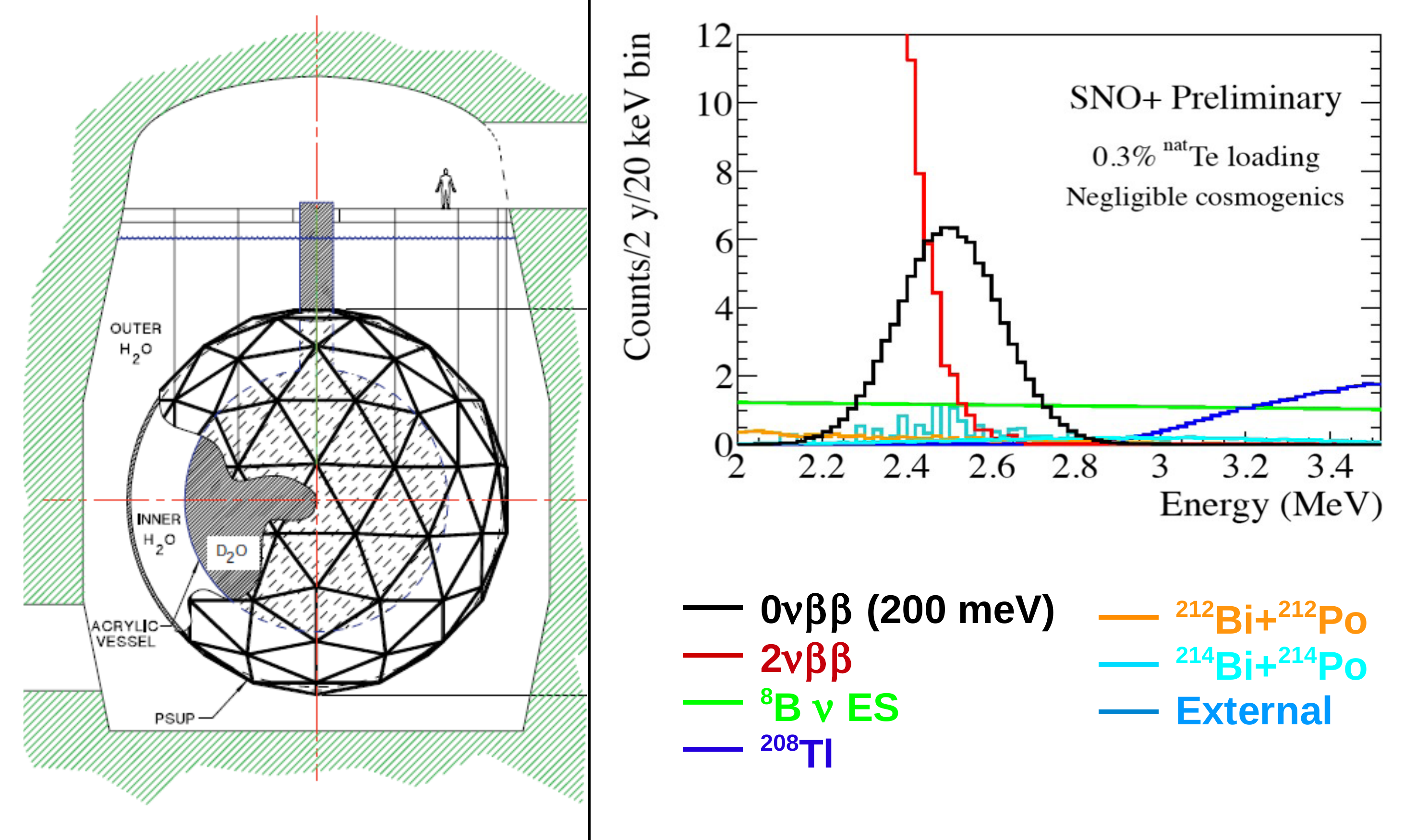}
\end{center}
\caption{Left: SNO detector (figure from~\cite{SNOdete}). Right: background prediction (figure from of~\cite{TAUPsno+}).}
\label{fig:SNO}
\end{figure}

The SNO experiment, located in the one of the deepest experimental sites (SNOlab, 6010 m.w.e.), was an imaging Cherenkov detector used in the first decade of the year 2000 for a successful campaign of solar neutrino measurements. 
The SNO detector (Fig.~\ref{fig:SNO}) consists of a 12~m diameter acrylic sphere filled with heavy water and surrounded by a shield of ultra-pure water (1700 tonnes) contained in a 32~m high, 22~m diameter tank.  A stainless steel geodesic structure supports \ca 9500 photomultipliers looking toward the the center of the acrylic sphere to read-out the Cherenkov light produced by neutrino interactions on deuterium. A smaller number of photomultipliers looking outwards are used to tag any particle producing Cerenkov light in the external water shield (5700 tonnes), and acting as a veto for cosmic rays and external background radiation. 

In November 2006, the experiment was terminated and heavy and light water were removed. At present, the SNO+ collaboration is modifying the detector by replacing the heavy water with about 780 tonnes of liquid scintillator (linear alkylbenzene with 2 g/L of PPO as wavelength shifter) loaded with a \BB candidate. The lower density of the scintillator with respect to water has required the installation of a rope net over the top of the acrylic sphere to anchors it to the floor. A purification system able to ensure U and Th concentrations in the scintillator similar to those reached in the Borexino experiment (10$^{-17}$ g/g of \udt and \thdt) is under construction~\cite{hartnell12,SNO+web}.

In a first proposal, $^{156}$Nd was the \BB isotope to be studied~\cite{SNO-2008}, but in April 2013 it was decided to start the first phase of the experiment with natural tellurium. Tellurium contains about 34\% of \tect, has a high transition energy and a much slower \BBd decay than $^{156}$Nd (nearly by two orders of magnitude). This choice has the advantage of being cheaper than the $^{156}$Nd one, mainly because neodymium isotopic enrichment is not obvious since cannot be done by centrifuge.
According to preliminary studies a 0.3\% loading of the liquid scintillator will be possible, corresponding to a mass of \tect of 800~kg. The goal of this phase is to reach a sensitivity that touches the IH region. If successful, a further step will consist in increasing the tellurium loading to 3\% (8~tonnes of \tectn) with the goal of covering the IH region. 
The sensitivity of the SNO+ experiment, in this first phase can be tentatively inferred from data and studies presented in the $^{156}$Nd proposals~\cite{SNO-2008,hartnell12}: the FWHM energy resolution \ca 240~keV (evaluation done on the basis of the scintillator photon yield: 400 photoelectron/MeV at 1 MeV), the fiducial volume is assumed to be 20\% of the actual volume (i.e. a \tect mass of 163~kg). The main background sources (as discussed in ~\cite{TAUPsno+}) are expected to be the \tect \BBd rate and the elastic interaction of solar neutrinos ($^8$B line). From the figure shown in~\cite{TAUPsno+} (reported in the right panel of Fig.~\ref{fig:SNO}) a counting rate integrated over the ROI of \ca 3\per 10$^{-4}$ \cpkky can be extrapolated. This corresponds to a 5 years sensitivity of \SFz =2.0\per 10$^{26}$ yr.

%%%%%%%%%%%% Outlook
\section{Summary and outlook}
\label{sec:sum}

We have reviewed the status and perspectives of the search for \BBzn.
Neutrinoless double beta decay is still the most promising probe to test lepton number violation and verify if neutrinos are Majorana particles.

The features of this challenge are more clear after the discovery of neutrino oscillations and the measurement of oscillation parameters.
Indeed, \BBz has turned into a sensitive probe for neutrino masses capable of providing relevant information on their absolute scale and ordering.

However, a precise nuclear physics knowledge is required in order to map the observed \BBz rates into neutrino mass constraints. 
 Actually, several calculations exist for \BBz nuclear matrix elements. 
They share common ingredients and differ in their treatment of nuclear structure. Unfortunately a relevant disagreement 
still exists between different calculations. This is of course a serious problem which has triggered in the past decade a strong effort to improve the situation.

From the experimental point of view, a good performance (high energy resolution and very low background), a proper scale (large number of \BBz candidate nuclei and long measure time), a favorable candidate and a proper experimental technique are the essential ingredients for a sensitive experiment.
These requirements are often conflicting, and no next generation proposal has succeeded so far in optimizing all of them (Table~\ref{tab:BBfutprm}). 
Indeed, most of the projects tend to excel in one or the other aspect, still missing the goal of getting the best sensitivity. 

In particular the high resolution calorimeters are facing an incredible effort to achieve the best performance but in most cases they cannot guarantee a proper scalability (indeed some of them have crossed the ZB boundary, while maintaining a good energy resolution).

\begin{sidewaystable}
\caption{\label{tab:BBfutprm} List of sensitivity parameters for some of the most advanced \BBz projects. 
B$_{iso}$ is the background per tonne of isotope mass in units of~counts/(keV\dot tonne\dot yr). The column labeled ``\emph{Perf.}'' reports the performance index (Eq.~(\ref{eq:sensitivityM})) in units of 10$^{-3}$ counts/(n$_{\beta\beta}$ \dot yr). The column labeled ``\emph{Sc.}'' reports the scale of the experiment (Eq.~(\ref{eq:sensitivityM})) in units of n$_{\beta\beta}$ (number of effective moles of isotope $\times$ yr).  
The status of the experiment, R (running), C (construction), D (development) is shown in the column labelled ``\emph{Status}''.
Sensitivities (in unit 10$^{25}$ yr) are evaluated according to Eqs.\ (\ref{eq:sensitivity}) and (\ref{eq:0sensitivity}) as appropriate, assuming 5 years running time. \amnu values (meV) are calculated using NME and phase space factors from \cite{ibm2} and \cite{kotila12} respectively. Asterisks label ZB conditions.
in the case of GERDA II we report two different sensitivities according to the two hypotheses discussed in Sec.~\ref{sec:semi}}
	\centering
	\begin{tabular}{@{}llclccccc@{}}
	\hline
			&Isotope 	&B$_{iso}$ &FWHM (keV) &$Perf.$	&$Sc.$	&$Status$&\SFz (5 yr)&\amnu\\
	\hline
	CUORE0\cite{cuore0}
		&$^{130}$Te	& 266	&5.6	&0.2	&66	&R	&1.5	&224\\
	CUORE\cite{qprop05,arnaboldi04,bandac08}
		&$^{130}$Te	& 36	&5	&27	&1390	&C	&21	&60\\
	GERDA I\cite{GERDA-0n}			
		&$^{76}$Ge	& 21	&4.8	&9.2	&119	&R	&9.4	&165\\
	GERDA II\cite{GERDA-dete,abt04,janicsko09}
		&$^{76}$Ge			&20/1.1	&3.2	&5.7/0.3	&328  	&C		&22/60*	&107/65*\\
	LUCIFER\cite{Lucifer}
		&$^{82}$Se			&1.9	&13	&2.7	&86  	&D		&16*	&76*\\		
	MJD\cite{Maj-white,MajD2011,TAUPmajorana,guiseppe08}
		&$^{76}$Ge			&0.9	&4	&0.4	&238	&C	&44*	&77*\\
	SNO+\cite{TAUPsno+}	
		&$^{130}$Te	&0.9	&240	&27	&1253	&D	&20 &62\\
	EXO\cite{EXO-0n}
		&$^{136}$Xe	&1.9	&96	&30	&482	&R &12	& 97\\
	SND\cite{SNEMO,SNEMO-angular,NEMOtaup}
		&$^{82}$Se	&0.6	&120	&18	 &23	&D	&3.3	& 166\\
	SuperNEMO\cite{SNEMO,SNEMO-angular,NEMOtaup}
		&$^{82}$Se	&0.6	&130	&20	 &366	&D	&13	& 85\\
	KamLAND-Zen\cite{KamZen2012,KamZen2013}
		&$^{136}$Xe	&7.4	&243	&243	&1320	&R	& 6.9	&127\\
	NEXT\cite{NEXT-100,gomez12}	
		&	$^{136}$Xe		&0.8	&13	&5.4	&165		&D	& 16 & 82\\ 
\hline
\end{tabular}
\end{sidewaystable}

On the contrary, the extremely massive scintillators are found to be very effective in reaching very low (external) background rates but are irreducibly limited on the performance side by poor energy resolution (which widens the ROI, thus increasing the \BBd background). 

\begin{figure}
\begin{center}
\includegraphics[width=0.9\textwidth]{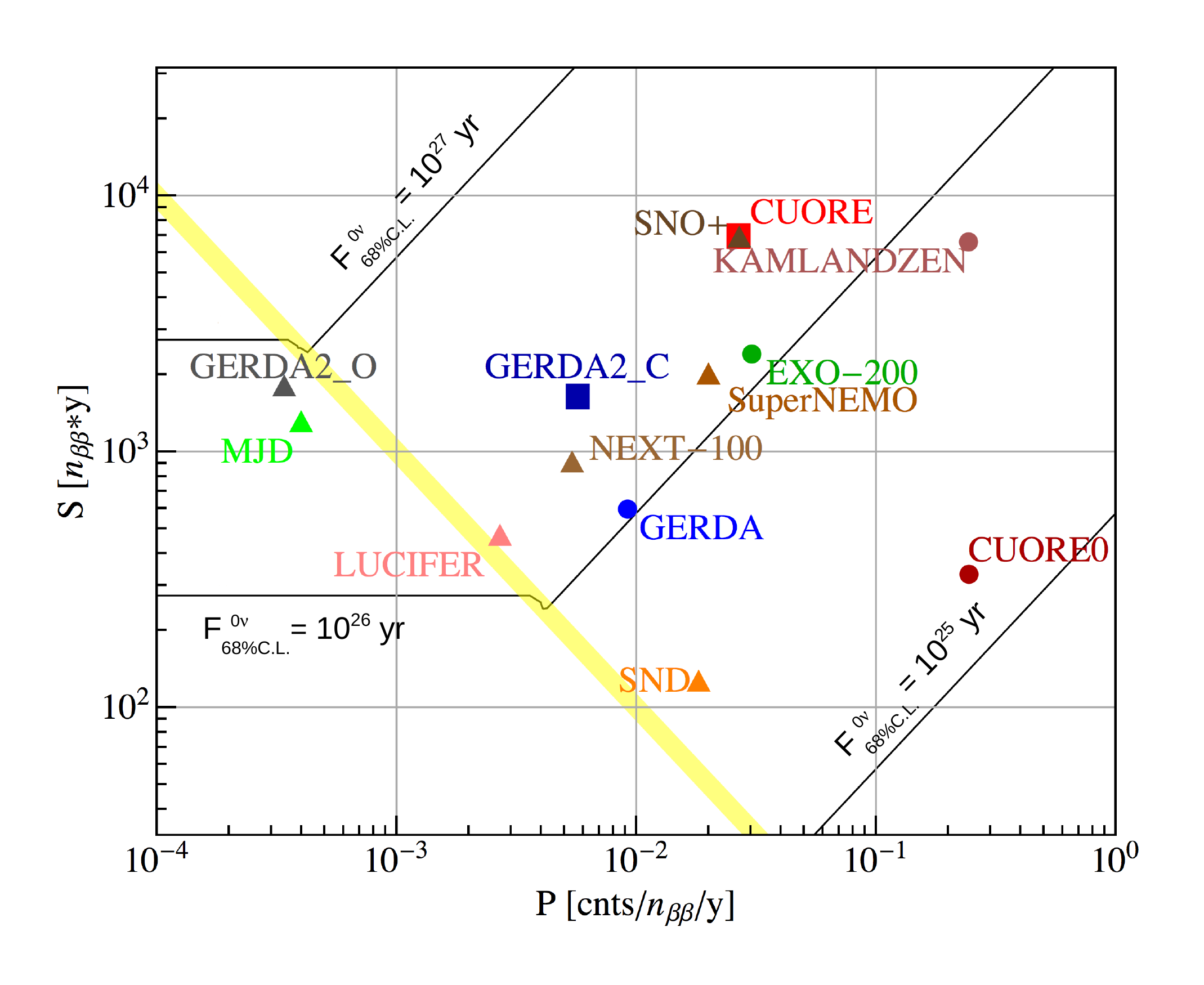}%
\caption{Comparison of the sensitivities of future \BBz projects (Figure from \cite{biassoni13}). Iso-sensitivities curves are shown in the \emph{Scale} vs. \emph{Performance} plane (see text for the definition of these two parameters). The yellow thick line highlights the transition to the zero background region (which is reached by Gerda II in the optimistic background configuration, MJD and Lucifer). Data are from Table.~\ref{tab:BBfutprm}. }
\label{fig:bcg}
\end{center}
\end{figure}

This situation is pictorially summarized in Fig.~\ref{fig:bcg} where it is apparent how future projects tend to align along the 10$^{26}$ yr \emph{iso-sensitivity} line, though spanning large intervals in \emph{Performance} and \emph{Scale}. These two parameters, defined in Sec.~\ref{sec:expmet} through Eqs.~(\ref{eq:sensitivityM}) and (\ref{eq:0sensitivityM}), measure respectively the number of background events in the ROI (\emph{Performance}=$P$=$\Delta \cdot B'$) per unit exposure and the exposure itself (\emph{Scale}=$S$=$n_{\beta\beta}\cdot T$) measured in number of \BB moles per year.
It is important to point out that the ZB condition is dynamic and depends on the interplay between \emph{Performance} and \emph{Scale} to maintain the $P \times S \lesssim 1$ condition. 

Then the common goal should be to approach the golden region characterized by $P$\dot $S$\ca 1 where the sensitivity increases in the fastest way along the $P$\dot $S$ = 1 direction~\cite{biassoni13}. 
Indeed by improving the performance one can succeed in entering the ZB region. 
Then the sensitivity can be improved linearly by increasing the detector mass until the ZB condition is no longer satisfied

This is a nice picture which can translate suddenly in a nightmare. Actually performance improvements cannot be maintained easily (if not at all) with larger scales and intermediate projects (demonstrators) are becoming a rule. 
Moreover, all the new generation experiments tend to sit far away (on opposite sides) from the golden region.

Demonstrators (SND, MJD, Lucifer) are paving the road to larger future projects while new ideas are being verified in a number of R\&D programs. 
The future of the \BBz experimental search depends critically on the richness and variety of the technologies under development. The most successful ones will turn quickly into real experiments characterized by improved sensitivities and capabilities.
 
Let us summarize the situation by considering just the very few projects characterized by the best conditions for impacting the future of \BBz research: CUORE, GERDA, EXO, SNO+ and KamLAND-Zen. An important impact is expected also from the demonstrators SND, the scintillating bolometers, MJD, EXO and NEXT, whose target is to assess the readiness and effectiveness of the respective techniques.
Altogether, these experiments represent the most advanced effort to guarantee the highest possible sensitivity study of the maximum number of different nuclei with different experimental techniques and approaches. 

The future of \BBz searches depends critically on the actual ordering of the neutrino masses. 
In case Nature has selected the quasi degenerate hierarchy (i.e. \amnu \ca 100--500~meV) then the \ges claim could be confirmed by GERDA. The signal could be cross-checked in \xe by EXO, KamLAND-Zen (if the background problems are solved) and NEXT (if the results with the prototypes are confirmed).  CUORE and SNO+ could detect \BBz in $^{130}$Te while a large scale array of scintillating bolometers could have chances to observe the signal in  $^{82}$Se or $^{100}$Mo. On the other hand SuperNEMO could get more insight into the decay mechanism looking at the single-electron energy and angular distributions in $^{82}$Se or $^{150}$Nd.
The redundancy of the candidates under study will reduce the uncertainties coming from NME calculations.

As mentioned above, this optimistic scenario is already in tension after the results of EXO-200. On the other hand GERDA-I results, expected shortly, will further clarify the situation.

In the case of the inverted hierarchy (i.e. \amnu \ca 20 - 50 meV) the observation of \BBz is still possible if \amnu is hidden just below the upper part of the error bars or if the projects under development will be able to achieve their planned sensitivity. 
CUORE (most likely enriched in \tect) or bolometric evolutions with improved reduction of the surface background have good chances to detect \BBz but nEXO, the extension of EXO-200 under discussion, could also succeed in \xen.
In case of success of their present phase, extensions of SNO+, KamLAND-Zen and NEXT, could have the chance to cross-check the result in \tect and \xen, while GERDA-III, after merging with MAJORANA, could observe a signal in \ges.

The discovery of \BBz for three or four isotopes is necessary for a convincing evidence. This should be possible thanks to the variety of projects and techniques under development. 

It is worth to stress that also the missed observation of \BBz could be very important for neutrino physics. 
Indeed, if the long baseline neutrino oscillation experiments would provide evidence for an inverted neutrino hierarchy, then a limit on \amnu below the inverted hierarchy band  would be a strong indication in favor of a Dirac nature of neutrino.

No present or future \BBz project seems to have any chance to probe the direct hierarchy region. The study of \amnu in the range of few meV needs new revolutionary strategies. R\&D activities are crucial to stimulate the new ideas needed to face this extreme challenge.

To conclude, \BBz searches are living a very exciting period characterized by a lot of enthusiasm for the possibility to finally observe this very rare decay. A lot of projects have been proposed either to exploit the capabilities of present technology or to pave the road to next generation experiments. Their sensitivity to \BBz half-lifetime is in the range of few 10$^{26}$ yr.

Long term predictions are not easy, but future generation experiments will unavoidably need a multi-tonne scale in the \BB isotope mass. It will then become difficult to maintain the present variety of experimental approaches. On the other hand, taking into account the past evolution of the \BBz experimental sensitivities, an improvement by an order of magnitude seems a likely frontier for future generation experiments on a scale of 10-20 years.

\end{document}